\documentclass{cimento}

\usepackage{amsmath}
\usepackage[usenames,dvipsnames]{xcolor} 
\usepackage{xcolor}
\usepackage{concmath}
\usepackage[T1]{fontenc}
\usepackage{graphicx}  
\usepackage{booktabs}
\usepackage{subfig}
\usepackage{textcomp}
\usepackage{multirow}
\usepackage{lineno}

\title{Ten Years of PAMELA in Space}

\author{The PAMELA collaboration \\

\noindent
O. Adriani\ETC\from{1}\from{2}, 
G. C. Barbarino\from{3}\from{4},
G. A. Bazilevskaya\from{5}, 
R. Bellotti\from{6}\from{7}, 
M. Boezio\from{8}, 
E. A. Bogomolov\from{9},
M. Bongi\from{1}\from{2}, 
V. Bonvicini\from{8}, 
S. Bottai\from{2}, 
A. Bruno\from{6}\from{7},
F. Cafagna\from{7}, 
D. Campana\from{4}, 
P. Carlson\from{10}, 
M. Casolino\from{11}\from{12},
G. Castellini\from{13}, 
C. De Santis\from{11}, 
V. Di Felice\from{11}\from{14}, 
A. M. Galper\from{15},
A. V. Karelin\from{15}, 
S. V. Koldashov\from{15}, 
S. Koldobskiy\from{15}, 
S. Y. Krutkov\from{9}, 
A. N. Kvashnin\from{5}, 
A. Leonov\from{15}, 
V. Malakhov\from{15}, 
L. Marcelli\from{11},
M. Martucci\from{11}\from{16}, 
A. G. Mayorov\from{15}, 
W. Menn\from{17}, 
M. Merg\`e\from{11}\from{16},
V. V. Mikhailov\from{15}, 
E. Mocchiutti\from{8}, 
A. Monaco\from{6}\from{7}, 
R. Munini\from{8}, 
N. Mori\from{2}, 
G. Osteria\from{4},
B. Panico\from{4}, 
P. Papini\from{2}, 
M. Pearce\from{10}, 
P. Picozza\from{11}\from{16}, 
M. Ricci\from{18}, 
S. B. Ricciarini\from{2}\from{13}, 
M. Simon\from{17}, 
R. Sparvoli\from{11}\from{16}, 
P. Spillantini\from{1}\from{2}, 
Y. I. Stozhkov\from{5}, 
A. Vacchi\from{8}\from{19}, 
E. Vannuccini\from{1}, 
G. Vasilyev\from{9}, 
S. A. Voronov\from{15}, 
Y. T. Yurkin\from{15}, 
G. Zampa\from{8} and 
N. Zampa\from{8}}

\instlist{
\inst{1} University of Florence, Department of Physics, I-50019 Sesto Fiorentino, Florence, Italy
\inst{2} INFN, Sezione di Florence, I-50019 Sesto Fiorentino, Florence, Italy
\inst{3} University of Naples ``Federico II'', Department of Physics, I-80126 Naples, Italy
\inst{4} INFN, Sezione di Naples, I-80126 Naples, Italy
\inst{5} Lebedev Physical Institute, RU-119991 Moscow, Russia
\inst{6} University of Bari, I-70126 Bari, Italy
\inst{7}  INFN, Sezione di Bari, I-70126 Bari, Italy
\inst{8}  INFN, Sezione di Trieste, I-34149 Trieste, Italy
\inst{9}  Ioffe Physical Technical Institute, RU-194021 St. Petersburg, Russia
\inst{10} KTH, Department of Physics, and the Oskar Klein Centre for Cosmoparticle Physics,
AlbaNova University Centre, SE-10691 Stockholm, Sweden
\inst{11} INFN, Sezione di Rome ``Tor Vergata'', I-00133 Rome, Italy 
\inst{12} RIKEN, Advanced Science Institute, Wako-shi, Saitama, Japan
\inst{13} IFAC, I-50019 Sesto Fiorentino, Florence, Italy
\inst{14} Agenzia Spaziale Italiana (ASI) Science Data Center, I-00044 Frascati, Italy
\inst{15} National Research Nuclear University MEPhI, RU-115409, Moscow, Russia
\inst{16} Department of Physics, University of Rome ``Tor Vergata'' I-00133 Rome, Italy
\inst{17} Universitaet Siegen, Department of Physics, D-57068 Siegen, Germany
\inst{18} INFN, Laboratori Nazionali di Frascati, Via Enrico Fermi 40, I-00044 Frascati, Italy
\inst{19} University of Udine, Department of Mathematics and Informatics, I-33100 Udine, Italy
}

\PACSes{\PACSit{96.50.S-}{Cosmic rays}
\PACSit{94.20.wq} {Solar radiation and cosmic ray effects}
\PACSit{95.55.-n}{Astronomical and space-research instrumentation}
\PACSit{95.55.vj}{Neutrino, muon, pion, and other elementary particle detectors; cosmic ray detectors}}

\begin{document}

\maketitle

\begin{abstract}
The PAMELA cosmic ray detector was launched on June 15$^{th}$ 2006 on board 
the Russian Resurs-DK1 satellite, and during ten years of nearly continuous data-taking 
it has observed new interesting features in cosmic rays (CRs). In a decade of operation 
it has provided plenty of scientific data, covering different issues related to cosmic
ray physics. Its discoveries might change our basic vision of the mechanisms
of production, acceleration and propagation of cosmic rays in the Galaxy. 
The antimatter measurements, focus of the experiment,  have set strong constraints
to the nature of Dark Matter. Search for signatures of more exotic processes (such 
as the ones involving Strange Quark Matter) was also pursued.

Furthermore, the long-term operation of the instrument had allowed a constant monitoring
of the solar activity during its maximum and a detailed and prolonged study of the solar
modulation, improving the comprehension of the heliosphere mechanisms. PAMELA had also 
measured the radiation environment around the Earth, and it detected for the first time
the presence of an antiproton radiation belt surrounding our planet.

The operation of Resurs-DK1 was terminated in 2016. In this article we will review
the main features of the PAMELA instrument and its constructing phases. Main part of
the article will be dedicated to the summary of the most relevant PAMELA results over
a decade of observation.
\end{abstract}

 \tableofcontents

\section{The PAMELA mission}

The PAMELA (a Payload for Antimatter-Matter Exploration and Light-nuclei Astrophysics) satellite 
experiment was designed to study the charged component of the cosmic radiation, focusing on antiparticles. 
PAMELA was launched with a Soyuz-U rocket on June 15$^{\text{th}}$ of 2006 from the Baikonur cosmodrome (Kazakhstan).
The apparatus has been hosted on the Russian Resurs-DK1 satellite, a commercial Earth-observation spacecraft.
At first the orbit was elliptical (altitude varying between 355 and 584 km) and semipolar (inclination of about
70$^{\circ}$) and with a period of about 94 minutes. In 2010 the orbit was set to be circular with an almost 
fixed altitude of about 550 km. The mission, which was planned to last 3 years, was prolonged and 
finally PAMELA lifetime was remarkably long, lasting for about 10 years.

This experiment represents the greatest effort of the WiZARD Collaboration which, starting 
almost 30 years ago with the New Mexico State University group led by the late Robert L. Golden 
(July 28$^{\text{th}}$ 1940 - April 7$^{\text{th}}$ 1995), has successfully built and flown several balloon-borne
experiments like MASS-89, MASS-91, Tramp-Si93, CAPRICE-94 and CAPRICE-98 and also satellites 
like NINA-1 (1998) and NINA-2 (2000) and small experiments aboard
the space stations MIR and ISS (the series of missions Sil-Eye and ALTEA).  
The WiZARD group has also pioneered some innovative detection techniques such as electromagnetic
calorimeters, superconducting magnet spectrometers and ring imaging \u{C}erenkov detectors in 
balloon flights. A representative selection of articles by the WiZard collaboration are summarized
in \cite{wiz1,wiz2,wiz3,Barbiellini,wiz4,wiz5,Boezio,wiz6,Berg,wiz7,wiz8,wiz9,wiz10}.

The PAMELA collaboration, an international team comprising Italian (Universities and  Istituto 
Nazionale di Fisica Nucleare I.N.F.N. Structures), German (Universitaet Siegen), Russian (Lebedev 
Physical Institute, Ioffe Physical Technical Institute, National Nuclear Research University MEPhI)
and Swedish (KTH Royal Institute of Technology) institutes, was formed in the late 1990 with the goal
to make a satellite-borne experiment with very high sensitivity and excellent particle identification
capability. After several years of detector and satellite developments, a compact permanent magnetic 
spectrometer with a microstrip silicon tracker, a set of time-of-flight scintillators, a 16 radiation
length thick electromagnetic calorimeter, an anticoincidence system and a neutron detector were proposed
and accepted as a piggyback experiment on a Russian Resurs satellite.

The PAMELA main scientific objectives are to measure the antiproton spectrum up to 200 GeV, 
the positron spectrum up to 200 GeV, the electron spectrum up to 600 GeV, the proton and helium 
nuclei spectra up to 1.2 and 0.6 TeV/n respectively and the nuclei spectra (from Li to O) up to 
$\sim$100 GeV/n but also to search for antinuclei (with a $\overline{\textnormal{He}}$/He sensitivity
of 10$^{-7}$), new forms of matter, e.g. strangelets and finally to detect possible structures in 
cosmic ray spectra from e.g. Dark Matter or new astrophysical sources. Furthermore PAMELA is well 
suited to conduct studies of cosmic-ray acceleration and propagation mechanisms in the Galaxy, solar
modulation effects, the emissions of Solar Energetic Particles (SEPs) inside the heliosphere and 
investigate the particles in the Earth's magnetosphere.

The  PAMELA instrument as a whole, including sub-detectors, is described in Section 2. The Resurs 
satellite and Ground Segment are presented in Section 3. Qualification tests of the apparatus are 
shown in Section 4. PAMELA main scientific results 
are presented in Section 5, divided in several subsections to better explain the different topics 
and data analysis methods.  Conclusions are finally drawn in Section 6.

\begin{figure}[t]
\begin{center}
\includegraphics[width=10.cm]{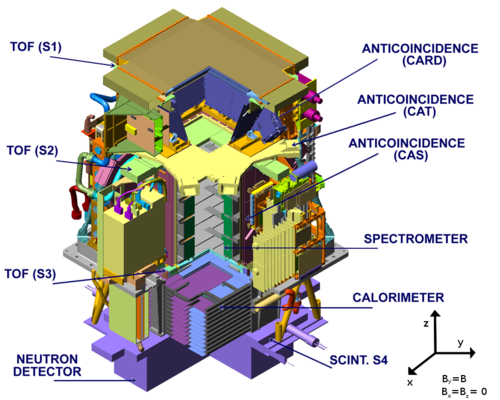}
\caption{The PAMELA instrument: a schematic overview of the
apparatus. } \label{pam1}
\end{center}
\end{figure}

\section{The PAMELA apparatus}
\label{sec:pamela}

The PAMELA apparatus is composed of the following subdetectors,
arranged as shown in Figure \ref{pam1}:
{ a Time-of-Flight system (ToF: S1, S2, S3)},  
{ an anticoincidence system (CARD, CAT, CAS)},
{ a permanent magnet spectrometer},
{ an electromagnetic calorimeter}, 
{ a shower tail catcher scintillator (S4)}, 
{ a neutron detector}.
The apparatus is $1.3$ m high, has a mass of $470$ kg  and
an average power consumption of $355$ W. The acceptance of the 
instrument  is $21.5$ cm$^2$sr and the maximum detectable 
rigidity is $1$ TV. Spillover effects limit the upper detectable 
antiparticle momentum to $\sim$190 GeV/c ($\sim$270 GeV/c) for
antiprotons (positrons). 

\vspace{.3cm}

PAMELA sub-detectors are described with more detail in the following paragraphs.
However, a fully comprehensive description can be found in \cite{astroparticle}. 

\subsection{The Time-of-Flight (ToF) system }

\label{tof}
\begin{figure}[t]
\begin{center}
\includegraphics[width=6cm]{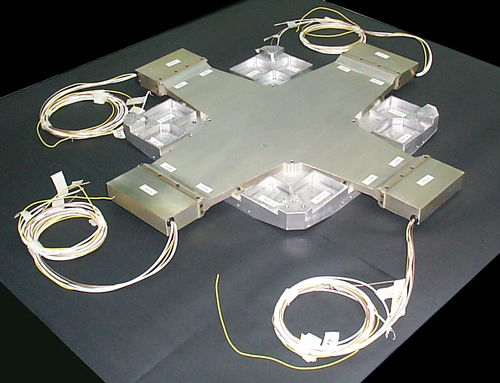} \includegraphics[width=70mm]{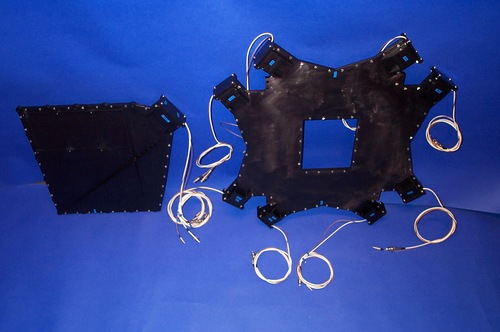}
\caption{Left: A picture of the S2 plane of the Time-of-Flight system. The sensitive area 
is $15\times18$ cm$^2$ segmented into $2\times2$ orthogonal bars. Right: An overview of 
the anticoincidence system. The CARD system is not shown but the design closely follows that
of CAS. The CAS scintillator (on the left) is approximately 40 cm tall and 33 cm wide. The hole
in the CAT scintillator (on the right) measures approximately 22 cm by 18 cm.} \label{fig:tof}
\end{center}
\end{figure}

The ToF system \cite{ost04a} comprises 6~layers of fast plastic scintillators
arranged in three planes (S1, S2 and S3),
with alternate layers placed orthogonal to each other. In Figure \ref{fig:tof}, 
left, an image of S2 is shown.
The distance between S1 and S3 is 78.3~cm. 
The sensitive area of each of the two S1 layers is
(330$\times$408)~mm$^{2}$ with the first layer divided into 8~bars
and the second layer divided into 6~bars. The total sensitive area
of the S2 and S3 planes is (150$\times$180)~mm$^{2}$ segmented
into 2$\times$2 and 3$\times$3 orthogonal bars, respectively. The
S1 and S3 layers are 7~mm thick while the S2 layers are 5~mm
thick. There are 24 scintillator bars in total. Both ends of each
scintillator bar are glued to a one-piece light guide which is
mechanically coupled to a photomultiplier 
by means of silicone optical pads of 3 and 6 mm. 
The S3 plane is mounted directly on the base plate of
PAMELA, while the other two planes are enclosed in light-proof
boxes suspended off the PAMELA structure. 

Time-of-Flight information for charged particles
passing between planes S1 and S3 is combined with track length
information derived from the magnetic
spectrometer  to determine particle velocities and reject albedo
particles. Ionisation (dE/dx) measurements in the scintillator layers 
allow the particle charge to be
determined at least up to Z$=$8. Coincidental energy deposits in combinations
of planes provide the main trigger
for the experiment. The segmentation of each plane allows
redundant studies of the trigger efficiency.

\subsection{Anticoincidence Systems}
\label{ac}

The PAMELA experiment contains two anticoincidence (AC)
systems \cite{ors05a}, as shown in Figure \ref{fig:tof}, right. The
primary AC system  consists of 4~plastic scintillators
(CAS) surrounding the sides of the magnet and one covering the top
(CAT). A secondary AC system consists of 4~plastic scintillators
(CARD) that surrounds the volume between the first two
Time-of-Flight planes. The AC systems use 8~mm thick plastic
scintillators read out by  PMTs.  Each CAS and CARD detector is read
out by two identical PMTs in order to decrease the possibility of
single point failure. The scintillators and PMTs
are housed in aluminium containers which provide light-tightness,
allow fixation to the PAMELA superstructure and ensure that a
reliable scintillator-PMT coupling is maintained. 

The efficiency of the large area CAS detectors has been studied
using an external drift chamber to map the spatial distribution of
incident cosmic ray muons. A detection efficiency for mips of
(99.91$\pm$0.04)\% was observed. The AC system has
also been tested by studying the backscattering of particles from the calorimeter during tests with high
energy particle beams. The robustness of the AC
system has been determined by studying the stability of the
scintillator-PMT coupling to variations in
temperature and the vibration spectra expected
during launch. 

\subsection{Magnetic Spectrometer}
\label{sec:spectrometer}

\begin{figure}[t]
\begin{center}
\includegraphics[width=6cm]{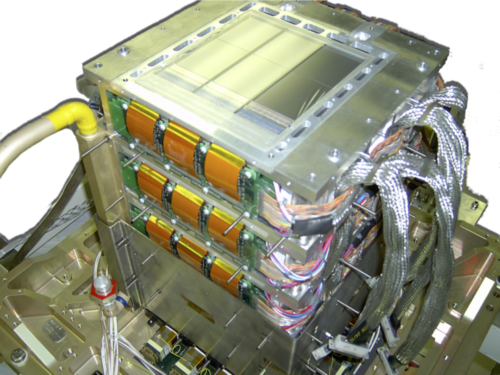}\hspace{.3cm}\includegraphics[width=6cm]{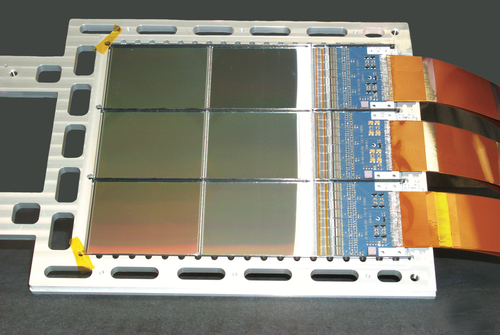}
\caption{Left: an overview of the magnetic spectrometer. Right: a
silicon plane comprising three silicon strip detectors, front-end
electronics and an ADC read-out board.} \label{hybrid}
\end{center}
\end{figure}

The central part of the PAMELA apparatus is a magnetic
spectrometer ~\cite{adr03} consisting of a permanent magnet and a
silicon tracker. The magnetic spectrometer is used to determine
the sign of charge and the rigidity (momentum /charge) of
particles up to $\sim$1~TV	. Ionisation loss measurements are
also made in the silicon planes, allowing absolute particle charge
to be determined at least up to Z=6.

The magnet is composed of five modules forming a tower 44.5~cm
high. Each module comprises twelve magnetic blocks, made of a
Nd-Fe-B alloy with a residual magnetisation of 1.3~T. The blocks
are configured to provide a uniform magnetic field oriented along
the y-direction inside a cavity of dimensions
(13.1$\times$16.1)~cm$^2$. The dimensions of the permanent magnet
define the geometrical factor of the experiment to be
21.5~cm$^2$sr. To allow precise rigidity measurements to be
obtained from the reconstructed particle trajectory, the magnetic
field has been measured with a Hall probe. The mean magnetic field inside the cavity is 0.43~T with a
value of 0.48~T measured at the centre. Any stray magnetic field
outside of the cavity can potentially interfere with the satellite
instruments and navigation systems. In order to attenuate the
stray field, the magnet is enclosed by ferromagnetic shielding.

Six equidistant 300~$\mu$m thick silicon detector planes are
inserted inside the magnetic cavity. The double-sided silicon
sensors provide two independent impact coordinates on each plane.
The basic detecting unit is the {\it ladder} which comprises two
sensors, (5.33$\times$7.00)~cm$^2$, assembled with a front-end
hybrid circuit, as shown in Figure \ref{hybrid}. Each plane is
built from three ladders that are inserted inside an aluminum
frame which connects to the magnet canister. In order to limit
multiple scattering in dead layers, no additional supporting
structure is present above or below the planes. Each high
resistivity n-type silicon detector is segmented into micro-strips
on both sides with p$^+$ strips implanted on the junction side
(bending-, x-view) and n$^+$ strips on the Ohmic side (y-view). In
the x-view, the implantation pitch is $25 \mu$m and the read-out
pitch is 50~$\mu$m. In the y-view, the read-out pitch is 67~$\mu$m
with the strips orthogonal to those in the x-view. The mip
efficiency for a single plane (including dead regions) is $90\%$.

Due to the random failure of a few front-end chips in the tracking system the
track reconstruction efficiency exhibit a significant time dependence. The efficiency was found to decrease over the years from a
maximum of $\sim 90\%$ in $2006$ to a $\sim 20\%$ at the end of $2009$. From $2010$ the efficiency was roughly 
constant at $20\%$ up to late $2015$.

\subsection{Electromagnetic Calorimeter}
\label{calorimeter}

\begin{figure}[t]
\begin{center}
\includegraphics[width=9cm]{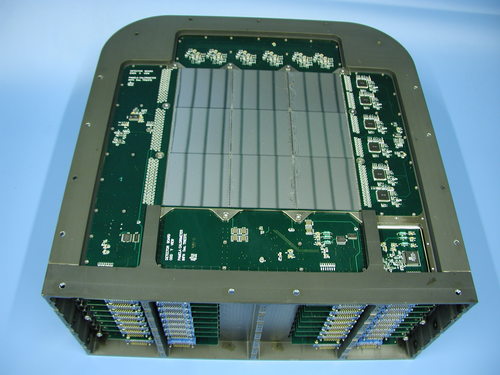} \caption{The PAMELA
electromagnetic calorimeter partially equipped with silicon and
tungsten planes. The device is $\sim$20~cm tall and the active
silicon layer is $\sim$24$\times$24~cm$^2$ in cross-section. } \label{calo}
\end{center}
\end{figure}

The sampling electromagnetic calorimeter comprises 44 single-sided
silicon sensor planes (380~$\mu$m thick) interleaved with
22~plates of tungsten absorber ~\cite{calo}. Each tungsten layer
has a thickness of 0.26~cm, which corresponds to 0.74~X$_0$
(radiation lengths), giving a total depth of 16.3~X$_0$
($\sim$0.6~nuclear interaction lengths).  The
(8$\times$8)~cm$^2$ silicon detectors are segmented into
32~read-out strips with a pitch of 2.4~mm. The silicon detectors
are arranged in a 3$\times$3 matrix and each of the 32~strips is
bonded to the corresponding strip on the other two detectors in
the same row (or column), thereby forming 24~cm long read-out
strips. The orientation of the strips of two consecutive layers is
orthogonal and therefore provide two-dimensional spatial
information ('views'). Figure \ref{calo} shows the calorimeter
prior to integration with the other PAMELA detectors.

The longitudinal and transverse segmentation of the calorimeter,
combined with the measurement of the particle energy loss in each
silicon strip, allows a high identification (or rejection) power
for electromagnetic showers. Electromagnetic and hadronic showers
differ in their spatial development and energy distribution in a
way that can be distinguished by the calorimeter. The calorimeter 
provides a rejection  factor of about 10$^5$ while keeping about 
90\% efficiency in selecting electrons and positrons. 

The calorimeter is also used to reconstruct the energy of the
electromagnetic showers. This has provided a measurement of the
energy of the incident electrons independent from the magnetic
spectrometer, thus allowing for a cross-calibration of the two
energy determinations. The calorimeter energy resolution is about 
$\sim 5.5\%$ for electromagnetic showers generated by particles 
with energy above 10 GeV up to an energy of several hundred GeV.

\subsection{Shower tail catcher scintillator}
\label{s4}

\begin{figure}[t]
\begin{center}
\includegraphics[width=6.5cm]{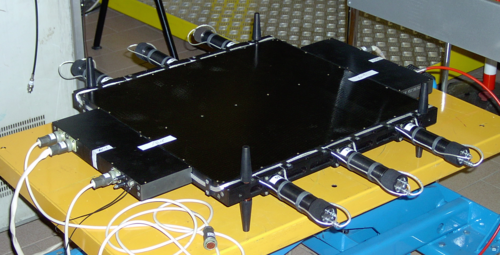}\hspace{8mm}\includegraphics[width=6cm]{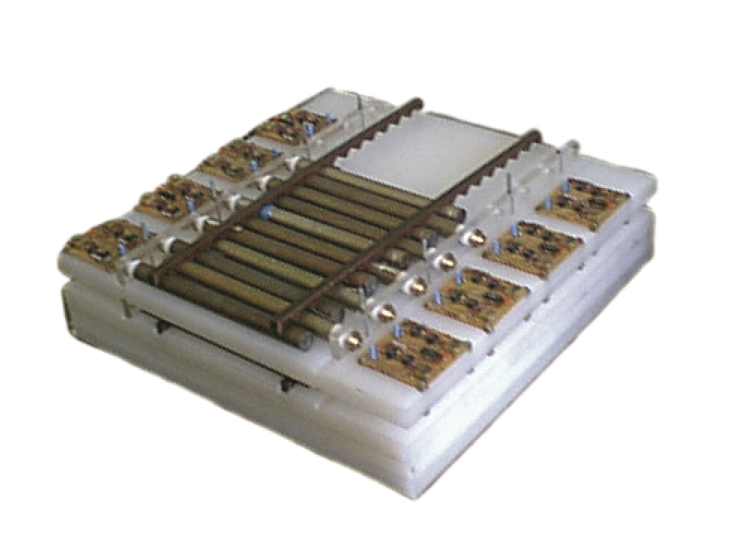}
\caption{Left: the shower tail catcher scintillator, S4, showing
the 6~PMTs used for read-out. Right: The neutron detector
partially equipped with $^3$He proportional counters.}
\label{nd_s4}
\end{center}
\end{figure}

The shower tail catcher scintillator (S4) improves the PAMELA
electron-hadron separation performance by measuring shower leakage
from the calorimeter. It also provides a high-energy trigger for
the neutron detector (described in the next section).
This scintillator is placed directly beneath the calorimeter.
It consists of a single square piece of scintillator of dimensions (482$\times$482)~mm$^2$
and 10~mm thick which is read out by six PMTs, as shown in Figure \ref{nd_s4} (left).
The detector has an overall  efficiency of ($99.97\pm 0.02$)\% and
a dynamic range of 1-1000~mip.

\subsection{Neutron detector}

The neutron detector \cite{nd} complements the electron-proton
discrimination capabilities of the calorimeter. The evaporated
neutron yield in a hadronic shower is 10--20 times larger than
expected from an electromagnetic shower. The neutron detector is
sensitive to evaporated neutrons which are thermalised in the
calorimeter. The detection efficiency (including thermalisation)
is $\sim$10\%. Joint analysis of the calorimeter and neutron
detector information allows primary electrons energies to be
determined up to several~TeV.

The neutron detector is located below the S4 scintillator and consists of 36 proportional counters,
filled with $^3$He and surrounded by a polyethylene moderator enveloped in a thin cadmium layer to 
prevent thermal neutrons entering the detector from the sides and from below. 
The counters are stacked in two planes of $18$ counters, oriented along the y-axis of the instrument. 
The size of the neutron detector is ($600 \times 550 \times 150$)~mm$^3$ and is shown in Figure \ref{nd_s4}, right.

\vspace{1cm}

More details about PAMELA instrument, including the data acquisition and the trigger system, are given in \cite{astroparticle}. 

\section{The Resurs DK1 satellite and the NTs OMZ ground segment}
\label{sec:dk1}

The Resurs DK1 satellite class is manufactured by the Russian space
company TsSKB Progress to perform multi-spectral remote sensing of
the Earth's surface and acquire  high-quality images in near
real-time. Data delivery to ground is realized via a high-speed
radio link.
The satellite  has a
mass of $\sim$6.7~tonnes and a height of 7.4~m. The solar array
span is $\sim$14~m. The satellite is three-axis stabilized with an axis orientation
accuracy of 0.2~arcmin and an angular velocity stabilization accuracy of 0.005$^\circ$/s.

\begin{figure}[t]
\begin{center}
\includegraphics[width=8.5cm]{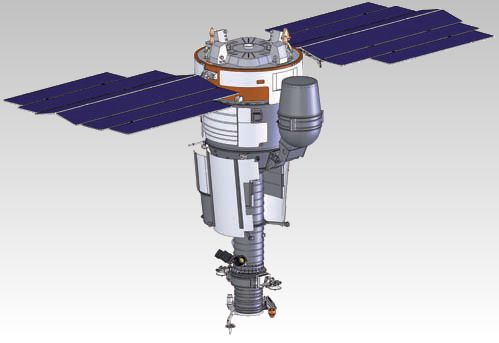} \includegraphics[width=4.4cm]{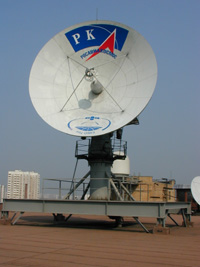}\caption{Left: A sketch of the
Resurs DK1 satellite which hosts the PAMELA experiment in a
Pressurized Container. Right: the reception antenna at NTs~OMZ.} \label{resurs-pc}
\end{center}
\end{figure}

On the Resurs-DK1 satellite employed for PAMELA, a dedicated Pressurized Container (PC)
was built and attached to the satellite to hold the instrument, as clearly visible in  
Figure \ref{resurs-pc} (left). The container is cylindrical in shape and has an inside
diameter of about 105~cm, a semi-spherical bottom and a conical
top. It is made of an aluminum alloy, with a thickness of 2~mm in
the acceptance of PAMELA. During launch and orbital
maneuvers, the PC is secured against the body of the satellite.
During data-taking it is swung up to give PAMELA a clear view into
space. 

PAMELA was successfully launched on June 15$^{\text{th}}$ 2006
from the cosmodrome of Baikonur in Kazakhstan by a Soyuz-U rocket.  
The orbit was non-sunsynchronous, with an altitude varying between
$\sim$ 350  km and   $\sim$ 600 km at an inclination of 70.0$^\circ$. 
In September 2010 the orbit was changed to a nearly circular one at an 
altitude of about 550 km, and it remained so until the end of the mission. 

PAMELA was first switched-on on the  June 20$^{\text{th}}$ 2006. 
After a brief period of commissioning, the instrument has been in a 
continuous data-taking mode since July 11$^{\text{th}}$ 2006. 

\begin{figure}[t]
\begin{center}
\includegraphics[width=13cm]{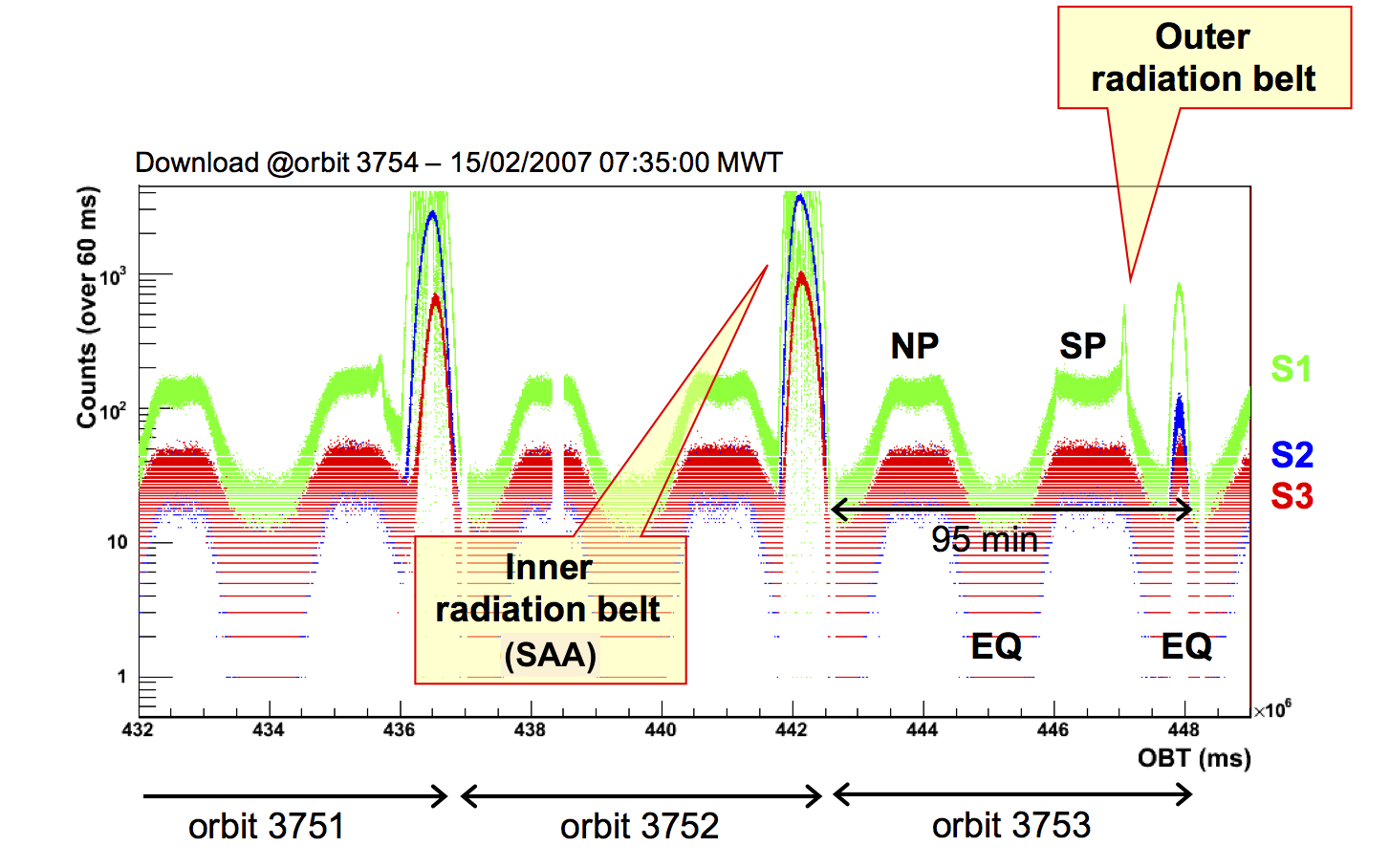} 
\caption{S1, S2 and S3 scintillation count rates registered for three consecutive orbits. During its revolution, 
PAMELA passes through different regions: polar regions (NP and SP),  equatorial 
regions (EQ), Inner and Outer Radiation Belts enriched with protons and electrons.
The highest peak is the South Atlantic Anomaly SAA.) } \label{trigger}
\end{center}
\end{figure}

Data downlinked
to ground has proceeded regularly during the whole mission. 
Figure \ref{trigger} shows PAMELA count rates as a function of time, for 
three sample orbits.  The rates are measured by S1, S2 and S3 scintillation counters. 
One orbit lasts about 94 minutes. During its revolution, PAMELA 
passes through different regions: polar regions (indicated as North Pole 
NP and South Pole SP in the Figure), with high count rate, equatorial regions
(EQ) with low count rate due to the geomagnetic cutoff which prevents low 
energy cosmic ray to approach the Earth, the Inner and Outer Radiation 
Belts enriched respectively with protons and electrons. In these regions 
the rate is maximum and saturates the onboard equipment; in particular, 
the most intense count rate is reached inside the South Atlantic Anomaly (SAA). 

\subsection{NTs OMZ ground segment}

The ground segment of the Resurs DK1 system is located at the
Research Center for Earth Operative Monitoring (NTs~OMZ) in
Moscow, Russia. 

The reception antenna at NTs~OMZ (Figure \ref{resurs-pc}, right) is a parabolic lector of 7~m
diameter, equipped with an azimuth-elevation rotation mechanism,
and has two frequency multiplexed radio channels. The Resurs DK1
radio link towards NTs~OMZ is active roughly 2-3 times a day. The
volume of data transmitted during a single downlink is currently
around 6 GBytes for a total of 15~GBytes/day. 

Data received from
PAMELA are collected by a data-set archive server. The server
calculates the downlink session quality (the error probability per
bit) and faulty downlink sessions can be assigned for
retransmission within $\sim$ 6 hours. The downlinked data are transmitted
to a server dedicated to data processing for instrument monitoring
and control, and is also written to magnetic tape for long-term
storage. All such operations are automatized to minimize the time
delay between the data reception and the extraction of monitoring
information.

After this first level of data analysis, both raw and first
processed data are moved through a normal Internet line to the
main storage centre in Easter Europe, which is located in MePHI
(Moscow, Russia). From this institution, by means of the GRID
infrastructure, raw and first level processed data are moved to
the main storage and analysis centre of the PAMELA collaboration,
located at CNAF (Bologna, Italy), a specialized computing centre
of INFN (Istituto Nazionale di Fisica Nucleare). Here data are accessible to all various institutions
within the PAMELA collaboration.

\section{Qualification tests}
\label{sec:qual}

As with any space mission, the PAMELA instrument had to overcome very demanding qualification 
tests before being able to go into orbit. In space missions, in fact, electronic components 
must be tested for radiation tolerance \cite{bos03}; the mechanics of the detector must be strong enough
to overcome the very high stresses due to the launch; thermal system must ensure perfect 
operation of the instrument despite the huge temperature range met in flight.  Finally, 
there must be no electromagnetic interference between the detector and the satellite's on-board instrumentation.

In this section the steps taken to qualify PAMELA for operation in space are reviewed.

\subsection{Mechanical and Thermal Qualification}

\begin{figure}[t]
\begin{center}
\includegraphics[width=6cm]{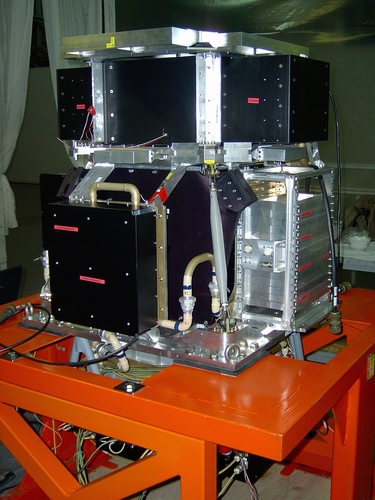} \includegraphics[width=5.6cm]{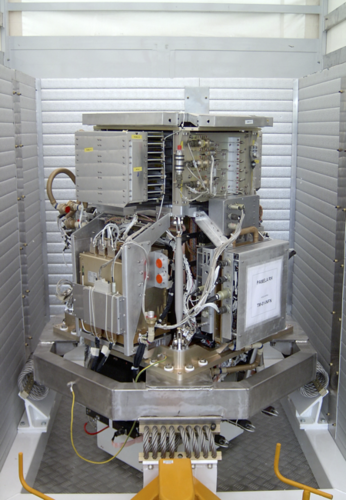}
\caption{Left panel: the PAMELA MDTM in the laboratories of TsSKB-Progress (December 2002). 
Right panel: the PAMELA Technological Model during transportation from
Italy to Russia (April 2004).} \label{iabg}
\end{center}
\end{figure}

The mechanical and thermal space qualification tests of the PAMELA
instrument were performed in the years 2002-2003. In order to
perform such tests, a mock-up of the entire instrument,
Mass-Dimensional \& Thermal Model (MDTM), was manufactured, shown in Figure \ref{iabg}, left. The
MDTM reproduces the geometrical characteristics of PAMELA (e.g.
dimensions, total mass, center of gravity, inertial moments) and
the basic thermal behavior. All particle detectors in the MDTM
were simulated by dummy aluminum boxes. The electronics systems
were non-functional and only reproduced the power consumption of
each subsystem.

In order to ensure that no damage occurs to PAMELA or the
spacecraft during any of the different operational phases of the
mission (transport, launch, orbital operations, unlocking of the
Pressurized Container, flight), the MDTM was exposed to
vibration spectra at mechanical loads exceeding those expected during the mission.
The MDTM vibration tests were performed at IABG Laboratories (Munich, Germany) in August
2002.
During the test it was verified that structural integrity was
maintained and that there was no change in the dynamic behavior of
MDTM (using resonance searches). The MDTM structure was subjected
to the required vibration loads along three orthogonal axes.
Additional transport, vibration and shock tests of the MDTM whilst
integrated into the Pressurized Container were performed at the
TsSKB-Progress Testing Center in May~2003. Additional
information about PAMELA mechanical space qualification can be
found in \cite{sparvoli1}.

\vspace{1cm}

The PAMELA thermal system consists of a $8.6$ m long pipe that
joins 4~radiators and 8~flanges. The task of this system is to
dissipate the heat produced by the PAMELA subsystems and transfer
it into the spacecraft. This transfer is performed by means of a
heat-transfer fluid pumped by Resurs satellite through the PAMELA
pipelines. The total heat release of PAMELA cannot exceed 360~W.

Thermal and vacuum tests of the PAMELA MDTM were performed in the
laboratories of TsSKB-Progress in April~2003. Six thermal modes of
operation were implemented, where the relevant parameters
which regulate the instrument thermal behaviour were varied between the design extreme
to simulate in-flight operations. Each mode persisted until a
steady state condition was reached.

The qualification test of the PAMELA thermal system showed that
all parameters of the system stayed within the design limits
(5$^\circ$C - 40$^\circ$C). During the Resurs DK1 orbit, the
operating temperature range of PAMELA varied between 7$^\circ$C
for the coldest systems and 38$^\circ$C for the warmest ones.
Additional information about PAMELA thermal space qualification
can be found in \cite{sparvoli1}.

\begin{figure}[t]
\begin{center}
\includegraphics[width=7.cm]{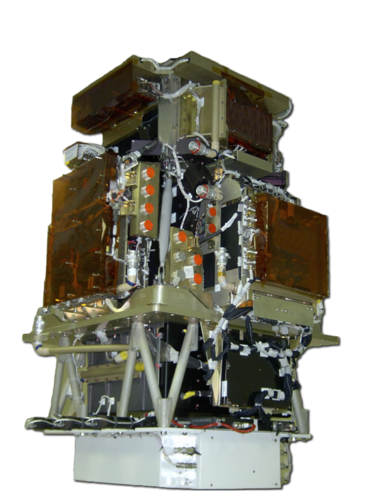}\hspace{2mm}\includegraphics[width=5.5cm]{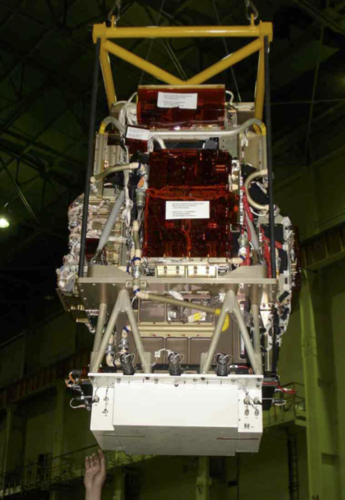}
\caption{The Flight Model instrument. Left:  a photograph taken just prior delivery to Russia
for integration with the Resurs DK1 satellite. Right: during integrations tests in Baikonur. } \label{pam}
\end{center}
\end{figure}
\subsection{Electrical tests}

To perform tests of the electrical interface between
PAMELA and the spacecraft, a second mock-up of the
PAMELA instrument was assembled. This "Technological
Model" was an exact copy of the Flight Model from the
point of view of electrical connections to the satellite and
for the readout electronics boards, with the particle detectors
substituted by dummies. 

The Technological Model
was shipped to TsSKB-Progress in April 2004 (see
Figure \ref{iabg}, right). The task of the Technological Model was to thoroughly
test the electrical interface to the Resurs DK1 satellite.
In addition, it was used to check that the residual
magnetic field from the PAMELA spectrometer did not
interfere with the Resurs instrumentation. These complex
tests proceeded in phases. A first test was performed in
Rome in December 2003, with the satellite emulated by a
Ground Support Equipment (EGSE) system. A second test
started in May 2004 at TsSKB-Progress and verified the
powering procedures. In October 2004 the PAMELA
Technological Model was fully integrated into the Resurs
DK1 to complete all remaining tests.

\subsection{Beam tests}

During its construction phase, PAMELA was tested three times
with beams of protons and electrons at the CERN SPS accelerator,
to study the performance of the sub-detectors with relativistic
particles. A preliminary integrated flight model set-up
consisting of the tracker and anticounting system and calorimeter
was exposed to protons (200 GeV $\div$ 300 GeV) and electrons (40
GeV $\div$ 300 GeV) at the CERN SPS in June 2002. Few months later 
the test was repeated with a more complete set-up.
Last test with
an almost complete Flight Model set-up has been carried out in
September 2003 again at SPS.

Because of the tight schedule of PAMELA integration, 
it was not possible to perform a beam test of the flight
model with light nuclei before its delivery to Russia in March
2005. For this reason, such a light-nuclei test was performed by
using prototypes of the PAMELA TOF and tracking system in a
dedicated mechanical arrangement, on February 2006 at the GSI
(Gesellschaft fuer Schwerionenforschung, Darmstadt, Germany)
beam accelerator.  The prototypes were exposed to $^{12}C$ beams. Targets
of aluminium and polyethylene were used to generate a
variety of fragmentation products. 

The  main aim of the test was the determination of the time
resolution of the TOF system and of the charge resolution of both 
the TOF and the tracking system for light nuclei. Results from this 
beam test are reported in \cite{gsi-campana}. 

\subsection{Integration of the PAMELA models and ground data}

Most of PAMELA sub detectors were built in the scientific labs of the 
PAMELA collaboration. The industrial side dealt of the mechanics of the
instrument, its thermal system and a significant portion of the electronics.

The construction of the instrument's parts has proceeded in parallel; once 
the different detectors were ready, they were delivered to the clean room of the 
University and INFN laboratories of Roma Tor Vergata, Rome, Italy. More specifically, 
two clean rooms (class 10000/ISO7 and 1000/ISO6) were built on purpose for PAMELA integration. 
The 3 versions of the instrument (Mechanical and Thermal, Electrical and Flight) were   
 assembled in these clean rooms, with the participation of the whole PAMELA team scheduled in shifts. 

In parallel with the mechanical and electronic integration,  the whole on-board 
software was implemented and tested in the clean rooms of the Tor Vergata laboratory. 
This work was done in close connection between researchers and industry. 

The integration of the Flight Model was completed at the beginning of 2005.
Prior the delivery to Russia (March 2005), 
the system was extensively tested with cosmic rays. 

\section{Review of PAMELA main scientific results} 

\begin{figure}[t]
\begin{center}
\includegraphics[width=1\textwidth]{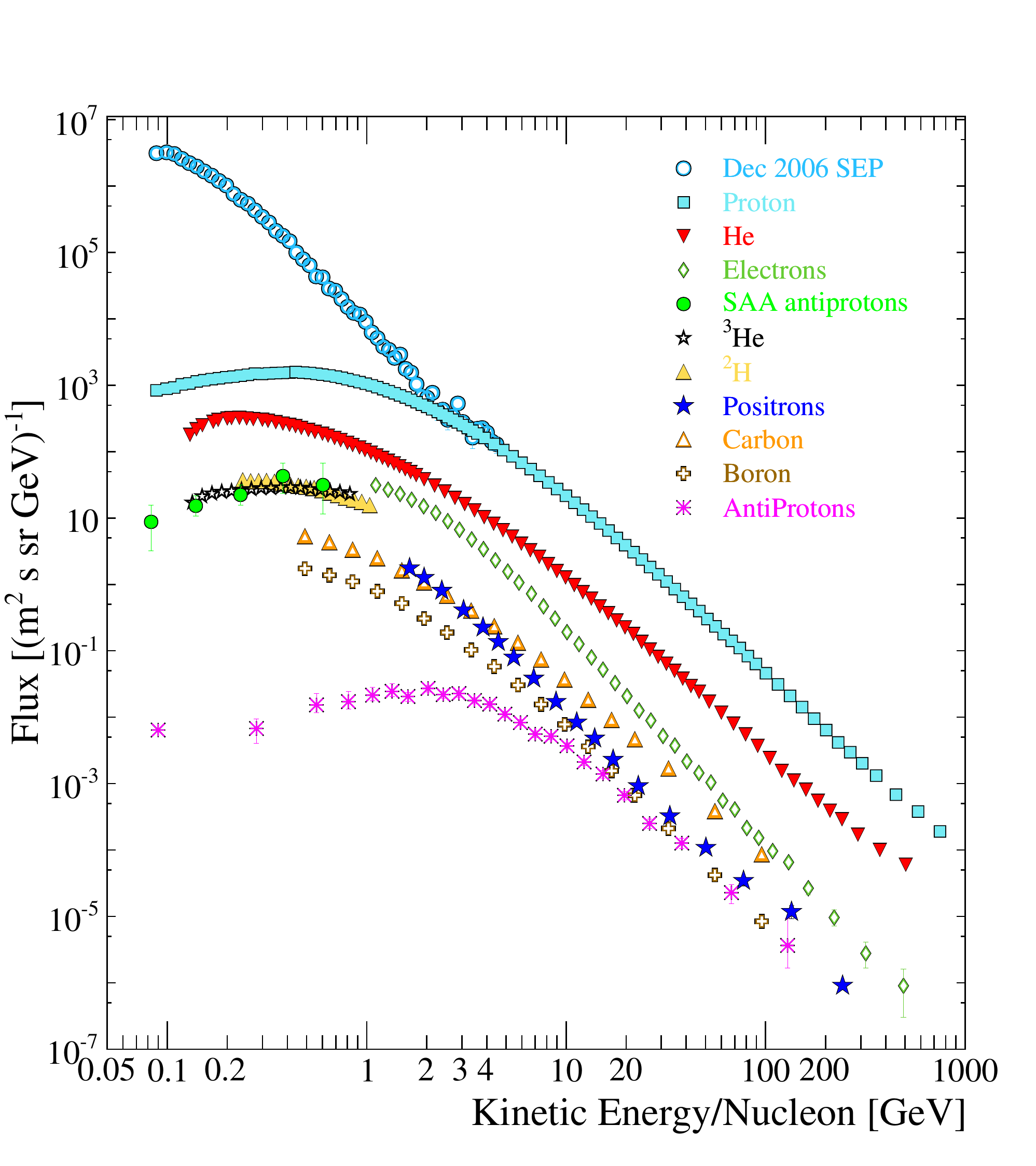}
\caption[]{Complete set of PAMELA measurements of cosmic rays of different origin: galactic particles and solar energetic 
particles, particles trapped in the Earth's magnetosphere. }
\label{tuttopamela}
\end{center}
\end{figure}

PAMELA contributed to open a new era of precision study of CR physics, 
providing the first comprehensive data set of individual particle spectra over a very broad energy interval.
Thanks to the data collected during ten years of operation, CR physics witnessed significant advancements both on 
the experimental and theoretical sides. Figure \ref{tuttopamela} shows the excellent capabilities of the instrument in the
detection of comic rays of different origin, in a wide energy interval. The energy spans $4$ decades and the flux extends over $12$
orders of magnitude.

In this following, PAMELA measurements 
will be reviewed. A compendium  of 
all PAMELA results after almost 8 year of operation has  already been presented in \cite{Adriani_PhysRep_2014};
in this second review  latest PAMELA measurements will be added, together with a comparison with 
complementary information by other running experiments. 

\subsection{Cosmic rays of galactic origin} 

In this section an overview of the PAMELA measurements related to the galactic CR components
is given and the comparison with other contemporary measurements is discussed.

\subsubsection{The antiparticle puzzle}

The study of the antimatter component of galactic CRs has attracted the scientific community since the first measurements of CRs 
antiparticles in the $60$s and $70$s \cite{Shong,Golden,Bogomolov}; this is an extremely intriguing scientific endeavor for the connection with fundamental physics
problems, such as the nature of cosmological dark matter or the asymmetry between matter and antimatter in our Universe. 
Antiparticles are produced by the inelastic interactions of CR nuclei with the interstellar medium.  
The production cross sections for antimatter particles is relatively low, so that they represent an unique observational channel to 
probe exotic phenomena due to the low astrophysical background. 
Until now only antiprotons and positrons have been detected in CRs.

The PAMELA instrument has been specifically designed to measure cosmic-ray antiproton and positron fluxes,
over a wider energy range and with a better accuracy than previous experiments. 

\begin{figure}
\begin{center}
\includegraphics[width=.9\textwidth]{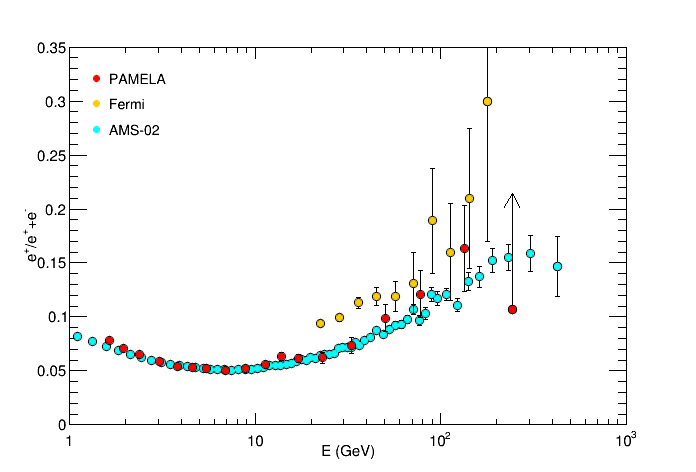}
\caption[]{Galactic cosmic ray positron fraction measured by PAMELA \cite{Adriani_poslast}, along with the recent 
AMS-02 \cite{AMS02-pos} and FERMI \cite{Ackermann12}  measurements. Only statistical errors are shown.}
\label{fig:poselera}
\end{center}
\end{figure}


The most striking result of PAMELA has been the so-called positron anomaly.
An unexpected rise in the positron-to-electron ratio at energies above 10 GeV has been clearly observed for the first 
time by PAMELA  \cite{Adriani_positrons, Adriani_poslast} (see Figure \ref{fig:poselera}). This is in contrast to expectation of 
all secondary production in standard models of CR propagation in the galaxy. 
These models, which treat the  CR propagation as a diffusive process, cannot explain 
PAMELA result if positrons have a pure secondary origin (e.g. 
\cite{Moskalenko09, Delahaye09}). 
  
A first confirmation of the results came few years later from the FERMI telescope which, by using the Earth magnetic field
to separate electrons and positrons, published a rising positron charge fraction above 20 GeV \cite{Ackermann12}, even though significantly
larger values than those measured by PAMELA.  

Conclusive confirmation of the positron anomaly came from the AMS-02 magnetic spectrometer that,  with  a significantly larger acceptance,
could measured the positron abundance up to 500 GeV  \cite{AMS02-pos}. 
The excellent agreement with PAMELA (see Figure \ref{fig:poselera}) dispelled any residual doubt about PAMELA 
positron results, therefore  providing a definitive evidence for a significant deviation from the standard scenario of 
origin and propagation of galactic electron and positron. 

\begin{figure}
\begin{center}
\includegraphics[width=.9\textwidth]{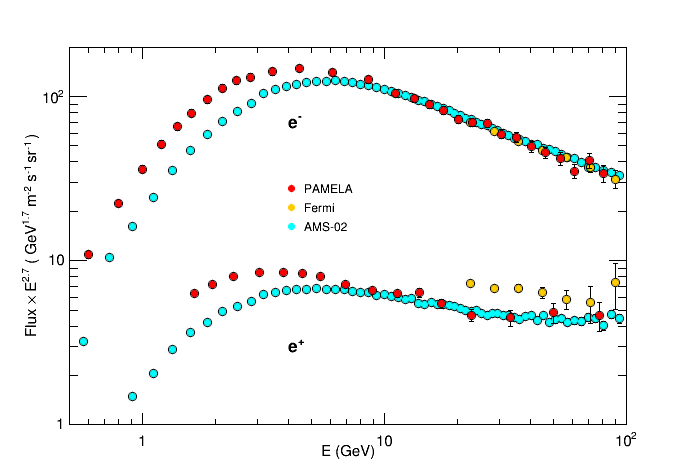}
\caption[]{Electron and positron energy spectra measured by PAMELA \cite{Adriani_elelast,Adriani_posfuture} along with other recent 
results from AMS02 \cite{AMS02-pos2} and FERMI \cite{Ackermann12}.  Only statistical errors are shown.}
\label{fig:posele}
\end{center}
\end{figure}

Though the particle ratio is extremely significant to detect possible deviations from standard positron and electron production,
the  most complete information comes from the individual fluxes of particles and antiparticles. 
The precise measurement of  the spectral shape of each component allows to disentangle different contributions. 
Figure \ref{fig:posele} shows the latest PAMELA results on the absolute fluxes of electrons
\cite{Adriani_elelast} and the preliminary
results on the new positron spectrum analysis \cite{Adriani_posfuture}. 
A re-analysis of electron data below 100 GeV yielded a  higher estimation of the absolute fluxes with 
respect to the  previously published results \cite{Adriani_ele}, most evident below $\sim$20 GeV. 
Above 10 GeV the results are in good agreement  with the AMS-02 \cite{AMS02-pos2} and FERMI \cite{Ackermann12} data, while the positron flux measured by  
FERMI \cite{Ackermann12}, also shown for comparison in Figure \ref{fig:posele}, is systematically higher than the other measurements. 
PAMELA results shown in the picture refer to data collected over the recent solar minimum activity period, from 
July 2006 to December 2009: the higher fluxes observed by PAMELA at low energies are consistent with this period of the solar cycle. 

From the analysis of the individual electron and positron spectra it emerges that, while the electrons mostly follow a power-law trend, the positron 
spectrum manifests a progressive hardening above 20 GeV, producing the increasing charge fraction observed in Figure \ref{fig:poselera}. 
Even if the exact  mechanism of CR propagation and secondary positron production is still not resolved (see eg. \cite{lipari}), the majority of 
the scientific community believes that the observed positron hardening  points to the existence of positron sources, whose nature 
is still a lively debated topic.  
Hundreds of models have been proposed as interpretations of the anomaly, either invoking annihilation/decay of DM particles 
\cite{DMPos,DMPos1,DMPos2,DMPos3} or nearby astrophysical
sources such as pulsars \cite{pulsars} or supernova remnants \cite{remnants,remnants1,remnants2}. 
However, the origin of the positron excess can not be identified/discerned solely on the basis of the spectral shape,   and additional 
information is needed. For example because of the local nature of astrophysical sources of positrons (and electrons), 
a certain amount of  anisotropy in the arrival directions  is expected.  
Anisotropy searches with PAMELA will be described in Sec. \ref{anisotrophy}.


\begin{figure}
\begin{center}
\includegraphics[width=.9\textwidth]{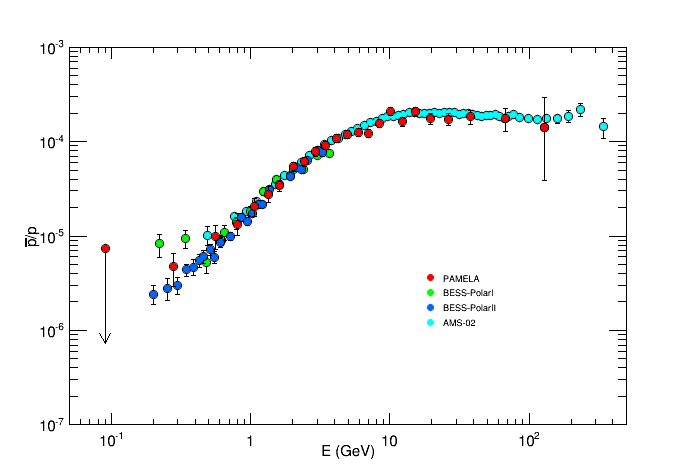}
\caption[]{Antiproton to proton ratio in galactic cosmic rays measured by PAMELA \cite{Adriani_pbar}
compared with AMS-02 \cite{AMS02-pbar} and BESS-Polar \cite{BESS-Polar1-pbar}  measurement. Only statistical errors are shown. 
}
\label{fig:pbarra}
\end{center}
\end{figure}
While the pulsar hypothesis predicts an overproduction of solely positron and electron couples, a significant constraints to the other models is
provided by the antiproton abundance. 
In fact, the same mechanisms of positron production that might take place in a DM halo or in supernova envelopes  would produce an excess of
antiprotons as well.

The PAMELA instrument is particularly suitable for antiproton identification. 
The antiproton measurement is extremely challenging, due to the overwhelming abundance of protons, combined with a  non-zero  
probability to misidentify their charge sign, which increases as the energy increases.
The PAMELA tracking system has been designed to limit the proton spillover background, so that antiprotons could be cleanly 
discriminated up to 200 GeV, as shown in reference \cite{Adriani_pbar}. 

\begin{figure}
\begin{center}
\includegraphics[width=.9\textwidth]{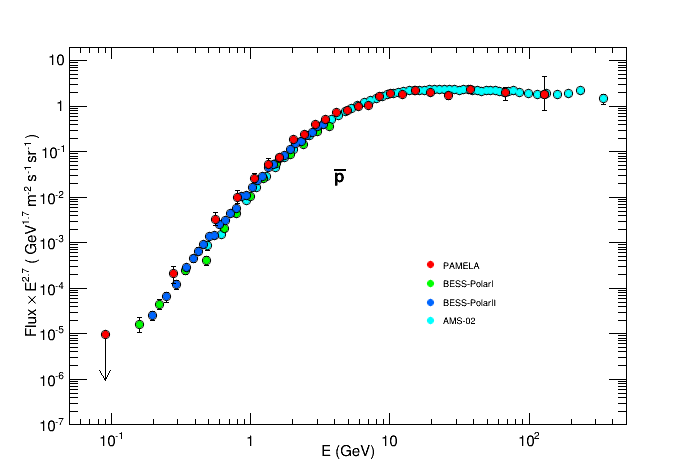}
\caption[]{Antiproton flux in galactic cosmic rays measured by PAMELA \cite{Adriani_pbar}, compared with AMS-02 \cite{AMS02-pbar} and BESS-Polar \cite{BESS-Polar1-pbar}  measurement. Only statistical errors are shown. }
\label{fig:pbarflu}
\end{center}
\end{figure}

Recently the antiproton abundance has been reported also by the AMS-02 collaboration \cite{AMS02-pbar}. 
The larger acceptance of AMS-02 allowed the extension of the antiproton measurement up to 450 GeV, with increased precision.
The excellent agreement between AMS and  PAMELA results can be appreciated in Figure \ref{fig:pbarra}, which shows the antiproton to proton ratio. 
 In the same figure 
the two BESS-Polar measurements are shown as well \cite{BESS-Polar1-pbar}. 
BESS-Polar has flown twice on long-duration balloons launched from Antartica, in 2004/2005 and 2007/2008 providing precise low measurements of the 
antiproton component. Below $1$ GeV differences in the antiproton to proton ratio may be due to charge-sing modulation effects \cite{BESSMOD} (see Section 
\ref{chargesign}).  
The corresponding measured absolute flux of antiproton from PAMELA, AMS-02 and BESS-Polar experiments is shown in Figure \ref{fig:pbarflu}.

Even if the subject is debate, no compelling evidence for a primary component emerges from 
the cosmic-ray antiproton spectrum (e.g. \cite{Cirelli}, see also \cite{evoli} for a recent review of antiproton secondary
production estimation). 
Furthermore this result excludes several models that might explain the observed positron anomaly.


After eight years from the first positron anomaly observation, the antiparticle component of galactic
cosmic rays remains an intriguing puzzle, still unsolved.
In order to establish the existence of new sources of antiparticles, the mechanism of secondary 
production in the interstellar medium has to be well understood and the models need to be better constrained. 

\subsubsection{The nuclear component}

PAMELA observation of the hydrogen and helium energy spectra produced another unexpected result.
Thanks to the high-resolution of the magnetic-spectrometer and a redundant system for element identification,
PAMELA performed a high precision measurement of the H and He
energy spectra up to $\sim$1 TeV  \cite{Adriani_Science}. 
These spectra showed  the first clear and unambiguous evidence for spectral structures below the knee, in the
form of a sudden hardening at a rigidity of about $230-240$ GV. Furthermore different slopes for H and He spectra 
were directly observed for the first time. 

\begin{figure}[p]
\begin{center}
\includegraphics[width=.98\textwidth]{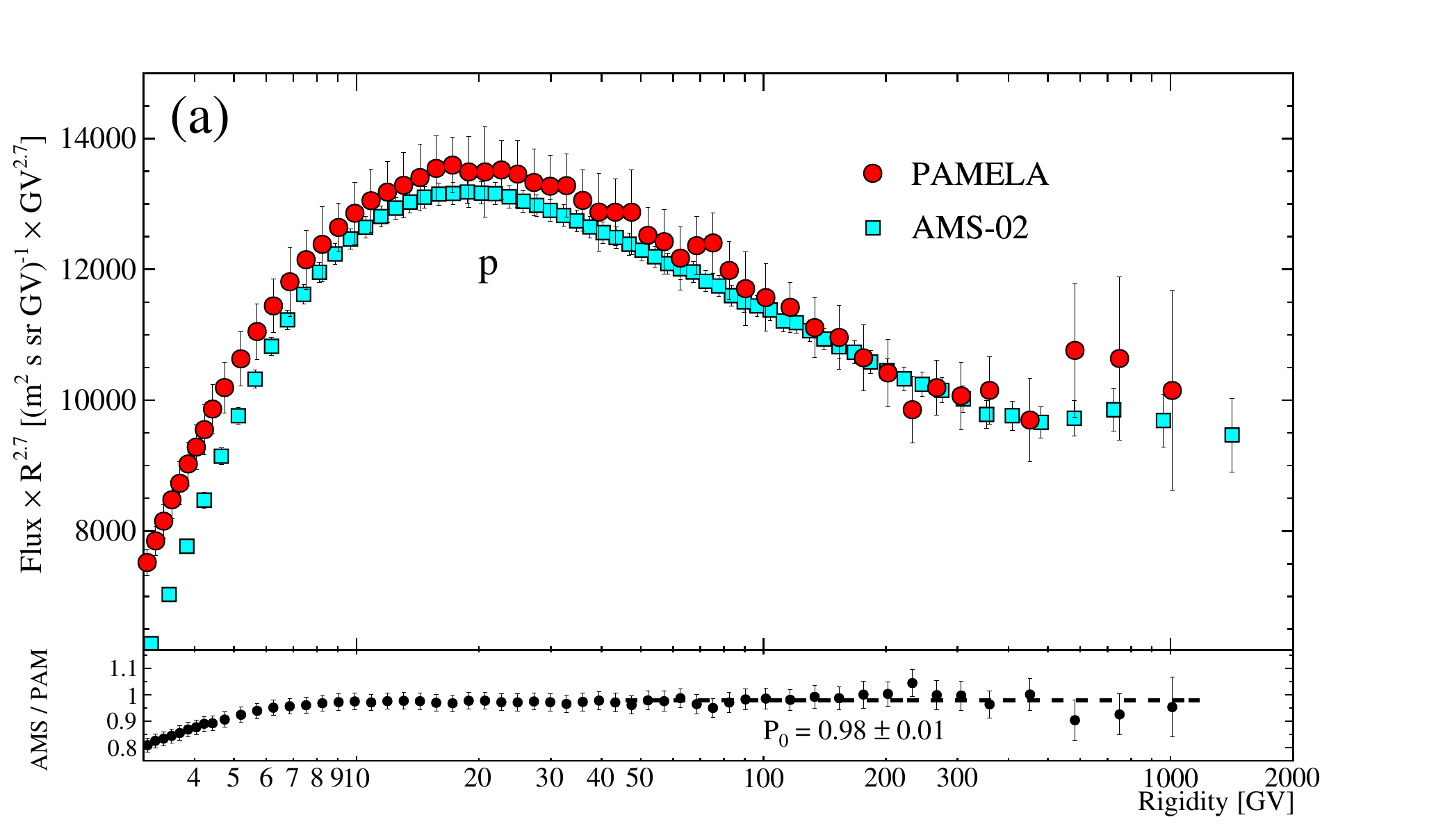}
\includegraphics[width=.98\textwidth]{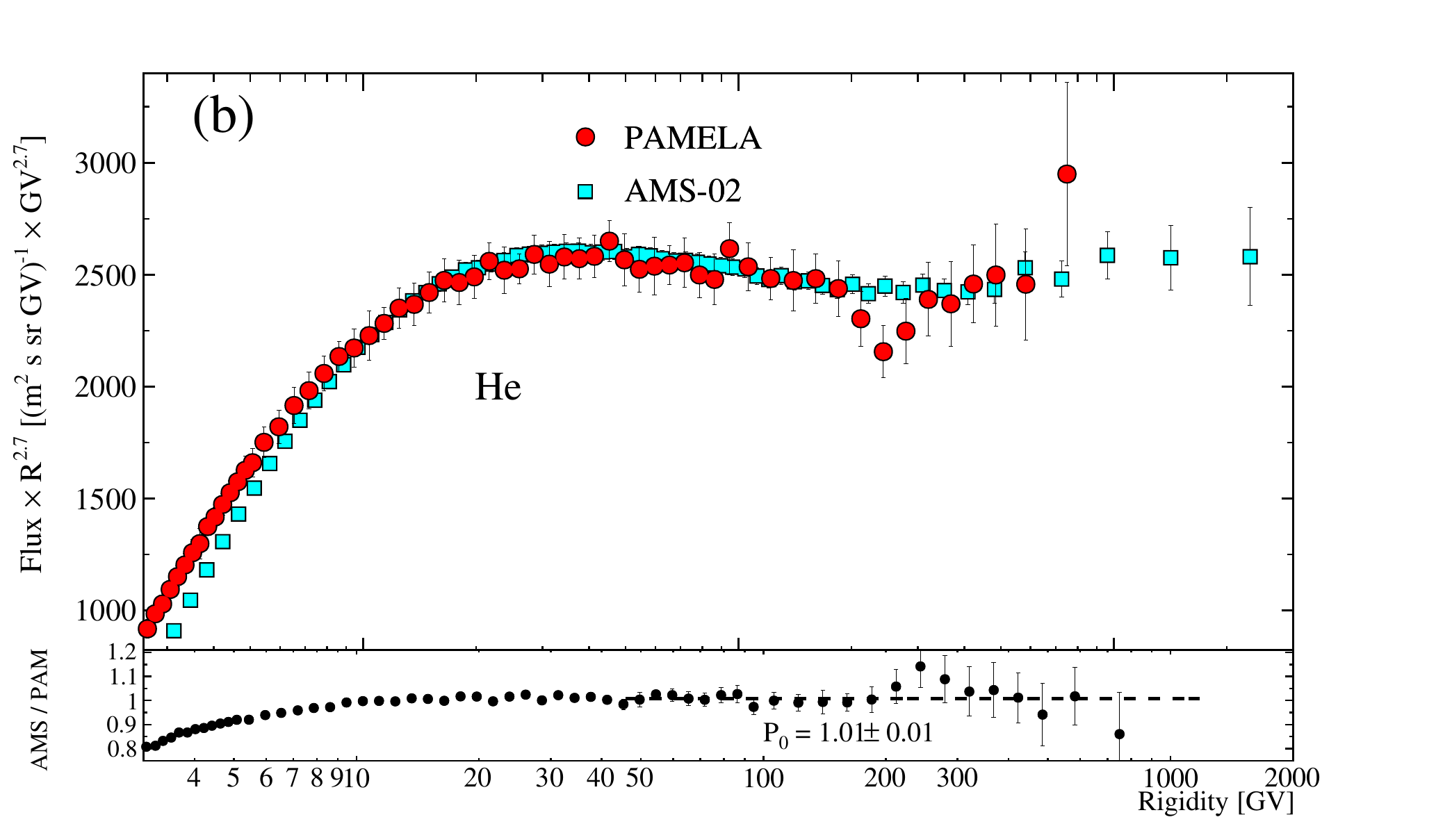}
\caption[]{(a) Top panel: comparison of the proton fluxes measured by PAMELA \cite{Adriani_Science} and
AMS-02 \cite{AMS02-H}. Lower panel: AMS-02 and PAMELA proton flux ratio. The value of P$_0$ results from a linear fit between 50 GV and 1 TV.
(b) Top panel: comparison of the helium fluxes measured by PAMELA \cite{Adriani_Science} and
AMS-02 \cite{AMS02-He}. Lower panel: AMS-02 and PAMELA He flux ratio. The value of P$_0$ results from a linear fit between 50 GV and 1 TV. }
\label{fig:H_He}
\end{center}
\end{figure}

After a preliminary contradicting result, the  AMS-02 experiment finally published the final spectra of H and
He \cite{AMS02-H,AMS02-He}, showing features consistent with those found by PAMELA. 
The comparison with the latest PAMELA results \footnote{H fluxes are taken from reference \cite{Adriani_Science}
but scaled by a factor $0.968$,  resulting from a more precise efficiency evaluation performed for the modulation 
study of PAMELA data \cite{proton_modulation}.} is shown in Figure \ref{fig:H_He} (a) for hydrogen and \ref{fig:H_He} (b) for helium.
The two measurements show an excellent agreement, at the level of few percent, above 50 GV. 

At rigidities below 50 GV the discrepancies among the measured fluxes are consistent, within the systematic
uncertainties, with the variations expected for the different solar activity encountered by the experiments. In 
Section \ref{prot_el_solmod} a new proton flux, estimated during the same period of AMS-02 
data taking, is presented. This new result shows an excellent agreement with AMS-02 results also in the energy region affected by solar modulation. 

\begin{figure}
\begin{center}
\includegraphics[width=.9\textwidth]{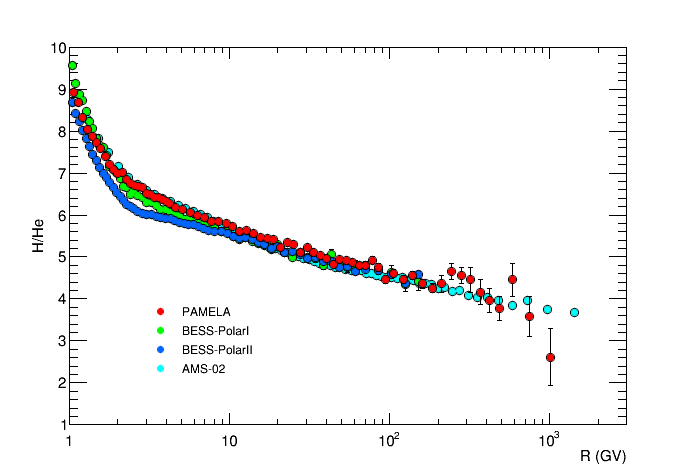}
\caption[]{H/He ratio  measured by PAMELA \cite{Adriani_Science} along with recent results by
AMS-02 \cite{AMS02-He} and BESS \cite{BESS-Polar}. Only statistical errors are shown. }
\label{fig:HHe}
\end{center}
\end{figure}

The ratio between hydrogen and helium nuclei, as a function of the measured rigidity, is shown in Figure \ref{fig:HHe}. 
PAMELA data reveal a clear slope in the H/He flux ratio. A good agreement between the various measurements is found. 
Differences can be ascribed to systematic uncertainties (not shown in the figure) even if effects due to solar modulation cannot be excluded.

Both the spectral hardening and the different slopes of the H and He fluxes challenge the standard paradigm
of cosmic-ray origin, acceleration and propagation/confinement within the Galaxy. 
All aspects of the paradigm are called into questions. 
The proposed explanations invoke acceleration mechanisms \cite{ptuskin}, diffusion effects 
\cite{blasi}, and different multi-source scenarios. A few of these models may explain 
the positron anomaly \cite{tomassetti1,tomassetti2}. 

\vspace{.7cm}


\begin{figure}[t]
\begin{center}
\includegraphics[width=.9\textwidth]{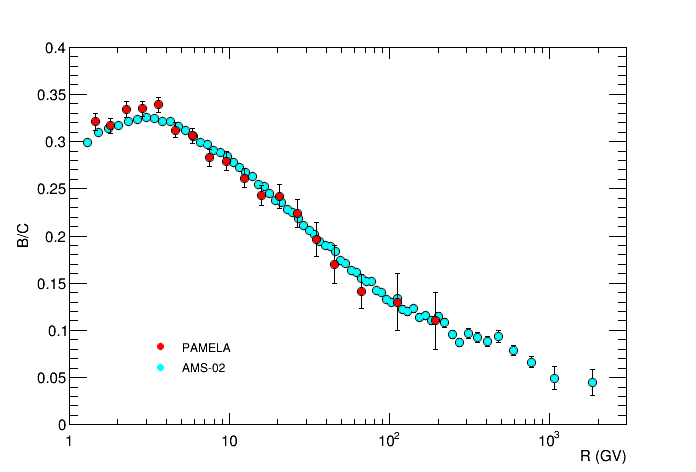}
\caption[]{B/C ratio  measured by PAMELA \cite{Adriani_BC} along with recent measurement from
 AMS-02 \cite{AMS_BC}. Only statistical errors are shown. }
\label{BC}
\end{center}
\end{figure}

In order to provide a satisfactory interpretation of all the observed CR anomalies, an important role 
is played by the secondary-over-primary ratios of nuclear species. 
Secondary nuclides result from inelastic interactions of heavier CR nuclei with the interstellar medium. 
On the opposite primary nuclides are almost exclusively accelerated in astrophysical sources and then 
injected into the interstellar medium. 
The relative abundance of secondary to primary nuclei is uniquely related to propagation processes and 
can be used to constrain the models, provided that  cross sections and decay chains for all the relevant 
nuclear processes are known.

Together with $^2$H and $^3$He, Li, Be and B in cosmic rays are the lightest and most abundant group of elements
of almost pure secondary origin.
Among them, B is mainly produced by fragmentation of C, which  originates  almost entirely from the acceleration sites. 
This, together with the larger statistics available  and the better knowledge of the cross sections for C and B nuclei, 
has made the B/C ratio the most sensitive observable for constraining the propagation parameters so far. 
Differently from B, a non-negligible fraction of Li and Be is of tertiary origin, i.e. 
produced by further fragmentation of secondary Be and B. 
This characteristic makes the knowledge of their abundance a complementary tool to tune propagation models,
helping in removing parameter degeneracy and giving a more detailed description of the galactic
propagation process.
On the other hand, the interpretation of Li and Be abundances requires the
knowledge of a complex chain of nuclear reactions, whose cross sections are still affected by large
uncertainties. 

PAMELA identification capability of individual elements is affected  by the saturation of the spectrometer tracking detectors,
which limits the maximum detectable rigidity to a few hundred GV for elements above Li.
Nevertheless, the redundant identification systems of PAMELA (ToF scintillators, Si layers of the calorimeter) allows to cleanly
discriminate elements at least up to oxygen.

The energy spectra of B and C have been measured over the kinetic energy range 0.44 - 129 GeV/n   \cite{Adriani_BC},
while the Li and Be analysis is still in progress. The B to C ratio is shown in Figure \ref{BC} along with the AMS-02 recent result \cite{AMS_BC}. 
An excellent agreement between PAMELA and AMS-02 measurements is observed.

\begin{figure}[t]
\begin{center}
\includegraphics[width=.52\textwidth]{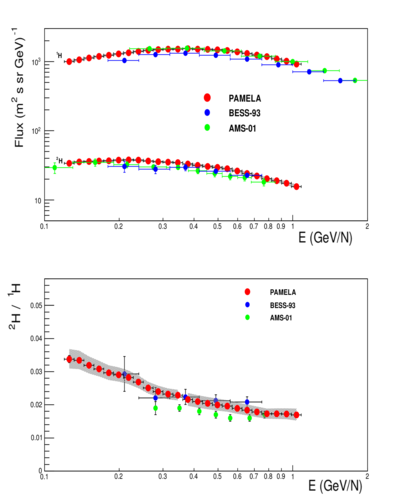}\includegraphics[width=.52\textwidth]{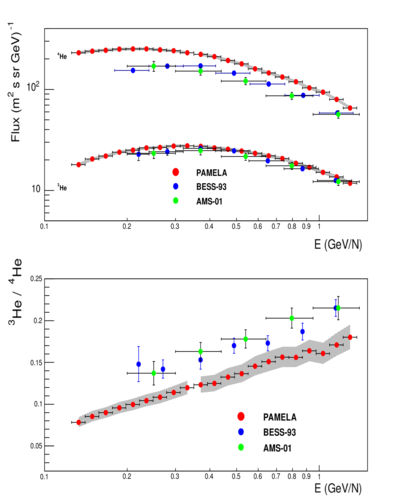}
\caption[]{Left: $^1$H and $^2$H absolute fluxes (top) and their ratio (bottom). The previous experiments are AMS-01,
BESS93. Error bars show the statistical uncertainty, and shaded areas show 
the systematic uncertainty. Right: $^4$He and $^3$He absolute fluxes (top) and their ratio (bottom). 
The previous experiments are AMS-01, BESS-93. Error 
bars show statistical uncertainty, and shaded areas show systematic uncertainty.}
\label{iso}
\end{center}
\end{figure}

\begin{figure}[h]
\begin{center}
\includegraphics[width=.5\textwidth]{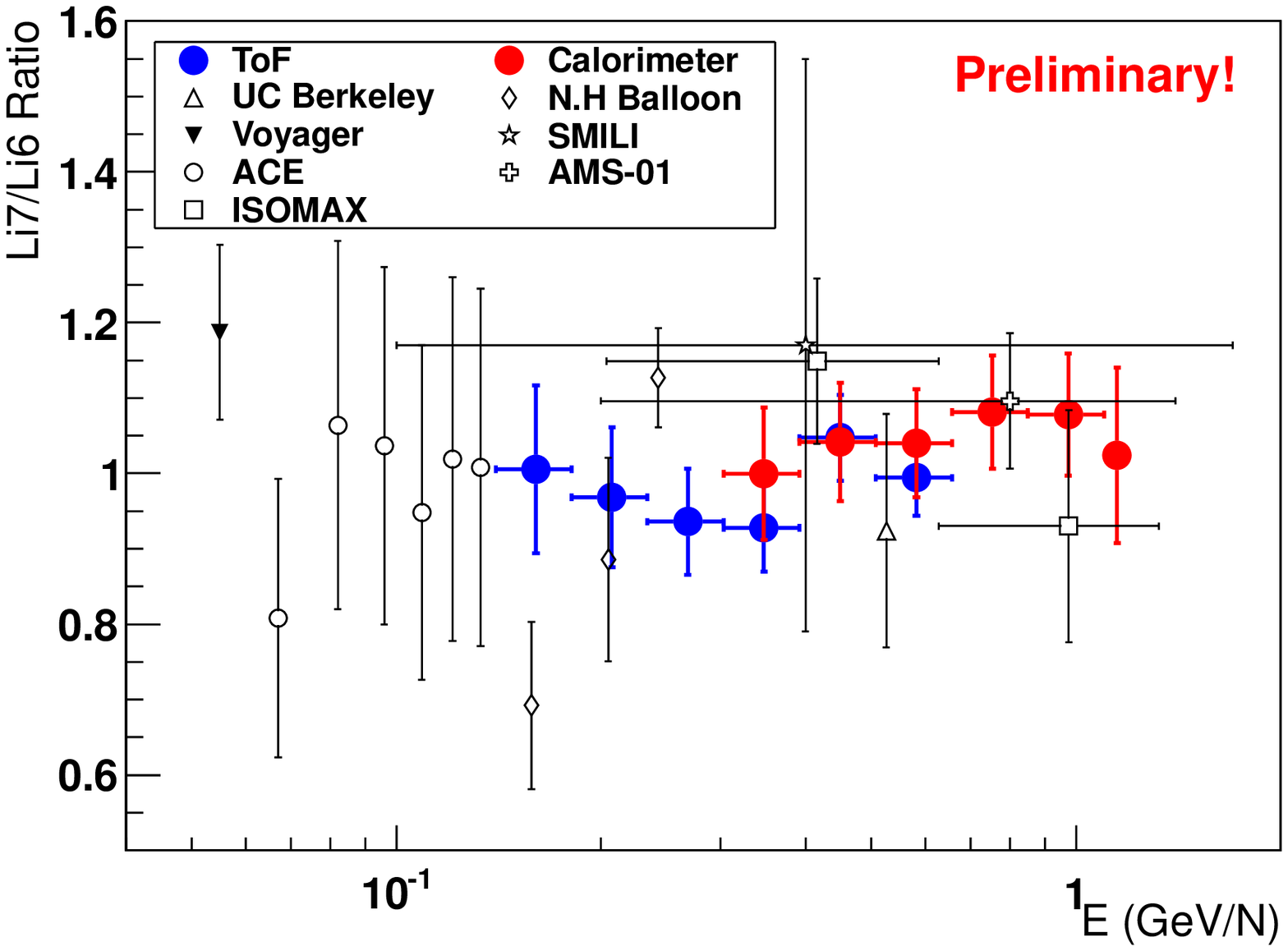}\includegraphics[width=.5\textwidth]{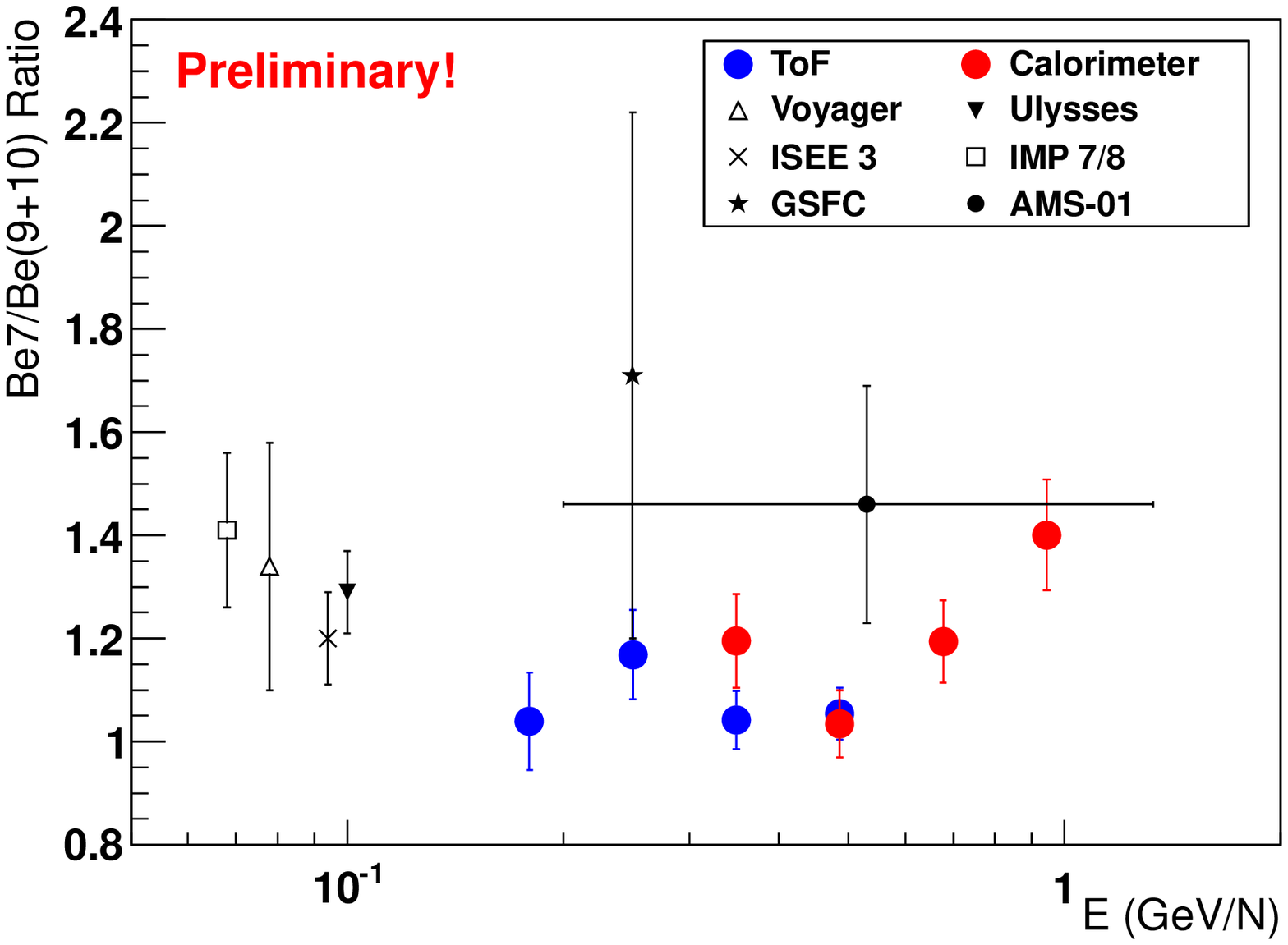}
\caption[]{Preliminary PAMELA results for the ratios $^7$Li/$^6$Li (left) and $^7$Be/$(^9$Be+$^{10}$Be$)$ (right).}
\label{menn}
\end{center}
\end{figure}

For non relativistic nuclei, roughly below 1 GeV/n, the PAMELA detector is able also to measure the isotopic 
composition of CR elements, up to Beryllium. 
The  $^2$H and $^3$He isotopes are both  produced by fragmentation of $^4$He, which is mostly of primary origin, and amount
individually to about $10\%$ - $20\%$ of the $^4$He flux. 
Figure \ref{iso} shows the fluxes of H and He isotopes in the energy range between $0.1$ - $1.1$ GeV/n and  $0.1$ - $1.4$ GeV/n respectively
\cite{Adriani_HHeiso} along with other recent measurements.  

Analogously to B, the $^2$H and $^3$He measured abundances can be used to test propagation models, with the significant
difference that they probe the propagation process of He rather than C.
With the precision reached by PAMELA and with future AMS-02 measurements, the resulting constraints to the propagation parameters 
are comparable and complementary with those derived from B/C measurements \cite{coste}.
A careful study of the $^2$H and $^3$He  allows to the investigation of the assumption that light nuclei experience
the same propagation processes than heavier nuclei, which have implications for the computation of other rare light species 
such as antiprotons and antideuterons.  

The secondary-to-secondary ratio, like  $^2$H/$^3$He  $^7$Li/$^6$Li and  $^7$Be/$(^9$Be+$^{10}$Be$)$, which are 
less sensitive to the astrophysical aspects of a given propagation model, provide a useful consistency check for the calculations.
At present,  the errors on the cross-section data represent the dominant source of uncertainty on the model predictions for 
rare isotopes \cite{tomassetti}, preventing a multi-component test of propagation models.

PAMELA is currently working on the determination of the Li and Be absolute fluxes, and on the 
determination of the secondary-to-secondary ratios like $^2$H/$^3$He  $^7$Li/$^6$Li and  $^7$Be/$(^9$Be+$^{10}$Be$)$. 
Figure \ref{menn} shows a few preliminary results. 

\subsection{Anisotropies in e$^+$ and e$^-$}
\label{anisotrophy}

\noindent
As discussed in the previous section, the most probable explanation for the positron anomaly requires an
additional source of cosmic ray positrons, like an astrophysical source (supernova remnants or pulsars) or a
contribution from dark matter decay or annihilation.
As high energy cosmic ray electrons and positrons (CREs) lose rapidly their energy, because of synchrotron radiation
emission and inverse Compton collisions with low-energy photons of the 
interstellar radiation field, their sources need to be relatively close at these energies.
Therefore, the detection of an anisotropy in the local CRE flux can be useful to distinguish  the positron origin.

The analysis of anisotropies in PAMELA has been performed on a sample of $\sim 2 \cdot 10^{4}$  electrons and  
$\sim 1.5 \cdot 10^{3}$ positrons in the rigidity range from 10 to 200 GV, 
selected in the period June 2006 - January 2010. For each event the particle arrival direction was reconstructed
using the trajectory inside the instrument and the satellite position and orientation on the orbit, with an accuracy of $\sim 2^{\circ}$
over the whole energy range. \\
The cosmic-ray background generated by an isotropic flux was calculated with two different approaches.
For positrons, the reference map was obtained from a sample of $\sim 4.5 \cdot 10^{5}$ protons, 
selected in the same period of time and rigidity range.
For electrons a MonteCarlo approach was used. About $10^{6}$ electrons in the
rigidity range 10-200 GV were randomly produced according to a uniform distribution
on the surface and in a 2$\pi$ solid angle. \\
Also the effects of the Earth's magnetic field were taken into account, back-tracing particles up to 25 Earth radii on the basis of 
numerical integration methods \cite{Bruno14}. 
The selection procedure and the analysis method are described in \cite{Adriani15}.
The analysis performed on electrons is reported in \cite{Panico15}. \\

\begin{figure}[t]
 \begin{center}
  \includegraphics[width=0.95\textwidth]{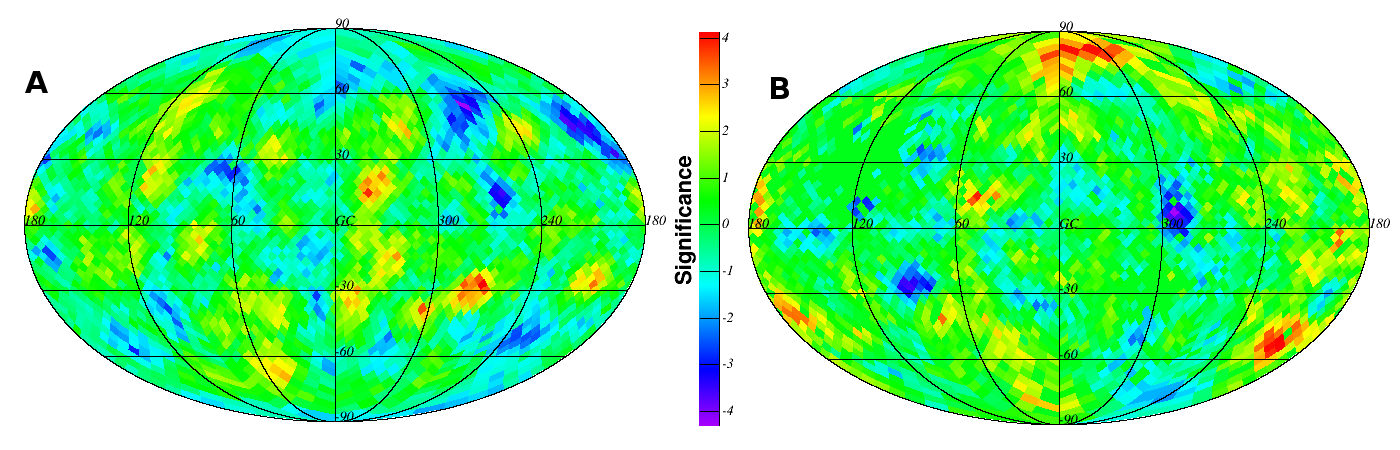}
  \caption[]{Significance maps for positrons (A) and electrons (B). The integration radius is 10$^{\circ}$. The Galactic reference system 
  is used and the color scale represents the obtained significance for each bin. }
  \label{fig:anisotropies}
 \end{center}
\end{figure}

Since the size of the anisotropy signal is unknown, the analysis were done on different angular scales. Signal and background
maps were integrated on four different angular scales, 10$^{\circ}$, 30$^{\circ}$, 60$^{\circ}$, 90$^{\circ}$. The content of each bin
is equal to the integrated number of events in a circular region around itself. For each integration radius, the significance 
was evaluated comparing signal and background maps. Results are consistent with isotropy at all angular scales considered. 
Figure \ref{fig:anisotropies} shows the significance maps obtained for positrons and electrons on an integration 
radius of 10$^{\circ}$. \\

\begin{figure}[h]
\begin{center}
\includegraphics[width=0.75\textwidth]{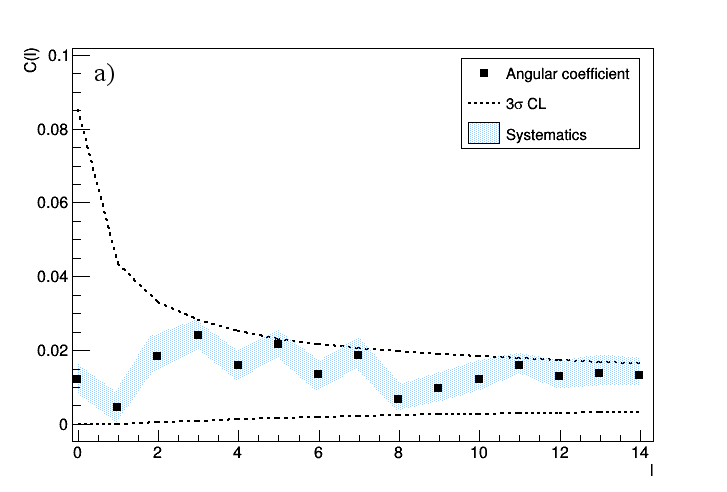}
\includegraphics[width=0.75\textwidth]{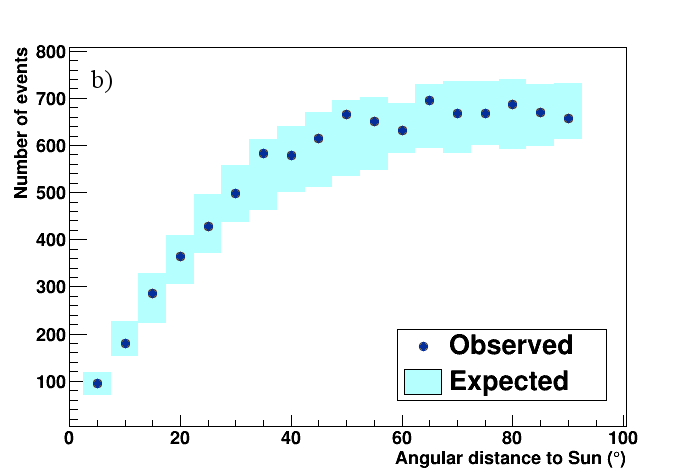}
\caption[]{a) Angular power spectra C(l) as a function of the multi-pole l for the positron signal over the proton background.
 b) Total number of electrons and positrons between 10 and 200 GV as a function of
the angular distance from the Sun. The gray boxes correspond to the 3$\sigma$ fluctuation respect
to an isotropic expectation.}
\label{fig:PowerSpecSun}
\end{center}
\end{figure}

The study of the angular power spectrum of arrival directions of the events provides information on the angular scale of the
anisotropy into the map. Therefore, the CR intensity has been expanded in spherical harmonics. Figure 
\ref{fig:PowerSpecSun} a) shows the result from modes l $= 1$ (dipole) up to l $= 14$.
The dotted lines represent the 5$\sigma$ bounds of the expected power spectrum from an isotropic sky. 
The grey band represents the systematic effects, that take into account the energy
and angular resolutions of the
instrument. No significant deviation from the isotropy was observed. The upper limit for the dipole amplitude 
is $\delta$=0.166 with a $95\%$ confidence level. \\

An excess of electrons and positrons was also searched for in the Sun direction \cite{Adriani15}. 
Figure \ref{fig:PowerSpecSun} b) shows the number of CREs within annuli centered on the Sun in the range 0$^{\circ}$-90$^{\circ}$, in 5$^{\circ}$ steps,
as a function of the angular distance from the Sun direction. 
The reference frame is the ecliptic coordinate system centered on the Sun and referred
to the J2000 epoch\footnote{J2000 is the current
standard epoch corresponding to the Gregorian date January 1, 2000 at
approximately 12:00 GMT. \cite{J2000}}.
Data are consistent with the isotropic expectation within a 3$\sigma$ interval.

\subsection{Large scale anisotropy}
\noindent
Galactic cosmic rays are found to have a faint and broad non-dipolar anisotropy across the entire
sky. Different experiments observed an energy-dependent large scale
anisotropy in the sidereal time frame with an amplitude of about 10$^{-3}$-10$^{-4}$ \cite{Nagashima98,Amenomori06,Aglietta09}.
In order to measure such a tiny effect, large instrumented areas and long-lasting data acquisition
campaigns are needed. Until now, only ground-based detectors had the required sensitivity at energies of $\sim 10^{13}$ eV. 

\begin{figure}[t]
\begin{center}
\includegraphics[width=0.75\textwidth]{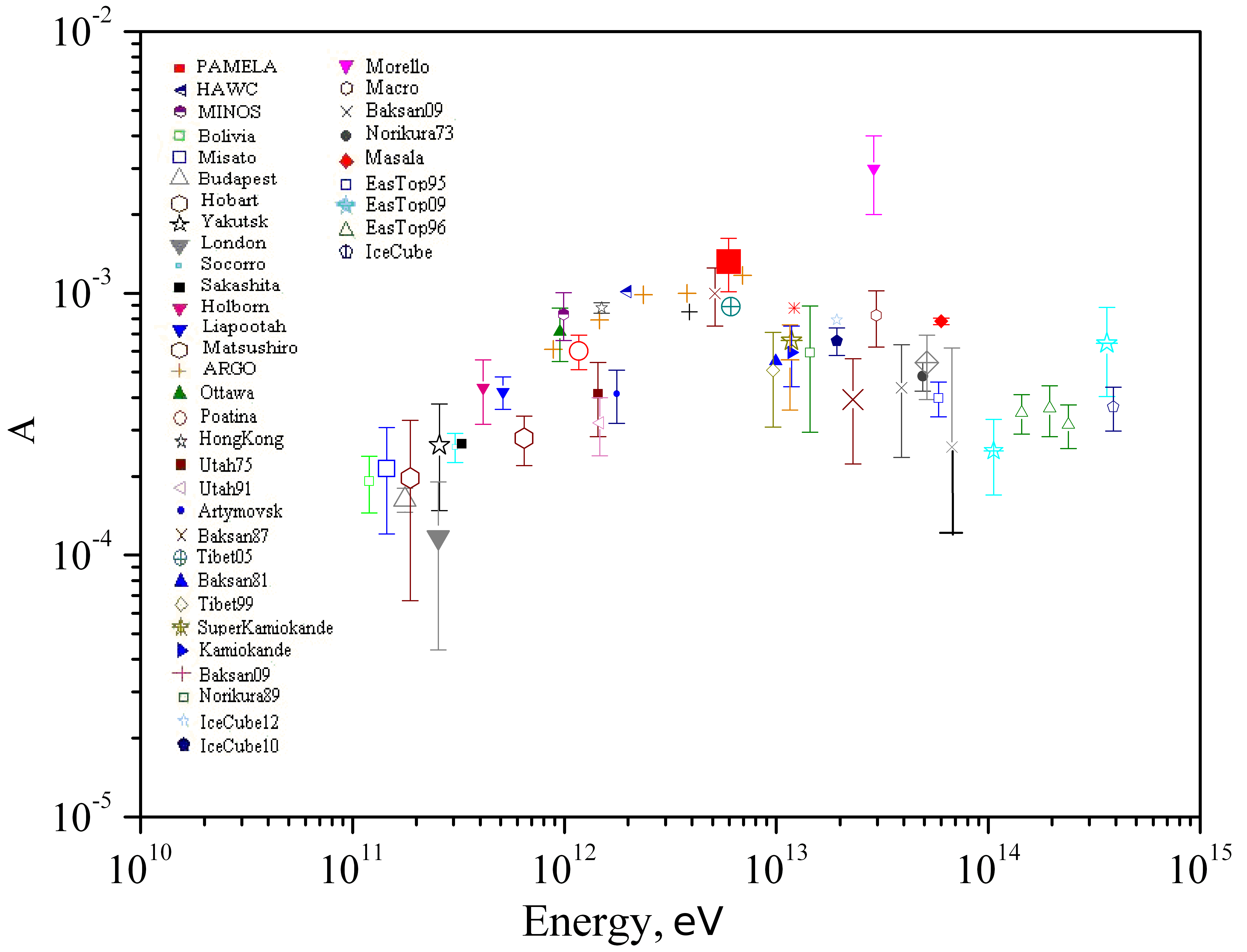}
\includegraphics[width=0.75\textwidth]{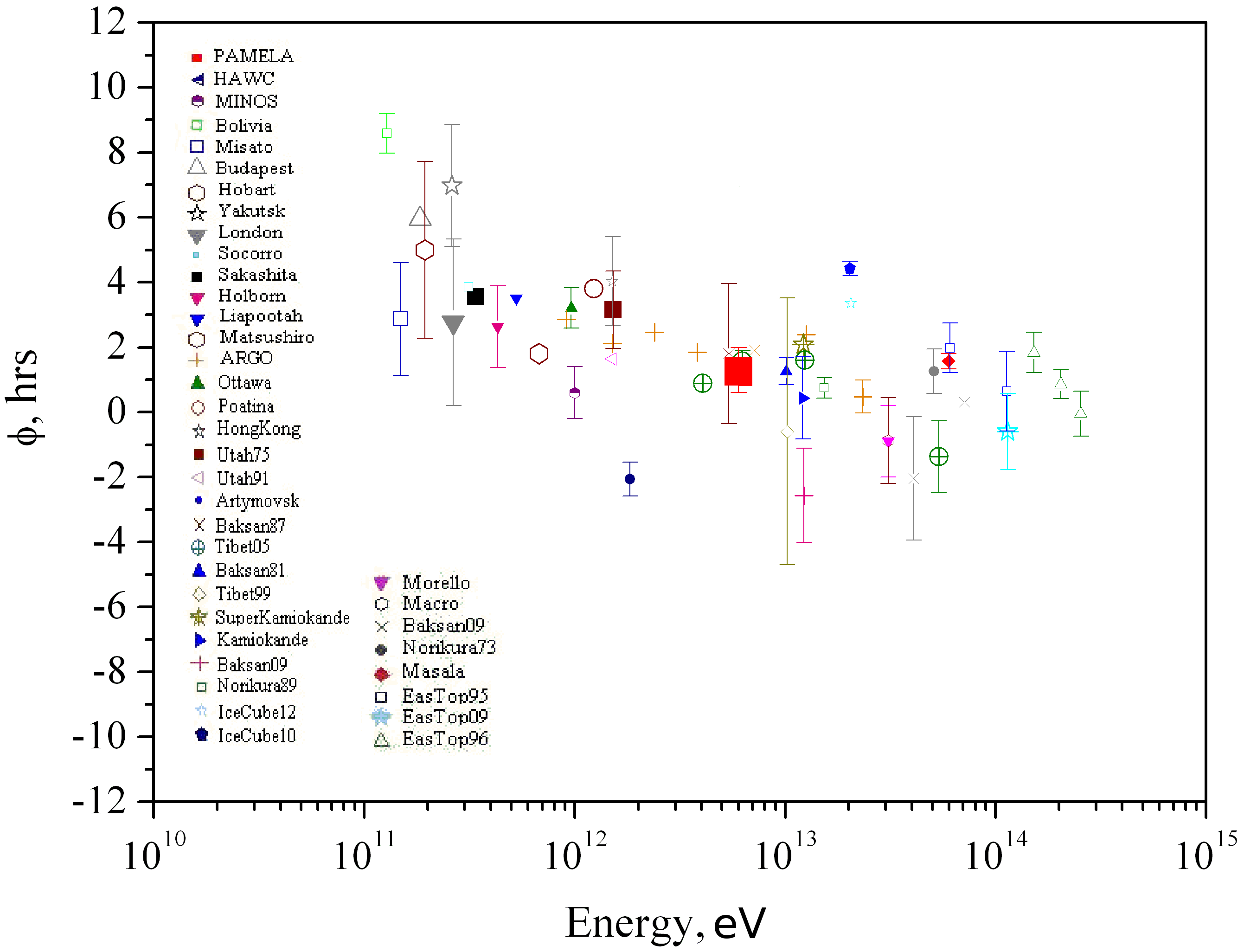}
\caption[Large scale anisotropy.]{Top panel: the amplitude of the cosmic ray large scale anisotropy obtained from the PAMELA data 
compared with a collection of other experimental results \cite{DeJong11},\cite{Benzvi13}. 
Bottom panel: the phase of the large scale anisotropy obtained by PAMELA data compared with other experimental results.}
\label{fig:results}
\end{center}
\end{figure}
Because of the small size of signal, uniform detector performance over instrumented area and over time are necessary as well as 
operational stability. The ground-based detectors typically suffer from large variations of atmospheric parameters as 
temperature and pressure, which translate into changes of the CR arrival rate.
PAMELA provides a reliable result on large scale anisotropy thanks to the minimization of these effects. 

In $1975$ Linsley proposed an analysis method based on the expansion of the experimental rate on the basis of spherical
harmonics \cite{Linsley75}. The counting rate has to be calculated within a defined declination band as a function of 
the right ascension. Fitting the result with a sine wave, the amplitude of the different harmonics and the corresponding 
phase is calculated.

The PAMELA calorimeter was used to reconstruct the particle energies in the range $1$ TeV/n - $20$ TeV/n with a good angular
resolution (0.3$^{\circ}$). Integrating the signal over an angular scale of 90$^{\circ}$ a sufficiently large sample was 
collected and the counting rate was calculated as a function of the right ascension. 
To minimize systematic effects, in the analysis the ratio of the signal and the background was used.
Applying the bootstrap method \cite{BOOTSTRAP}, the following results for the amplitude A and the phase $\Phi$ of the anisotropy
were obtained: 
\begin{itemize}
 \item A=(13 $\pm$ 2) $\cdot 10^{-4}$;
 \item $\Phi$=70$^{\circ} \pm$ 20 $^{\circ}$.
\end{itemize}
The values are in agreement with those reported in literature \cite{DeJong11,Benzvi13}, and are shown in Figure \ref{fig:results}.

\subsection{$\overline{{He}}/He$ and search for strange quark matter (SQM)}

The explanation of the observed baryon asymmetry, i.e., the almost complete absence of antimatter in the visible part of the Universe, 
is one of the most important problems in cosmology. 
Detection of antinuclei in the cosmic radiation would have a profound impact on our understanding of the Universe.
Theoretical arguments based
on constraints from gamma-ray sky surveys \cite{gamma1,gamma2} argue that the distance to any hypothetical domains of antimatter
must be roughly comparable to the horizon scale. The generally accepted theory to explain a baryon asymmetric Universe
involves mechanisms of baryo-synthesis that generate an asymmetry in an initially baryon symmetric Universe. This model,
while generally accepted, is not yet supported by experimental evidence: neither baryon non-conservation nor large levels
of CP-violation, both a required ingredient for a baryon asymmetry \cite{107}, have been observed.

While antiparticles such as positrons and antiprotons are produced by interaction of cosmic rays with the interstellar matter, 
the detection of heavier antinuclei (Z $>$ 2) could hint to the existence of antistellar
nucleosynthesis in antimatter domains or lumps of antimatter (e.g. \cite{108}). On the contrary the detection of antihelium may hint to the
existence of residual antimatter from the Big-Bang nucleosynthesis. Antihelium-3 was produced for the first time at
the IHEP accelerator in $1970$ \cite{109} and antihelium-4 was produced  at Brookhaven in $2011$ \cite{110}. The production cross section due
to cosmic ray interactions is very small and the expected flux compared to the helium flux is no more than 10$^{-12}$--10$^{-14}$
\cite{111,112,113}.

The search for antihelium in the PAMELA data was conducted in a similar way to that for helium.
Particular attention was paid to the treatment of the tracking system information: the deflection was reconstructed using
the whole, top and bottom part of the tracking system and consistency between the measured values was required. In this
way events with incorrectly measured deflection due to scattering in one of the planes of the tracking system were rejected,
thus eliminating contamination from helium nuclei reconstructed with an incorrect sign of curvature.

\begin{figure}[t]
\begin{center}
\includegraphics[width=.85\textwidth]{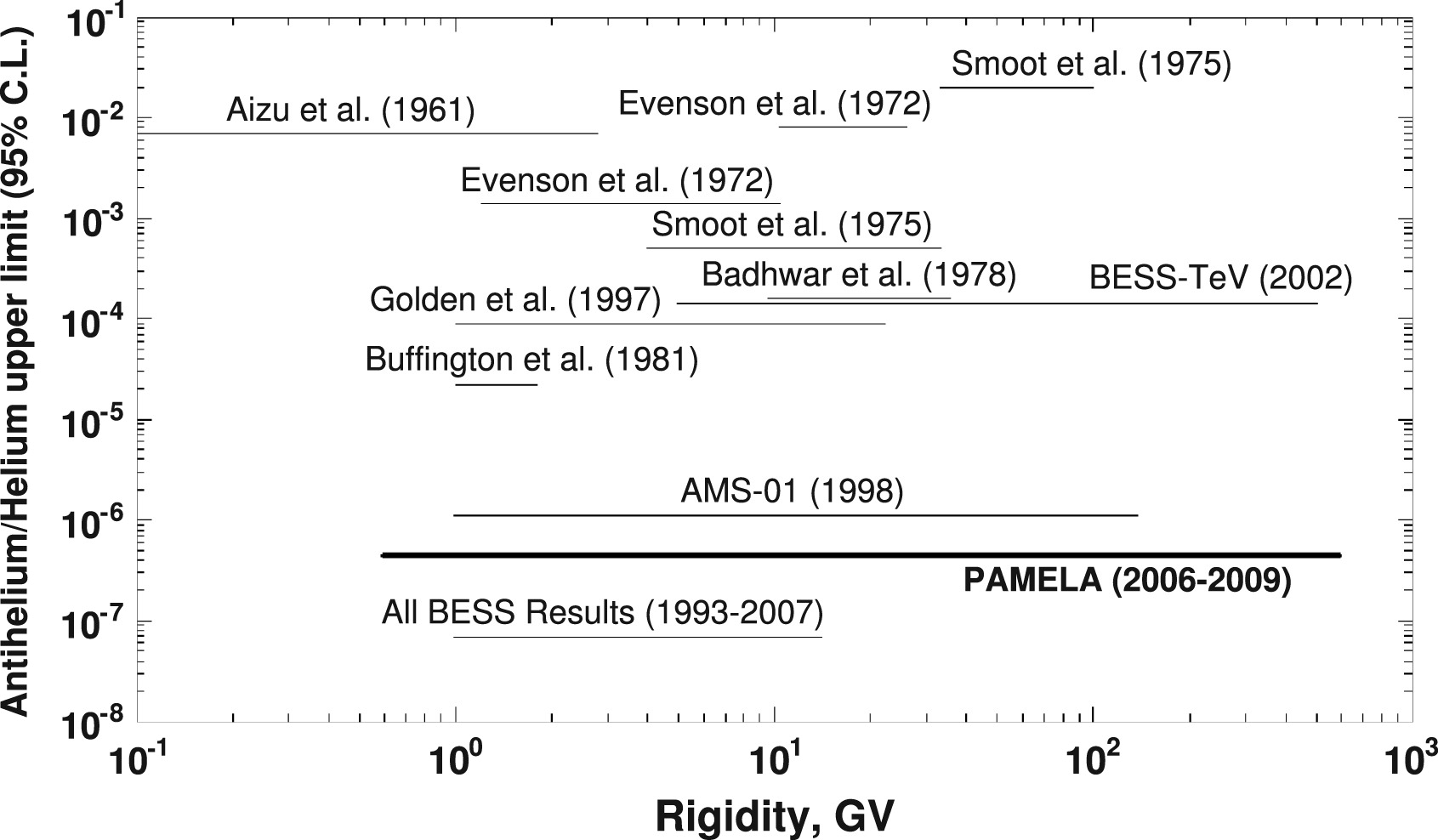}
\caption[]{Upper limit on the $\overline{\textnormal{He}}$/He ratio from PAMELA and other experiments.  }
\label{fig:exotic1}
\end{center}
\end{figure}

No helium events with negative rigidities were found among about $6.3 \times 10^{6}$ events with charge $|Z| \geq 2$  selected in the
rigidity range from 0.6 to 600 GV. Figure \ref{fig:exotic1}  shows PAMELA antihelium limit along with other experimental measurements \cite{22,115,116,117,118,119,120,121}. 
The upper limit with a $95\%$ confidence level has been estimated in accordance to the procedure adopted by the other
experiments. PAMELA result spans the largest energy range explored by a single experiment to date.

More details about the antihelium analysis can be found in  \cite{Mayorov_antihelium}.

\vspace{1cm}

The existence of a different state of hadronic matter other
than the ordinary nuclear matter, called strange quark
matter or SQM, was proposed in the $1980$s  \cite{1}. 
This hypothesized matter would be composed by an almost equivalent 
number of u, d, and s quarks, so it would be electrically neutral. 
Many models suggest that the presence of strange quarks could make SQM 
stable or metastable and much denser than ordinary matter \cite{2}.  
It is also suggested that the neutrality condition may
be approximate, allowing SQM particles to have a small
residual electrical charge; therefore a SQM candidate would have a 
low electric charge and a mass ranging from the minimum stable mass 
\cite{3}, up to values of baryon number A $\simeq$ 10$^{57}$ \cite{2}. 
Heavier objects appear more stable but some models suggest 
that the stability can also relate to light particles.  

SQM could be produced in the Big Bang \cite{3}, be part of
baryonic dark matter \cite{8}, or be present in astrophysical 
objects like neutron stars or "strange stars"; then it could be 
injected in the Galaxy as small fragments (called \textit{strangelets})
after stellar collisions and reach Earth where they could be detected.

\begin{figure}[t]
\begin{center}
\includegraphics[width=.53\textwidth]{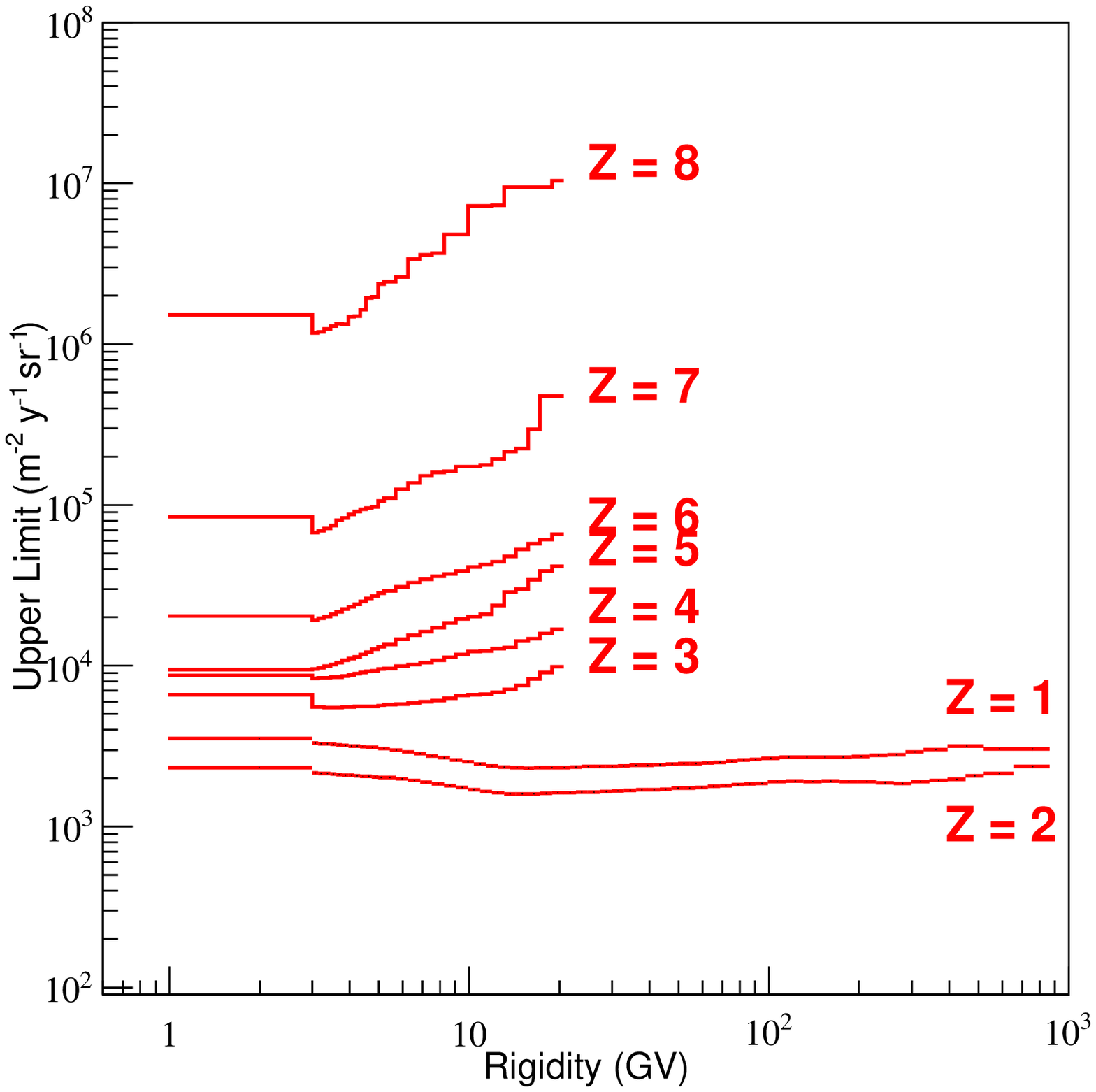}\includegraphics[width=.53\textwidth]{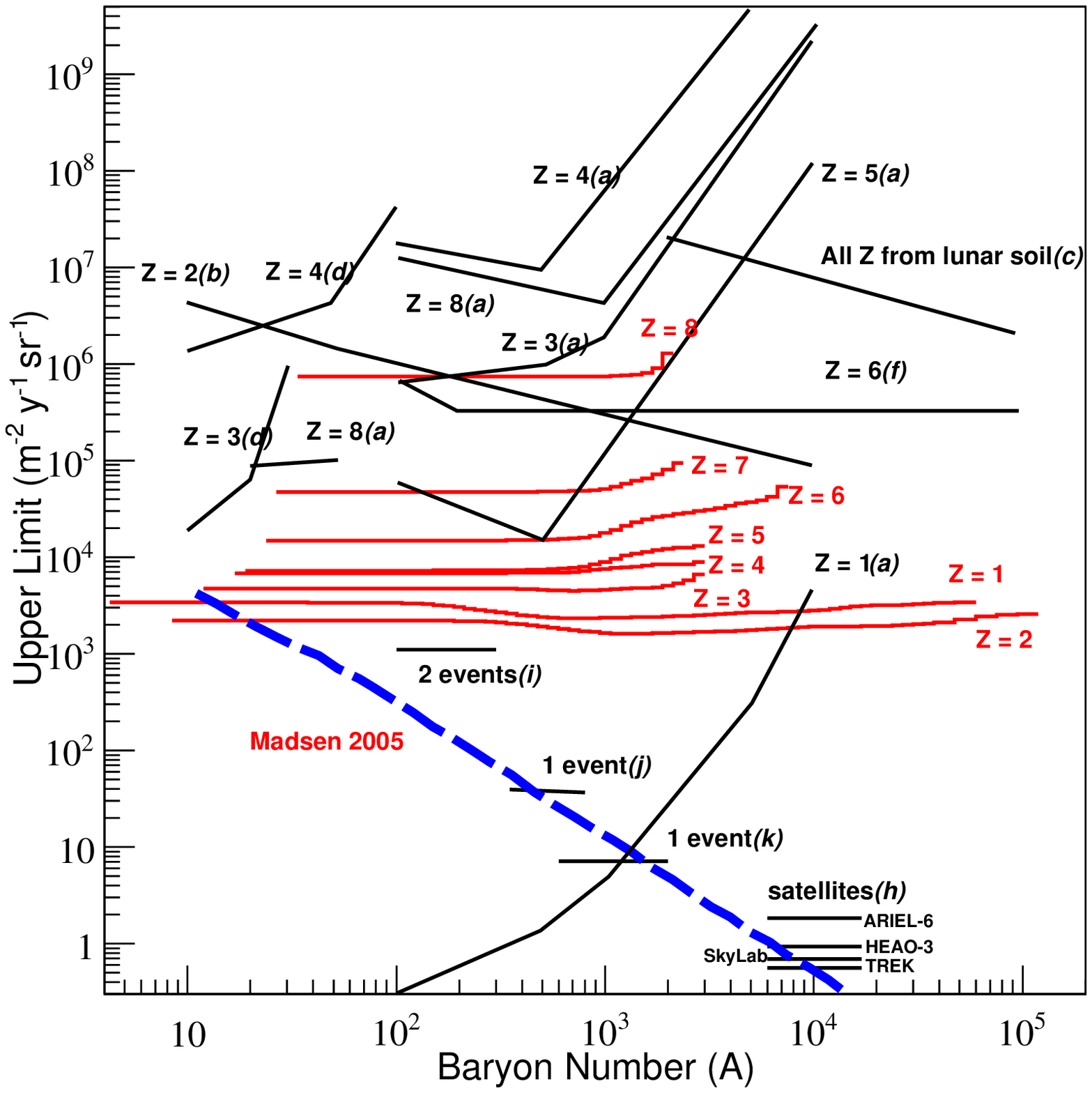}
\caption[]{Left: SQM integral upper limit in terms of rigidity as measured by PAMELA
 for particles with 1$\leq$Z$\leq$8. Right: integral upper limits in terms of baryon number A as measured
by PAMELA for particles with $1 \leq$ Z $\leq 8$ (red curves) together with previous experimental limits for strangelet,
translated into flux limits (black curves). The blue dashed line is the predicted flux of strangelets from \cite{3}.
The solid black lines represent previous
experimental limits for strangelets which are translated into flux
limits. }
\label{fig:exotic2}
\end{center}
\end{figure}

PAMELA was particularly suited for a SQM search because it could 
provide measurements of particle charge Z, velocity  $v$ and magnetic
rigidity $R$. The quantity $A/Z =  R/m_p\beta\gamma$, with $m_p$ mass of 
the proton, $\beta = v/c$ with $c$ the speed of light and $\gamma$ the 
Lorentz factor, was evaluated and used to 
characterize elements of both ordinary and exotic origin. Ordinary matter 
have $1 \leq A/Z \leq 3$ while SQM would exhibit higher $A/Z$ ratios.
The data set collected showed no candidate for particles with $1 \leq Z \leq 8$,
thus a upper limit was set.   
The overall SQM/matter flux ratio limit for such particles
is $1.2 \times 10^{-7}$. 
Moreover, the high precision measurements and high statistics allow to set both differential
 and integral upper limits as a function of rigidity for several species as shown in Figure \ref{fig:exotic2} (left panel).
The PAMELA upper limit as a function of baryon number $A$ is shown in the right panel of Figure \ref{fig:exotic2},
compared with results from other instruments.

In conclusion, SQM was searched in the PAMELA data from 
July 2006 to December 2009. No anomalous $A/Z$ particles were found
for $Z \leq 8$ particles in the rigidity range $1 \leq R \leq 1.0 \times 10^3$ GV and mass range 
$4  \leq A  \leq 1.2 \times 10^5$ and the subsequent upper limits could constrain or rule out models 
of SQM production/propagation in the Galaxy.
All details about this analysis are given in \cite{sqm}. 
 
\subsection{Solar modulation of GCR}

As previously discussed, galactic CRs are believed to be accelerated in astrophysical sources like 
supernovae remnants. Subsequently they propagate through the Milky Way undergoing diffusion on the 
irregularities of the Galactic magnetic field and loosing energy through different physical mechanisms. 
As a consequence, the power law spectrum generated at the source is distorted and the spectral
index gets harder when measured near Earth. Moreover, before reaching the Earth, CRs propagate inside the
heliosphere, the region formed by the outflow of plasma expanding radially from the Sun called solar wind ~\cite{Solar_Wind}.
The solar wind carries through the heliosphere the sun magnetic field creating the heliospheric magnetic 
field (HMF) ~\cite{GRL:GRL4201}.

The interaction of CRs with the solar wind heavily influences their propagation through the heliosphere
which is described by the standard CR transport equation derived by Parker in 1965 ~\cite{munini:Parker1}.
As a consequence of the propagation inside the heliosphere, the CR intensity measured at Earth decreases with 
respect to the local interstellar spectrum, i.e. the CR intensity as measured just outside the heliospheric
boundary~\cite{Potgieter2014_LIS}. This effect is significant up to $\sim 50$ GV, with the largest effects at the lowest energies. 
In addition, long-term changes in the solar activity, i.e. the 11-year solar cycle, produce time variations in the near-Earth
CR intensity. The solar activity is monitored observing the sunspot number, as shown in 
Figure \ref{fig:Time_Dependent_Proton} (lower panel). During solar minima, as from mid $2006$ to the end of $2009$ (23rd minimum),  
a lower number of sunspots is observed with respect to solar maxima, as from $2010$ to $2014$ (24rd maximum). 
Since during solar minima the solar wind velocity and the HMF are weaker, the CR are less modulated 
and their near-Earth intensity  is higher with respect to solar maxima period.
This behavior is illustrated in Figure \ref{fig:Time_Dependent_Proton} (middle panel) where the neutron monitor counts 
(normalized to mid 2006) measured by the Oulu neutron monitor\footnote{Cosmic rays colliding with molecules in the
atmosphere produce air showers of secondary particles including neutrons. The neutron monitor count rate is thus
proportional to the intensities of the CR flux at Earth. } is shown. Comparing the two lower panels of Figure 
\ref{fig:Time_Dependent_Proton} the anti-correlation between solar activity and CR intensity is noticed. For 
a more exhaustive discussion on the cosmic ray solar modulation see ~\cite{lrsp-2013-3}.

Precise measurements of the time-dependent CRs spectra are essential
to understand the cosmic ray propagation through the heliosphere. Furthermore, the experimental and theoretical investigation of
this system provides information that can be easily applied to larger astrophysical
systems. The possibility of performing in-situ measurements makes the interplanetary
medium the ideal environment to test the theory of
propagation of charged particles in magnetic fields under conditions
which approximate typical cosmic condition.
Hence very useful information for understanding
the origin and propagation of cosmic rays in the Galaxy can be derived.

Moreover, a precise knowledge of the solar modulation effects on 
galactic cosmic-ray are crucially important 
in the context of indirect search of light ($\sim10$ GeV/c$^2$ mass) dark matter 
candidates~\cite{Bottino_2012, Cerde_o_2012, Hooper_2015}.
Is essential to disentangle effects related to
a possible primary contribution (dark matter) and those related to the secondary
standard production of antiparticles by CR and their transport to the Earth.
Low energies data ($<30$ GeV) can be fully exploited only with a
precise understanding of the expected background,
including its variation with 
solar activity and its dependence on the particles sign of charge.
This can be achieved with the observation of accurate low energy
 spectra for protons, electrons, positrons, and even antiprotons, on an
almost continuous timescale. Lastly, understanding 
the effects and time dependence of solar modulation is
significant also for space weather since the amount of CRs reaching the
Earth can be predicted.

\subsection{Proton and electron modulation}
\label{prot_el_solmod}

The PAMELA experiment, as a result of its long flight duration and the capability to measure particles down to very 
low rigidities ($70$ MV for electrons), represented an excellent detector for the study of the cosmic ray
solar modulation. PAMELA was the first detector with the ability to measure almost continuously, with high
statistics and in a wide energy range the fluxes of many cosmic ray species covering nearly a full solar cycle,
from mid $2006$ to the beginning of $2016$. 

\begin{figure}
\centering 
\includegraphics[width=1.08\textwidth]{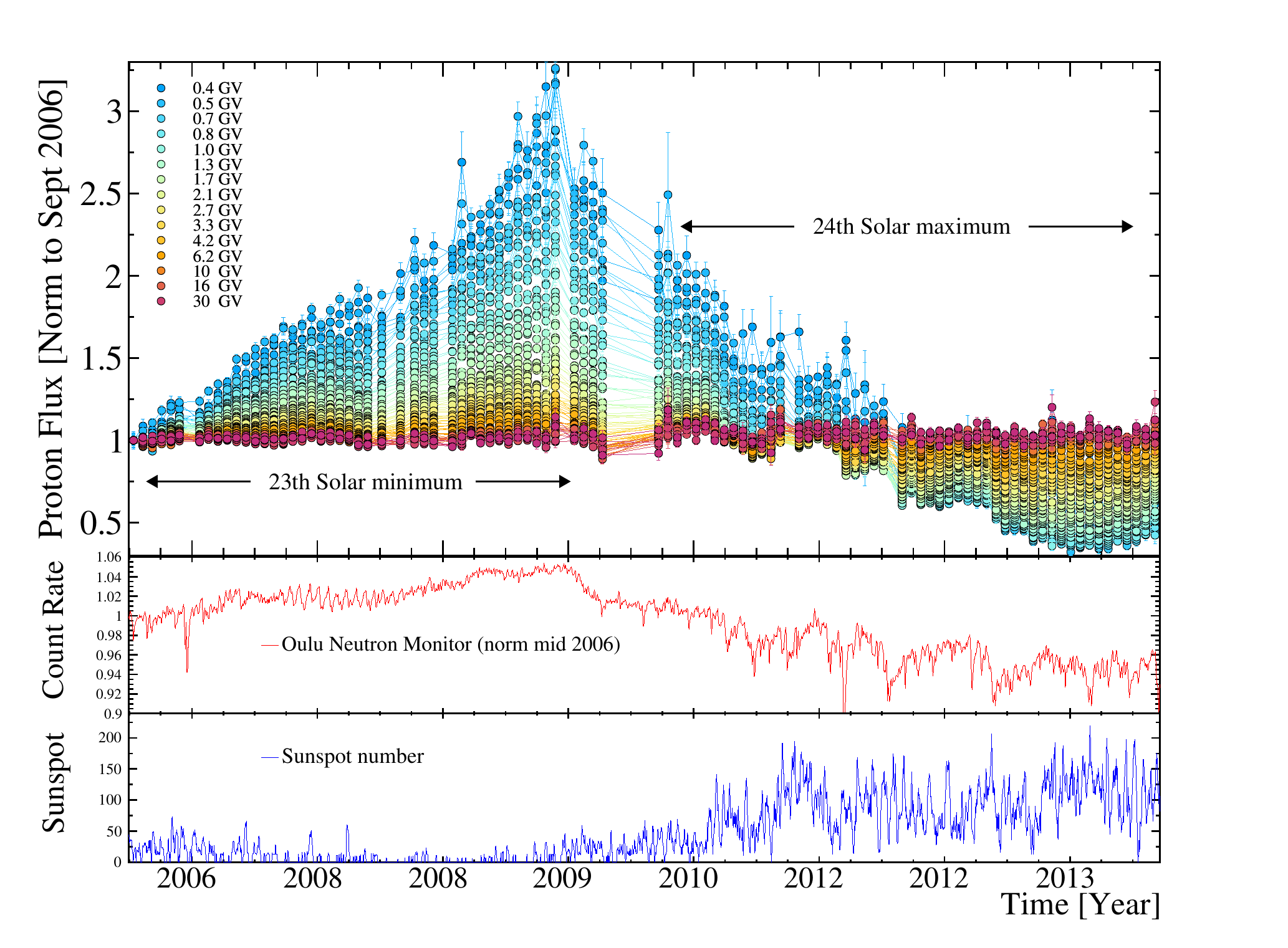}%
\caption[Time variation of proton flux.]{Top panel: time-dependent proton intensities (normalized to July 2006) measured by PAMELA between July 
2006 and May 2014. Each point represents a $\sim 27$ days time period. Holes are due to missing data from satellite or to the 
off-line exclusion of sudden transient from the Sun as solar flares. 
Mid panel: the Oulu neutron monitor count rate (normalized to July 2006) between mid 
2006 to May 2014. Data taken from http://cosmicrays.oulu.fi. Bottom panel: average daily number of sunspot. Data taken 
from http://www.sidc.be/silso/datafiles.}
\label{fig:Time_Dependent_Proton}
\end{figure}

\begin{figure}[t]
\centering 
\includegraphics[width=.9\textwidth]{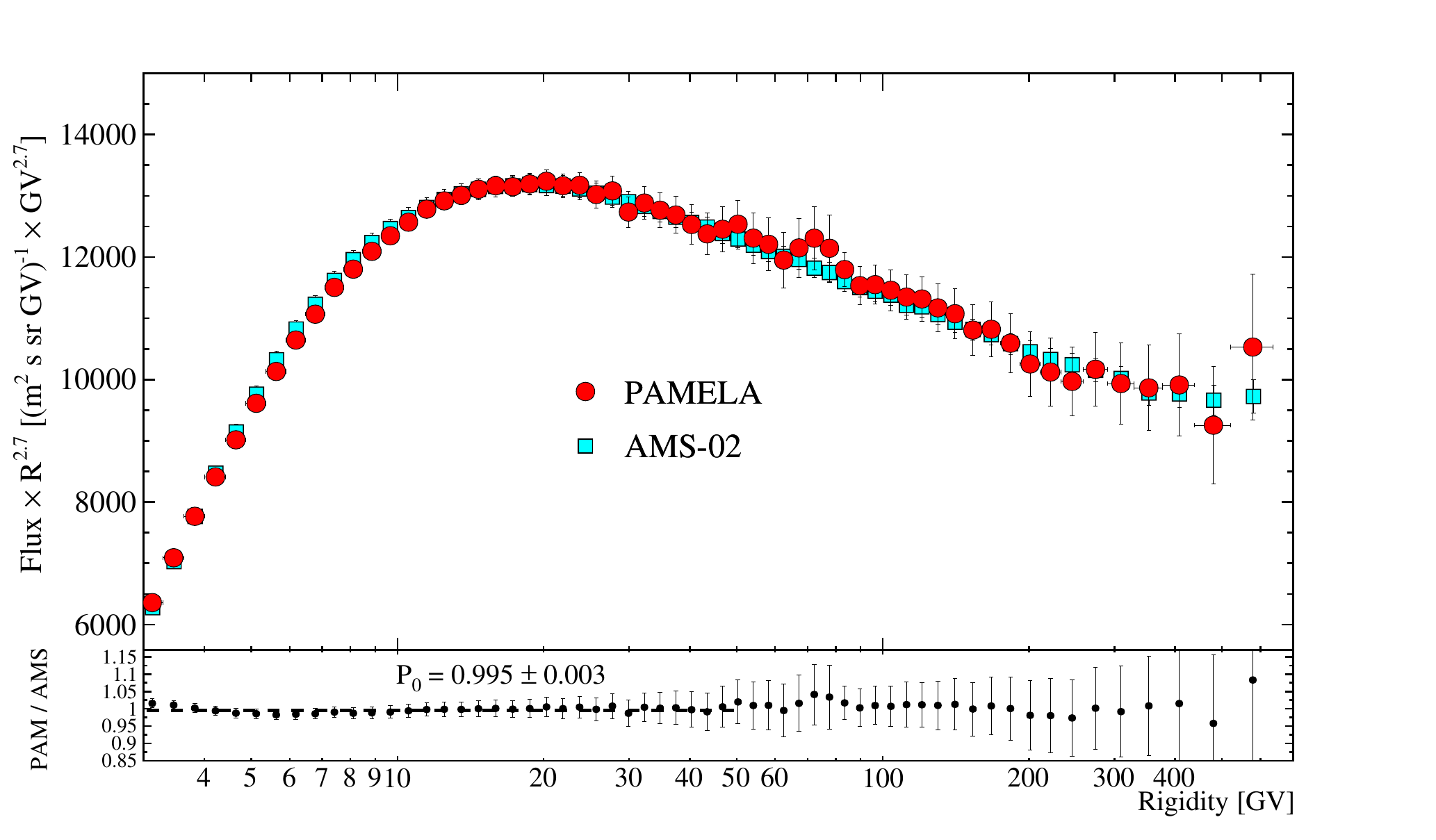}%
\caption[AMS02-PAMELA proton flux.]{The PAMELA proton flux evaluated between May $2011$ and November $2013$ compared with recent AMS-02 measurement. 
Only statistical errors are showed. 
The agreement between PAMELA and AMS-02 is excellent along the whole energy range. P$_0$ is the result of a linear fit 
on the flux ratio between 3 and 50 GV.}
\label{fig:AMS02_PAMELA_Proton}
\end{figure}
PAMELA results on the time variation of the proton spectrum during the $23$rd solar minimum,
from mid $2006$ to the end of $2009$, have already been published  ~\cite{proton_modulation}. 
In this article an extension of these analysis up to September $2014$ is presented. 
The efficiencies were estimated for each time period, while issues such as the Maximum Detectable Rigidity (MDR)
efficiency together with the alignment of the spectrometer were not relevant to the modulation regime
affecting significantly higher energies. Days in which solar events or 
Forbush decrease effects were present were excluded from the data. 
To control the residual time dependences the data were normalized at high energy ($50$ - $100$ GeV).
The reported errors are only statistical. 

Figure \ref{fig:Time_Dependent_Proton} (top panel), shows the proton intensity at different rigidities (normalized to July $2006$) 
measured between  the beginning of the data-taking, in July 2006, and September 2014. More than  2 $\times$ 10$^{8}$
protons were collected. The high statistic allowed to sample the proton fluxes over a Carrington Rotation (period of 27.27 days).
Holes in the presented data are due to periods of non operativity of the PAMELA instrumentation or the satellite.
Furthermore time periods during solar events were excluded from the analysis. 
The time-profile of the proton intensity resembles the neutron monitor count, with a peak in late
$2009$ and a gradual decrease up to early $2014$, when the maximum of solar cycle $24$th was reached. The 
major modulation effects are experienced at the lowest rigidity ($0.4$ GV) with an increase of about 
a factor $2.6$ with respect to July $2006$. 
At higher rigidities the solar modulation effects decrease and above $30$ GV 
the proton flux is time-independent within the experimental uncertainties.  
The $23$rd solar minimum activity and the consequent minimum modulation
conditions for CRs was unusual (e.g. see \cite{Potgieter2014}). It was expected that the new
solar cycle would begin in early $2008$. Instead solar minimum modulation
conditions continued until the end of $2009$ when PAMELA measured the highest cosmic ray proton spectrum from 
the beginning of the space age. From $2010$, as the solar activity started to increase, the proton intensity 
showed a decrease up to the beginning of $2013$, when the maximum activity of the $24$th solar cycle was reached.
After mid $2014$, the lowest energy protons showed again an increasing trend as a consequence of the 
solar activity decrease. 

As showed in Figure \ref{fig:H_He}, above $50$ GV the 
agreement between PAMELA and AMS-02 proton fluxes is within $2\%$. 
However, below this rigidity, the measurements differs 
due to the solar modulation effects. In order to compare the proton fluxes
below $50$ GV, a new analysis was done with the PAMELA data 
collected between May $2011$ and November $2013$, i.e. the time period 
corresponding to AMS-02 published spectrum. Figure \ref{fig:AMS02_PAMELA_Proton} 
shows this new proton spectrum compared with the AMS-02 results. Now also below $50$ GV 
the agreement is excellent as shown from the value of P$_0$ 
which represents the result of a linear fit to the flux ratio 
between $3$ and $50$ GV. The agreement with AMS-02 is 
a good indication about the reliability of the solar modulation 
proton analysis. 

In the context of solar modulation, simultaneous measurement of different particle species provide a 
better understanding of the physical mechanisms which rules the propagation of cosmic rays
inside the heliosphere. 
Measures of oppositely charged particles are extremely useful in order to study 
charge-sign-dependent solar modulation effects~\cite{Potgieter_2014}.
Besides protons, PAMELA measured the time-dependent electron 
fluxes between $70$ MeV and $50$ GeV during the $23$rd solar minimum  ~\cite{0004-637X-810-2-142,Munini}.
Since electrons represent only $1 \%$ of the cosmic radiation, the collected statistics allowed the fluxes to be measured only 
for a six months time intervals. A total of seven fluxes were obtained. 

\begin{figure}
\centering 
\includegraphics[width=.9\textwidth]{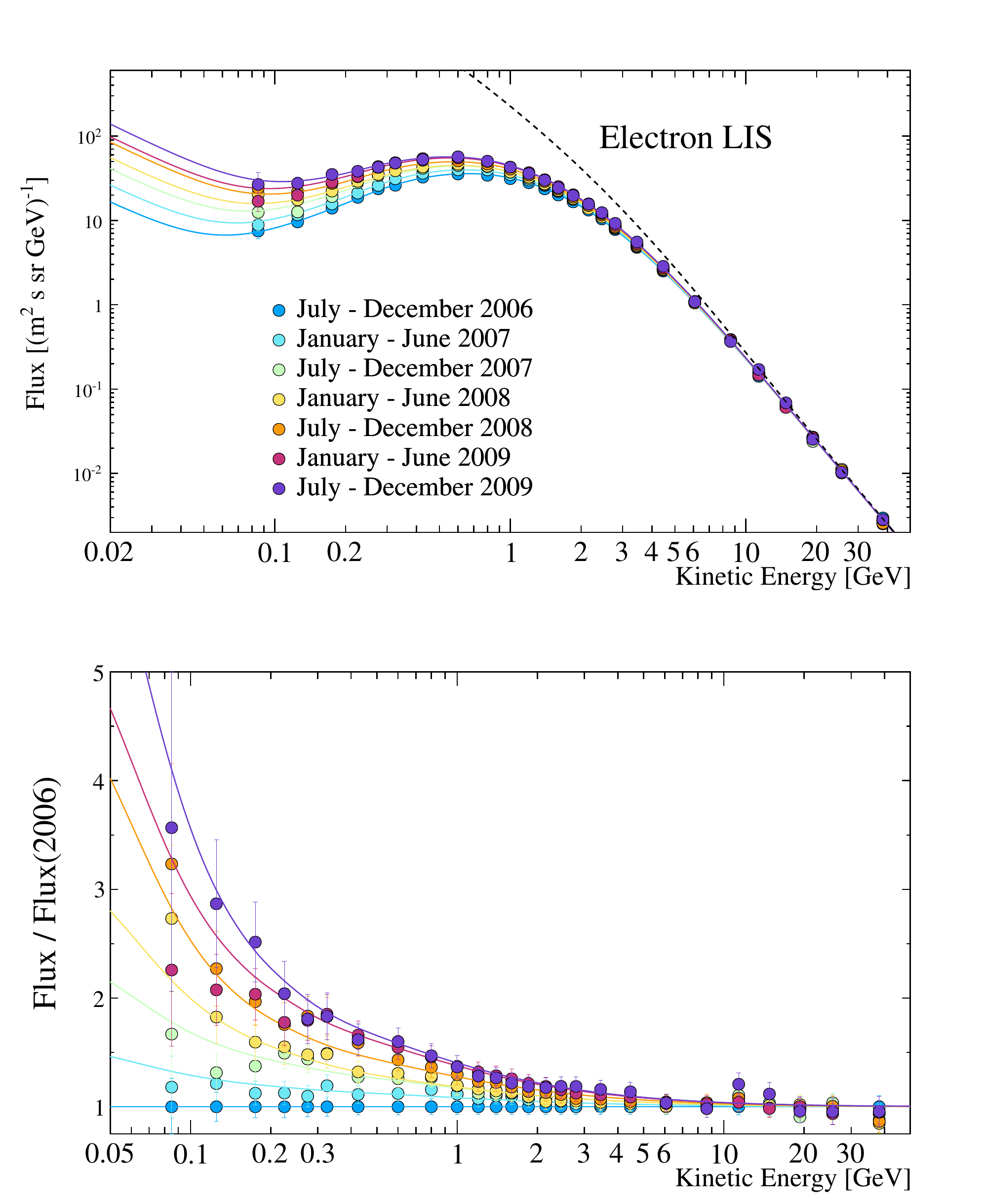}%
\caption[Time-dependent PAMELA electron spectra.]{Top panel: the half-year electron fluxes measured by PAMELA between mid 2006 to late 2009. The black 
dotted line represents the electron local interstellar spectrum. The colored lines come from a 3D numerical solution of the Parker equation applied to 
reproduce the PAMELA data. Bottom panel: 
the increase of the electron fluxes with respect to 2006. Colored line are the ratio of the 3D numerical solution with respect to 2006.}
\label{fig:Time_Dependent_Electron}
\end{figure}

A set of cuts were developed in order to select electrons from PAMELA data. 
Several calorimetric selections based on the topological development of the 
electromagnetic shower were defined. These selections allowed an almost complete rejection
of the contaminations represented by galactic antiprotons and negative 
pions\footnote{Pions are produced locally from the interaction of primary
CR with the aluminum dome above the instrument}.
Selection efficiencies were evaluated from flight data when the sub-detector 
redundancy allowed to select a clean sample of electrons, otherwise simulation 
was used. 
In order to take into account any possible temporal 
variation of the PAMELA apparatus, a set of efficiencies was evaluated for each 
six months time period. 
Eventually, to take into account energy losses inside the instrument and 
the finite tracker resolution, a Bayesian unfolding procedure 
based on simulation was applied to the spectra ~\cite{munini:deagostini}. 
For a detailed description of the experimental analysis see ~\cite{munini:phd}.

The results on the electron fluxes are presented in Figure \ref{fig:Time_Dependent_Electron} (top panel)
where the half-year electron fluxes are presented together with the calculated electron local interstellar spectrum 
(dashed black line)  ~\cite{0004-637X-810-2-141}. 
Below about $20$ GeV the solar modulation effect are clearly evident and the spectrum measured at Earth is
significantly distorted with respect to the local 
interstellar one. 
The bottom panel of 
Figure \ref{fig:Time_Dependent_Electron} shows the flux increase with respect to $2006$. As expected, a decrease in the solar activity 
corresponds to an increase in the electron intensity. The colored line on both panels are the results of a 3D numerical solution 
of the Parker equation ~\cite{0004-637X-810-2-141}.
From a modeling perspective, solar minima represent optimal periods to study the solar modulation of cosmic rays since the heliospheric 
magnetic field structure is well ordered.
 In contrast, during solar maxima, the sun activity change faster, the heliospheric environment 
  is chaotic and from a modeling point of view it is challenging to reproduce such conditions. 
The solutions presented in Figure \ref{fig:Time_Dependent_Electron} were obtained tuning the model free parameter, like the 
diffusion tensor, in order to reproduce the PAMELA data. In this way, the energy and time dependence of the 
diffusion tensor and the other parameters were obtained, providing important information about propagation mechanisms.
For example, it is interesting to notice how the flattening and the 
subsequent increase in the electron fluxes below $100$ MeV is due 
to a flattening of the diffusion coefficients which become energy-independent below $200$ MeV. Contrarily, the proton flux still decreases 
below $100$ MeV because the adiabatic energy losses dominate the propagation ~\cite{Potgieter2014}. For more details about this numerical
model see ~\cite{munini:vos}. 

\subsection{Charge-sign dependent solar modulation}
\label{chargesign}

\begin{figure}
\centering 
\includegraphics[width=.9\textwidth]{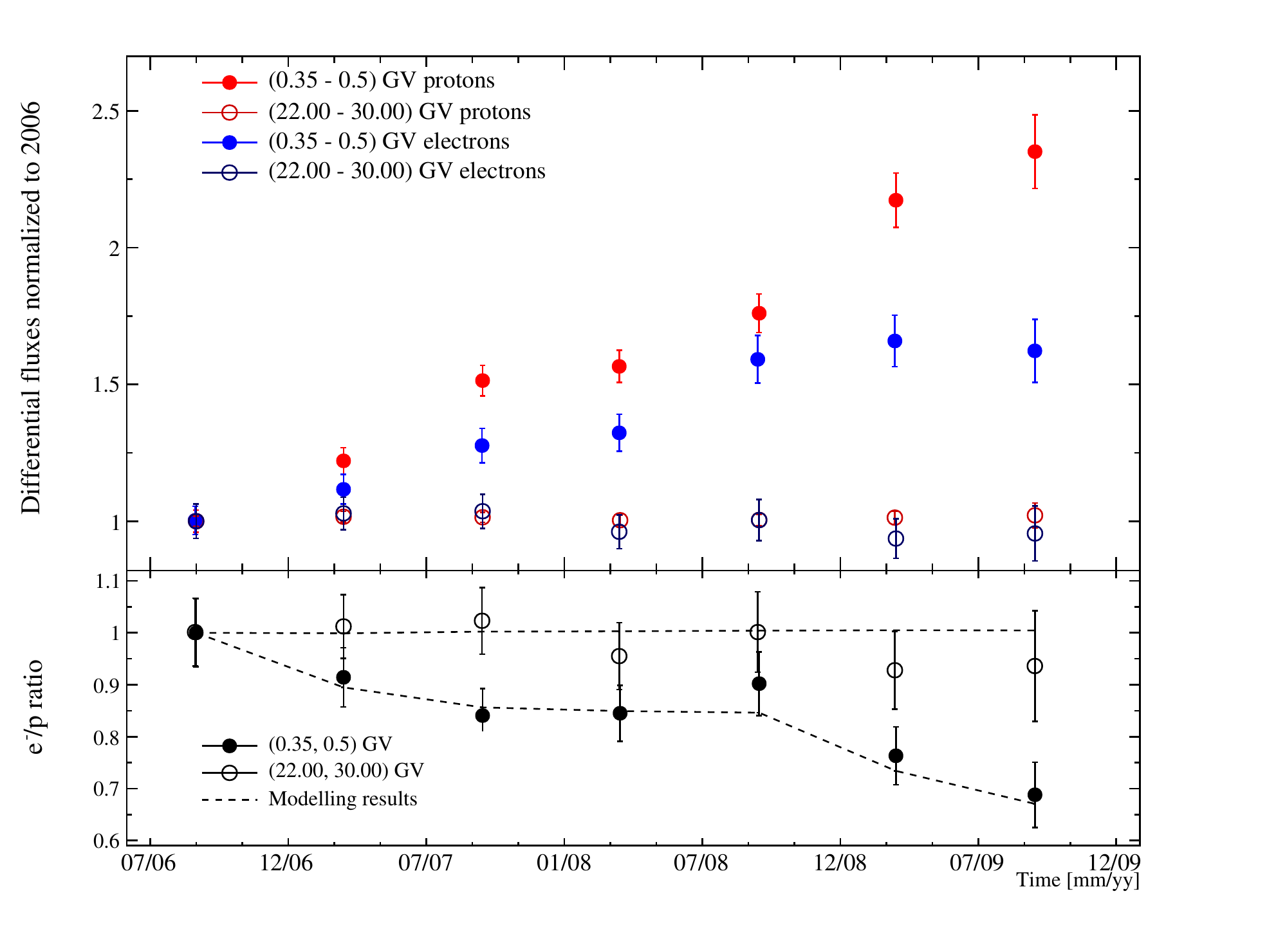}%
\caption[Proton and electrons temporal evolution at solar minimum.]{Temporal evolution of proton (red)
and electron (blue) intensities as measured by PAMELA from July 2006 and December 2009. Upper panel: proton and 
electrons at low rigidity (filled circles), in the interval $(0.35 - 0.5)$ GV, and at high rigidity (open circles), $(22.00 - 30.00)$ GV.
Lower panel: electron to proton ratio measured for the same period and rigidities, together with corresponding model results (dashed lines).}
\label{fig:pos_prot_ratio}
\end{figure}

On top of the time dependence of the solar modulation, a charge sign dependence is also present. This is due to 
drift motions experienced by cosmic ray traveling in the expanding solar wind, 
sensing the gradients and curvatures of the global HMF, and the presence of the heliospheric 
current sheet (HCS) which is a thin neutral sheet where the oppositely directed open
magnetic field lines from the Sun meet. The effect is evident in cosmic ray observations as a 22 year periodicity, reflecting the magnetic polarity changes 
of the Sun magnetic field: every 11 years, at solar maximum, the polarity of the sun magnetic field reverses, changing the sign of 
the projection of the dipole term of the sun magnetic field A~\footnote{In the complex sun magnetic filed,
the dipole term nearly always dominates the magnetic field of the solar wind.}. 
When magnetic field lines are pointing outward in the northern heliohemisphere, so called $A>0$ polarity epochs,
 positively charged particles are expected to drift into the inner heliosphere mainly over the solar
poles and out along the HCS. In $A<0$ epochs drift patterns reverse, with positively charged particles 
reaching the Earth mainly through the equatorial regions. Negatively charged particles follow opposite drift 
patterns during the same polarity epochs. Thus, during the same polarity epoch, oppositely charged particles 
traverse different regions of the heliosphere and sense different modulation conditions~\cite{Potgieter_2013},
representing a valuable tool to study drift effects in solar modulation. 
Solar minimum activity periods are particularly suited for charge-sign dependent solar modulation studies, 
being drift effects expected to be at their largest with very few solar-generated transients disturbing 
the modulation region.

New evidence for charge-sign dependent solar modulation was provided by studying the
different response to solar modulation conditions during solar minimum  
of cycle $23$ of the proton and electron fluxes measured by PAMELA ~\cite{ep_ratio}. 
Figure \ref{fig:pos_prot_ratio} shows the temporal evolution of proton and electron intensities 
from July 2006 to December 2009.  
The upper panel presents particle fluxes, normalized to the values measured in 2006, for two 
selected rigidity intervals. 
At low rigidity, between $0.35-0.5$ GV, protons clearly show a significantly steeper recovery trend respect to electrons, 
increasing by a factor $\sim2.4$ over the whole period, significantly more than electrons that increased by only a factor $\sim1.6$. 
In fact, with reduced solar activity, the waviness of the HCS is also reduced,  
impacting relatively more on protons, that reach the Earth drifting mainly through the equatorial regions. 
These more favorable modulation conditions imply a faster increase of proton intensities, while impacting relatively less 
on electrons, which mostly escape the changes of the HCS mainly drifting inward from the polar regions of the heliosphere. 
Differences are emphasized in the lower panel, 
where data are presented in terms of electron to proton ratios at the same rigidities. 
The effect dissipates with increasing rigidity, 
with no difference  within the experimental uncertainties already at $~11 $ GV, e.g. see the high rigidity interval $22 - 30$ GV.  
As previously mentioned, the evolution of the spectra of both electrons and proton was studied with a comprehensive 3D 
modulation model. The results of the computation in terms of electron to proton ratios are
shown on top of the data (dashed black lines). The observed effect can only be reproduced if all modulation processes, i.e. also drift effects are
included ~\cite{ep_ratio_modeling}. 

\begin{figure}
\centering 
\includegraphics[width=.9\textwidth]{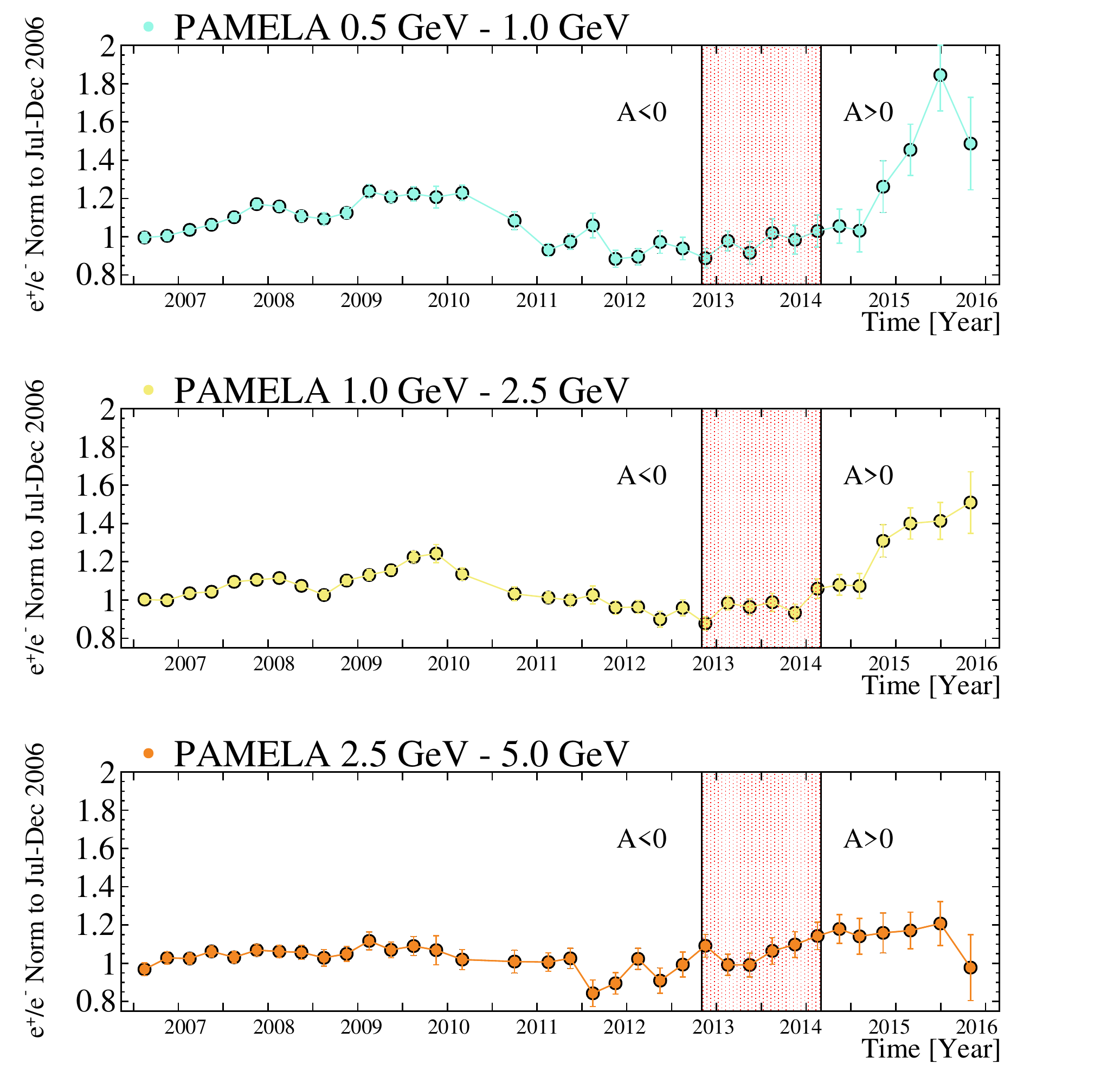}%
\caption[Positron to electron ratio.]{Positron to electron ratio as measured by PAMELA from July 2006 to December 2015. 
Data were normalized to 2006. Each panel refers to a different low energy interval, form top to bottom: $(0.5, 1.0)$~GeV, $(1.0, 2.5)$~GeV,
and $(2.5,5.0)$~GeV. The red shaded area indicates the period of polarity reversal, whit no well defined HMF polarity.  }
\label{fig:pos_el_ratio}
\end{figure}

Strong evidence of charge-sign dependent solar modulation
was also provided by PAMELA measurements of the positron to electron ratio, performed 
between July 2006 and December 2015~\cite{Adriani_2016_elpos_mod}. These data, shown in Figure \ref{fig:pos_el_ratio}, present the first 
clear indication of the evolution of drift effects
during different phases of the solar activity and the dependence on particle rigidity.
The three panels show the temporal evolution of the $e^{+}/e^{-}$ ratio as measured in intervals of about $3$ months
at different energies. The red shaded area represents the period during which the 
process of polar field reversal took place \cite{reversal}. 
Data clearly show a temporal dependence of the positron to electron intensity ratio, and 
can be interpreted in terms of drift, using the 
tilt angle of the HCS as the most appropriate proxy of solar activity in this context. 
During the solar minimum activity period, from 2006 to 2009, low energy positron intensities increased more than 
the electrons ones, about $20\%$ in the two intervals from $0.5$~GeV and $2.5$~GeV, and $\sim10\%$ for the 
third interval up to $5.0$~GeV. The consequently increasing positron to electron ratio can be interpreted in terms of 
the progressively reduced impact of the waviness of the HCS (decreasing tilt angle) on positrons
that, similarly to protons, drift towards the Earth mainly through the equatorial regions of the heliosphere.
After 2009 the solar activity started to increase, with the $e^{+}/e^{-}$ ratio decreasing until the middle of 2012.
 During this period the tilt angle increased, again impacting on positrons 
more than on electrons, producing a faster decrease of the positron 
flux respect to the electron one. Such decrease of the $e^{+}/e^{-}$ continued until the 
continuously increasing solar activity influenced both particles equally, leading to a steady positron to electron ratio (until mid 2013).
In 2014, approximatively four months after the polarity inversion, a sudden rise was observed in the data up to late 2015: 
the first two panels of Figure \ref{fig:pos_el_ratio}
show a larger increase of positrons respect to electrons of about $80\%$ and $50\%$, respectively.
This sudden rise appears to be a consequence of the polarity reversal of the HMF and of 
the changed drift patterns. 
Interestingly, PAMELA data allow to monitor charge sign dependent
 solar modulation effects also during the turbulent transition phase, lasted about 1 year 
during which the polarity reversal was completed, 
and show a ratio that slowly increased already in 2013 until the middle of 2014. 

PAMELA data are currently under analysis to study the temporal variation of additional particle species, 
focusing in particular on the solar modulation of helium fluxes.

\subsection{Solar Particle Events} 

The problem concerning the mechanism and site of Solar Energetic Particles (SEPs) acceleration remains an open question.  
SEPs are a population of particles emitted by the Sun with energy ranging from a few tens of keV 
to a few GeV which are associated with solar flares and Coronal Mass Ejection (CME). These events frequently inject large amounts of nuclei into 
space, whose composition varies from event to event and it is heavily linked to the production mechanisms that take place. 
Whether the Sun accelerates particles at low altitudes through magnetic reconnection or higher in the outermost layers of its 
atmosphere (like the corona) through coronal mass ejection-driven shocks, or perhaps an admixture of the two, is still unclear 
\cite{Mathews_1990,galina}. This kind of uncertainty involves both low energy particles measured \textit{in situ} 
and the higher energy populations which lead to particularly energetic phenomena called Ground Level Enhancements (GLEs).  
These are produced when solar protons in the $\sim$GeV range start a nuclear cascade through the Earth's atmosphere 
that can be observed by detectors at ground level, such as Neutron Monitors, as an increase above the background produced by 
ordinary galactic cosmic rays \cite{Reames_2013}. 
GLEs are very rare (only 71 have been registered so far) but very important because provide a good opportunity
to detect matter ejected from the Sun that reaches the Earth within tens of minutes. How
these particles are accelerated to GeV energies has been a matter of study since 1940s when these phenomena were first observed.

\begin{table}[t]
\centering
\begin{tabular}{lll}
\toprule
\bf{Solar Particle Event} & \bf{Associated Flare Class} \\
\midrule
{\bf 2006} December 13$^{\text{th}}$ & X3.4/4B \\
\quad \quad \ \ December 14$^{\text{th}}$ & X1.5 \\
\hline
{\bf 2011} March 21$^{\text{st}}$ & M3.7 \\ 
\quad \quad \ \ June 7$^{\text{th}}$ & M2.5/2N \\
\quad \quad \ \ September 6$^{\text{th}}$ & / \\ 
\quad \quad \ \ September 7$^{\text{th}}$ & / \\ 
\quad \quad \ \ November 4$^{\text{th}}$ & /\\ 
\hline
{\bf 2012} January 23$^{rd}$ & M8.7 \\
\quad \quad \ \  January 27$^{\text{th}}$ & X1.7/1F \\
\quad \quad \ \  March 7$^{\text{th}}$ & X5.4 \\
\quad \quad \ \  May 17$^{\text{th}}$ & M5.1/1F \\
\quad \quad \ \  July 7$^{\text{th}}$ & X1.1 \\
\quad \quad \ \  July 8$^{\text{th}}$ & M6.9/1N \\
\quad \quad \ \  July 19$^{\text{th}}$ & M7.7 \\
\quad \quad \ \  July 23$^{rd}$ & / \\\hline
{\bf 2013} April 11$^{\text{th}}$ & M6.5/3B \\
\quad \quad \ \  May 22$^{nd}$ & M5.0 \\
\quad \quad \ \  September 30$^{\text{th}}$ & C1.3 \\
\quad \quad \ \  October 28$^{\text{th}}$ & / \\
\quad \quad \ \  November 2$^{nd}$ & / \\
\hline
{\bf 2014} January 6$^{\text{th}}$ & C2.1\\
 \quad \quad \ \  January 7$^{\text{th}}$ & X1.2 \\
\quad \quad \ \  February 25$^{\text{th}}$ & X4.9 \\
\quad \quad \ \  April 18$^{\text{th}}$ & M7.3 \\
\quad \quad \ \  September 1$^{\text{st}}$ & / \\
\quad \quad \ \ September 10$^{\text{th}}$ & X1.6 \\
\hline 
{\bf 2015} June 21$^{\text{st}}$ & M2.0 \\
\quad \quad \ \  June 25$^{\text{th}}$ & M7.9 \\
\bottomrule
\end{tabular}
\caption{ List of Solar Particle Events registered by PAMELA (2006-2015)}
\label{tab:a}
\end{table}

\begin{figure}[t]
\centering
\includegraphics[width=0.9\textwidth]{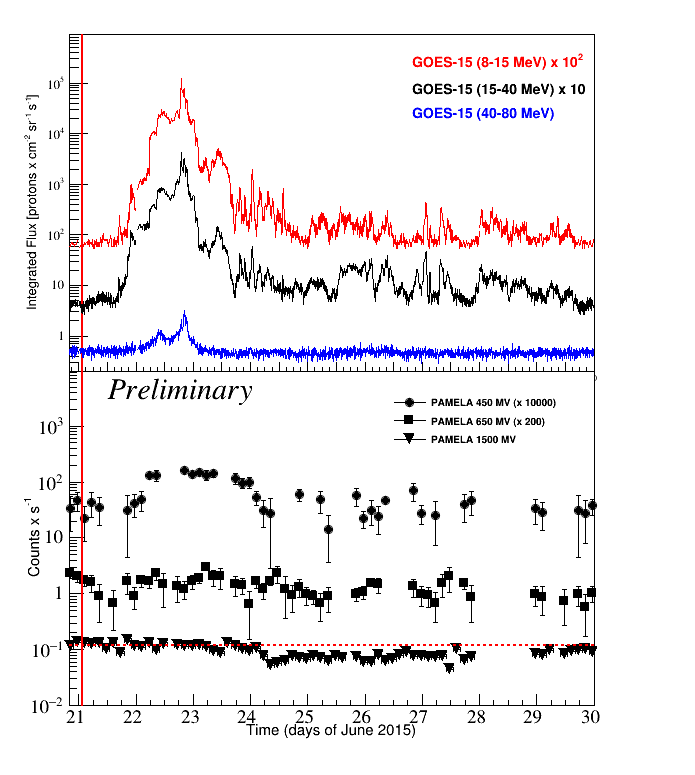}
\caption{Data for the  June $21^{\text{st}}$ $2015$ SEP event. Top panel: the GOES proton fluxes as a function of time in three energy intervals is presented. 
In the bottom panel PAMELA counts per second are shown for three different rigidities. The vertical line shows the 
maximum time of the flare on the Sun, while the horizontal line shows the preliminary counting rate measured few days before the 
event in an energy range around 1500 MV. The Forbush decrease created by the incoming CME is clearly visible in PAMELA data.}
\label{fig:time}
\end{figure}

Other than pursuing scientific knowledge, the study of SEPs is crucial because of the effects that these particles could have in 
the daily human life. Indeed the $>$40 MeV SEP component deflected by the geomagnetic field towards highest latitudes
may temporarily smother the ionosphere and interfere with radio communications damaging satellites or man-made spacecraft.

In this picture PAMELA fills the largely unexplored energy gap between the particles detected in space (below few hundreds of MeV) and 
 particles detected at ground level (above few GeV). The orbit of the spacecraft allows the PAMELA detector to
measure a wide range of energies starting from few tens of MeV.
First PAMELA observation of SEPs occurred in late 2006 with the December 13$^{\text{th}}$/14$^{\text{th}}$ events which also represent 
the first direct measurement of SEPs in space with a single instrument in the energy range from $\sim$80 MeV/n
to $\sim$3 GeV/n. Solar helium nuclei (up to 1 GeV/n) and protons (up to $\sim$2 GeV/n) were recorded. 
A study on the comparison between PAMELA data and Neutron Monitors, GOES and Ice Top was carried out, together with
a deep study on the spectral shape fitting. More details can be found in  \cite{Adriani_flare_2011}. 

Since then PAMELA collected a large amount of  data from other solar events.
In Table \ref{tab:a} a list of Solar Particle Events measured by PAMELA in the period 2006-2015 is reported together
with the class of the associated flare on the Sun \cite{NOAA}. 
\begin{figure}[t]
\centering
\includegraphics[width=0.9\textwidth]{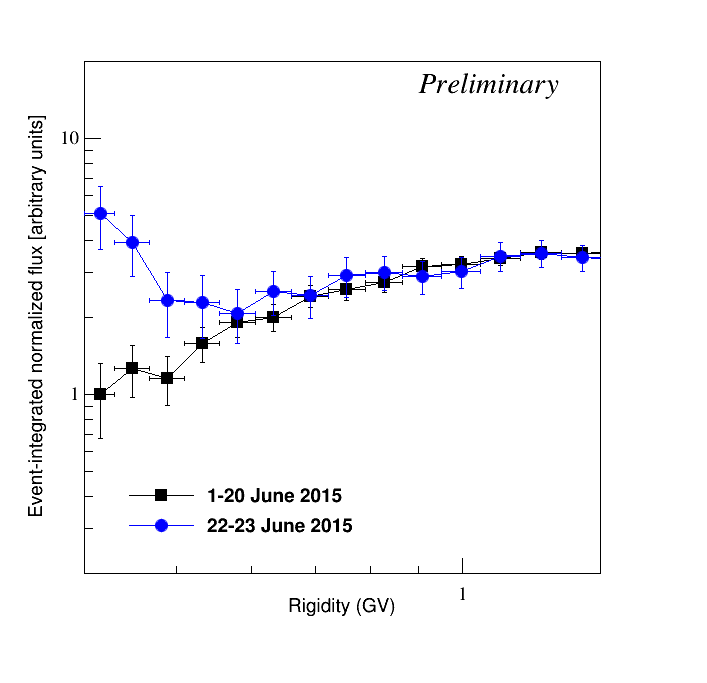}
\caption{Normalized event-integrated proton flux in the interval between June 22$^{\text{nd}}$ and June 23$^{\text{rd}}$ 2015 (blue circles) 
as a function of rigidity, together with the background proton flux from June 1$^{\text{st}}$ to June 20$^{\text{th}}$ 2015 (black squares).}
\label{fig:diff}
\end{figure}
In particular, the event of June 21$^{\text{st}}$ 2015 was of great interest. In fact, it was observed by a large number of
instruments, from magnetograms to ground stations. PAMELA has contributed to the study of the particles associated 
to the full-halo CME originated from the active region NOAA 12371, which caused a strong goemagnetic storm. 

Figure \ref{fig:time} (bottom panel) shows the preliminary results on the 
rate of protons measured by PAMELA in three rigidity intervals 
(from 450 MV to about 1500 MV) collected every 3 hours during the June 21$^{\text{st}}$ 2015 SEP event. For an easier comparison, the integrated
proton flux data from the GOES-15 satellite in three lower energy channels is also shown in the top panel. The 
intensity profiles show a relatively slow rise, as the SEP event originated from the central 
portion of the solar disk. Moreover, the PAMELA rate shows a very low energy extension, falling
into background above $\sim$600 MV (black squares in the bottom panel of Figure \ref{fig:time});
this means that a small number of particles above this rigidity have reached the distance of 1 AU maybe because the event 
was not powerful enough to accelerate particles at this energies. The preliminary PAMELA counting rate around 1500 MV
shows a clear Forbush decrease on June 24$^{\text{th}}$ 2015 caused by the halo CME passing through Earth. After June 
29$^{\text{th}}$ 2015 the effect of the CME ceased and the rate rose up again to the normal condition.

Figure \ref{fig:diff} shows the event-integrated proton flux as a function of rigidity measured by
PAMELA between June 22$^{\text{nd}}$ and June 23$^{\text{rd}}$ $2015$ (blue points) 
compared to the galactic flux collected during the first 20 days of June (black points). The galactic flux measured at $420$ MV were 
 normalized to $1$. This normalization was applied to the flux measured during the SEP event. 
 As stated before, the rise above the background
is evident below $\sim$600 MV, while above this limit the flux measured during the SEP event and the background flux overlap.
  
\subsection{Magnetospheric effects of high-energy SEPs}

\begin{figure} [t]
\centering
\includegraphics[width=.9\textwidth]{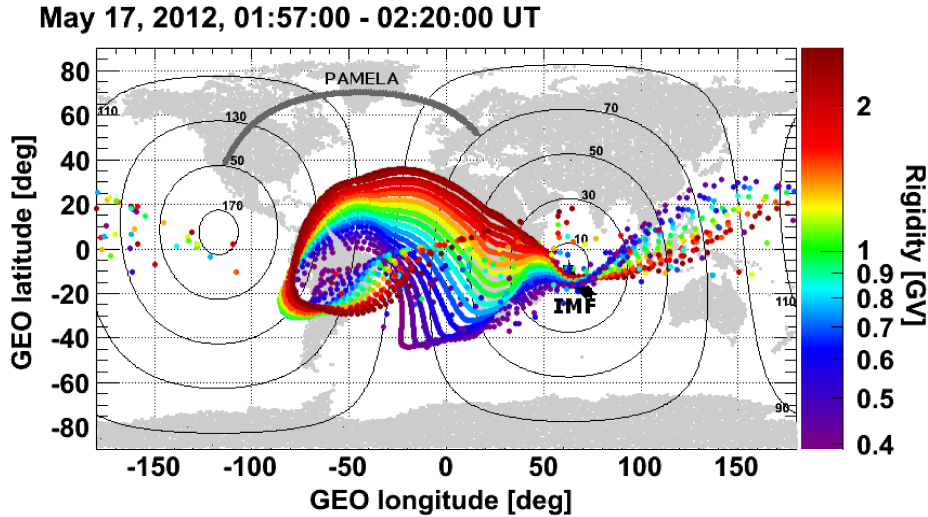}
\caption{Reconstructed asymptotic directions of view with rigidities between 0.4 and 2.5 GV, for PAMELA during the orbit 
0157-0220 UT on the  May 17$^{\text{th}}$ 2012 in geographic coordinates. PAMELA trajectory is shown in gray. Colour coding associates red to 
higher rigidities and blue to lower ones. The direction of the IMF (black dot) obtained from the Omniweb database together 
with contours of constant pitch angle (black curves) relative to the direction of the IMF itself are also shown.} 
\label{fig:pitchangle_1}
\end{figure}

\begin{figure}[t]
\centering
\includegraphics[width=.8\textwidth]{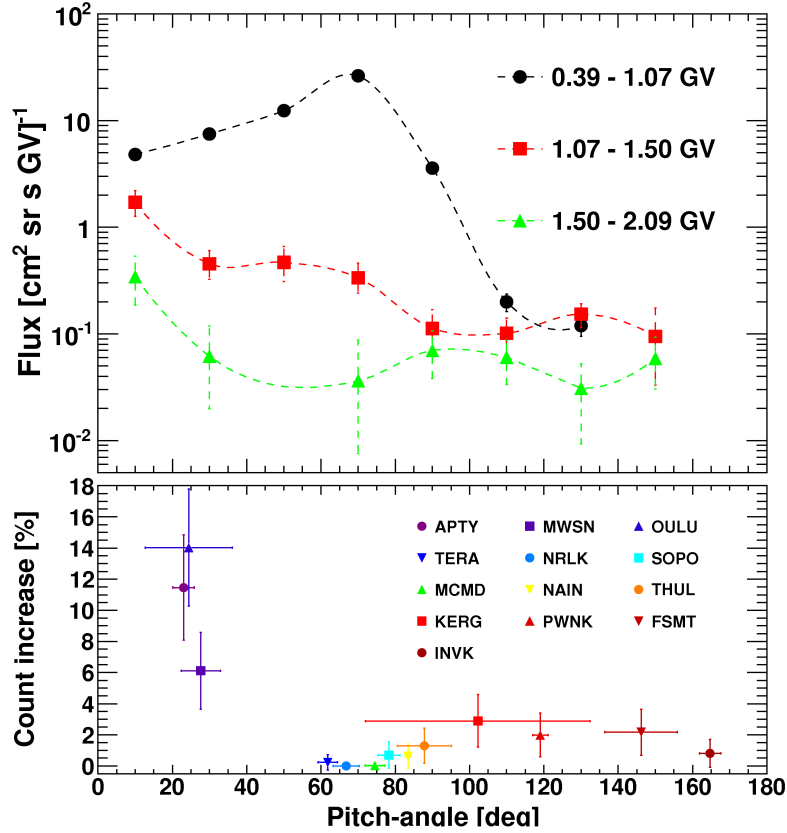}
\caption{Top panel: scaled intensities observed by PAMELA in three rigidity ranges during the May 17$^{\text{th}}$ 2012 event.
Bottom panel: count increase of the worldwide
neutron monitor pitch angle distribution averaged between 0157 to 0220 UT.}
\label{fig:pitchangle3pads}
\end{figure}

A proper understanding of the SEP acceleration has to account for the significant effects introduced 
by the particle transport in the heliosphere \cite{Vashenyuk_2006,McCracken_2008}. 
In fact, the energetic particles coming from the Sun are subject to processes
like adiabatic focusing, scattering by Alfv\`en waves, or reflections at 
various magnetic structures. Moreover, all these mechanisms may be entangled 
together and act on SEP propagation in different ways or at different times 
of the propagation obscuring the effects of acceleration. 
To understand these effects the spectra have to be measured in the widest 
energy range. 

PAMELA experiment produce the first direct evidence \cite{Adriani_ApjL_2015} of an evolving
pitch angle\footnote{The angular distribution of the moving particles about 
the magnetic field direction.} distribution resulting from transport through local magnetic 
perturbations within the Earth's magnetosheath during the May 17$^{\text{th}}$ 2012 
event which was the first GLE of solar cycle $24$.
The presence of anisotropic particle populations in these special kind of events 
provide useful information of the acceleration phase.
GLEs often consist of a very rapid, anisotropic onset associated with a well-connected
beam of outward and anti-Sunward moving particles aligned with the interplanetary 
magnetic field (IMF). This onset were interpreted as a result of magnetic focusing 
with little scattering and little velocity dispersion \cite{Earl_1976}. 
As the event goes on, however, the distribution becomes more isotropic and decays
(due to pitch-angle scattering) as the particles diffuse into the heliosphere.
Therefore, the study of the pitch-angle distribution is important to better 
understand the nature of the particles arriving from the Sun.
PAMELA experiment is able to measure the incident trajectory of the detected particles employing 
a combination of trajectory reconstruction and its highly-precision silicon tracking
system.  The particle tracing methods of \cite{smart_shea_2000} were used to determine
the asymptotic direction for each incident particle from which the particle pitch angle
with respect to the magnetic field was deduced. The method relied on both an internal
field model based on the International Geomagnetic Reference Field (IGRF) maps \cite{IGRF} and an external field model based on the study of
Tsyganenko \cite{Tsyganenko_2007}, while solar wind and IMF parameters were obtained 
from the high-resolution Omniweb database (https://omniweb.gsfc.nasa.gov).
PAMELA results shows a constantly changing, but narrow ($\sim$20$^{\circ}$ field of view) swath
of pitch angle along its orbit. To illustrate the instrument sensitivity, Figure 
\ref{fig:pitchangle_1} shows the asymptotic directions in geographic coordinates (GEO) resulting from the instrument 
field of view for rigidities ranging from 0.4 to 2.5 GV for the polar pass that 
first registered the event (0157 UT - 0220 UT). 

As the satellite moves through its orbit, it acts as a moving  observatory, with a sensitivity to pitch angle that depends on spacecraft  
location (or time). The ability to measure the particle intensity with a varying sensitivity to  pitch angle and over a large pitch-angle 
provided an unprecedented view of the spatial evolution of high-energy solar particle events. PAMELA data has been collected over $\sim 20$ minutes, 
time period over which the incoming GLE beam was assumed as stationary. 
\begin{figure}[t]
\centering
\includegraphics[width=4in]{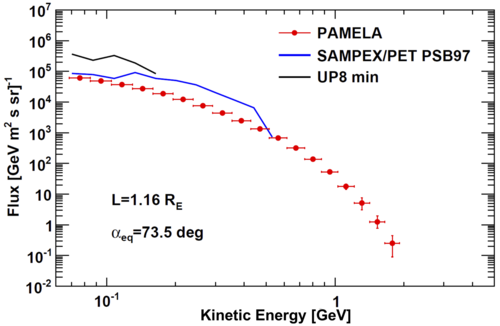}  
\caption{PAMELA trapped proton energy spectrum \cite{ref:TrappedProt2015} for illustrative $\alpha_{\text{eq}}$ and $L$-shell values, compared with the predictions from the UP8-min \cite{ref:AP8} and the PSB97 \cite{ref:PSB97} models.}
\label{fig:trapped_protons}
\end{figure}
The dependence on pitch angle for three different rigidity ranges is shown in Figure \ref{fig:pitchangle3pads} top panel. PAMELA results show two populations 
simultaneously with very different pitch angle
distributions. A low-energy component (0.39-1.07 GV) confined to pitch angles $\leq 90^{\circ}$ and a high-energy component
(1.50-2.09 GV) that was beamed with pitch angles $\leq 30^{\circ}$. The component with intermediate energies (1.07-1.50 GV) suggests a
transition between the low and high energies, exhibiting a peak at small pitch angles and a cutoff at 90$^{\circ}$. At rigidities $\geq$1 GV,
the particles were mostly field-aligned and are consistent with neutron monitor observations (Figure \ref{fig:pitchangle3pads}, bottom panel).

The rigidity dependence of the particle pitch angle distribution suggests that the low-energy component underwent significant
scattering while the highest-energy SEPs reached the Earth undisturbed by dispersive effects. Since both populations were measured at Earth 
simultaneously and very early during the event, the scattering must take place locally, probably by the Earth magnetosheath.

\subsection{Magnetospheric observations}
Thanks to its identification capabilities and the semi-polar orbit, PAMELA was able to carry out comprehensive observations of the 
geomagnetically trapped and albedo cosmic-ray populations in the near-Earth environment, over a wide range of energies and latitudes. 
These measurements are supported by an advanced trajectory tracing analysis based on a realistic modeling of the Earth's magnetosphere
\cite{ref:Bruno2016}, which enables the separation of particles of interplanetary and atmospheric origin, the study of the sub-cutoff
components (trapped and albedo), and the investigation of geomagnetic storm effects.

\subsubsection{Geomagnetically trapped particles}

\begin{figure}[t]
\centering
\includegraphics[width=3.6in]{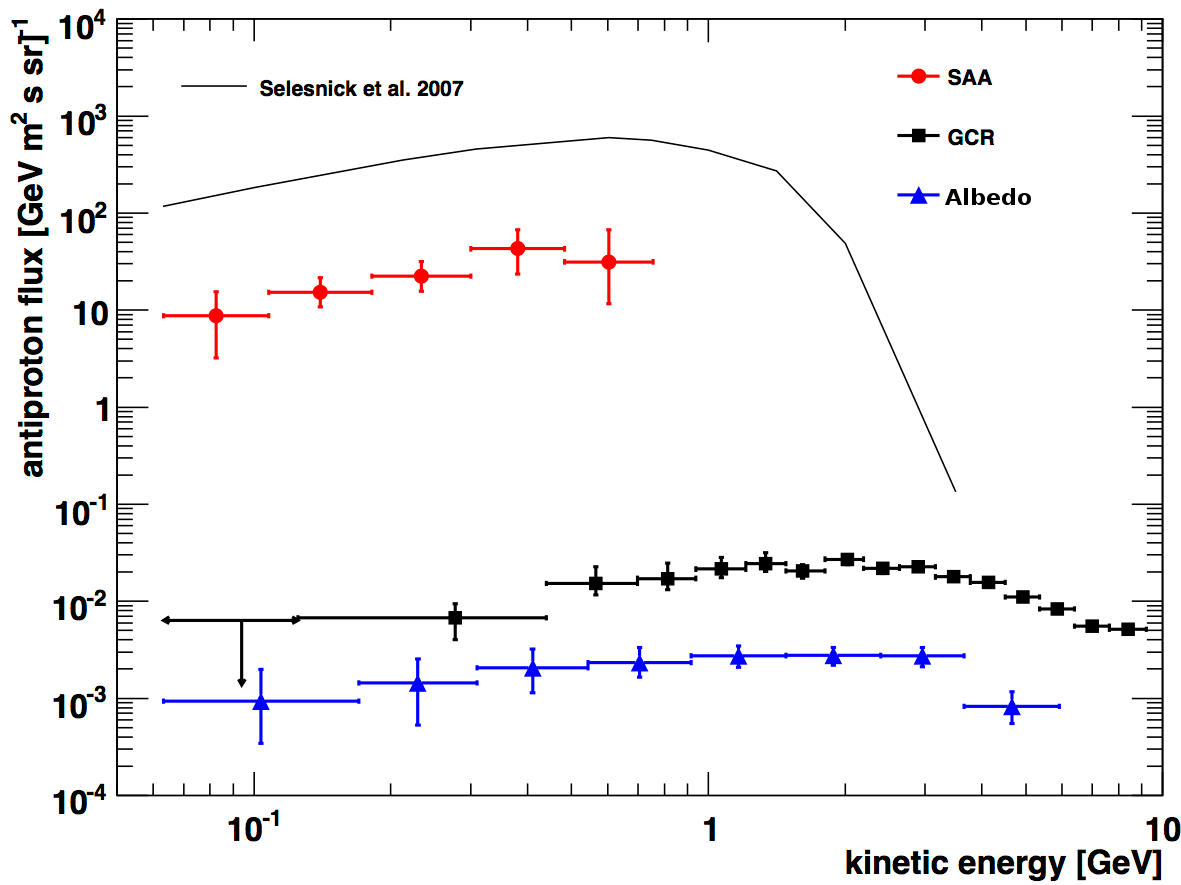}
\caption{The geomagnetically trapped antiproton spectrum measured by PAMELA in the SAA \cite{ref:TrappedPbars2011}. 
For a comparison, a trapped antiproton theoretical calculation \cite{ref:Selesnick2007}, the mean albedo antiproton spectrum 
and the galactic antiproton spectrum measured by PAMELA are also shown.}
\label{fig:trapped_pbarelepos}
\end{figure}

\begin{figure}
\centering
\includegraphics[width=4.7in]{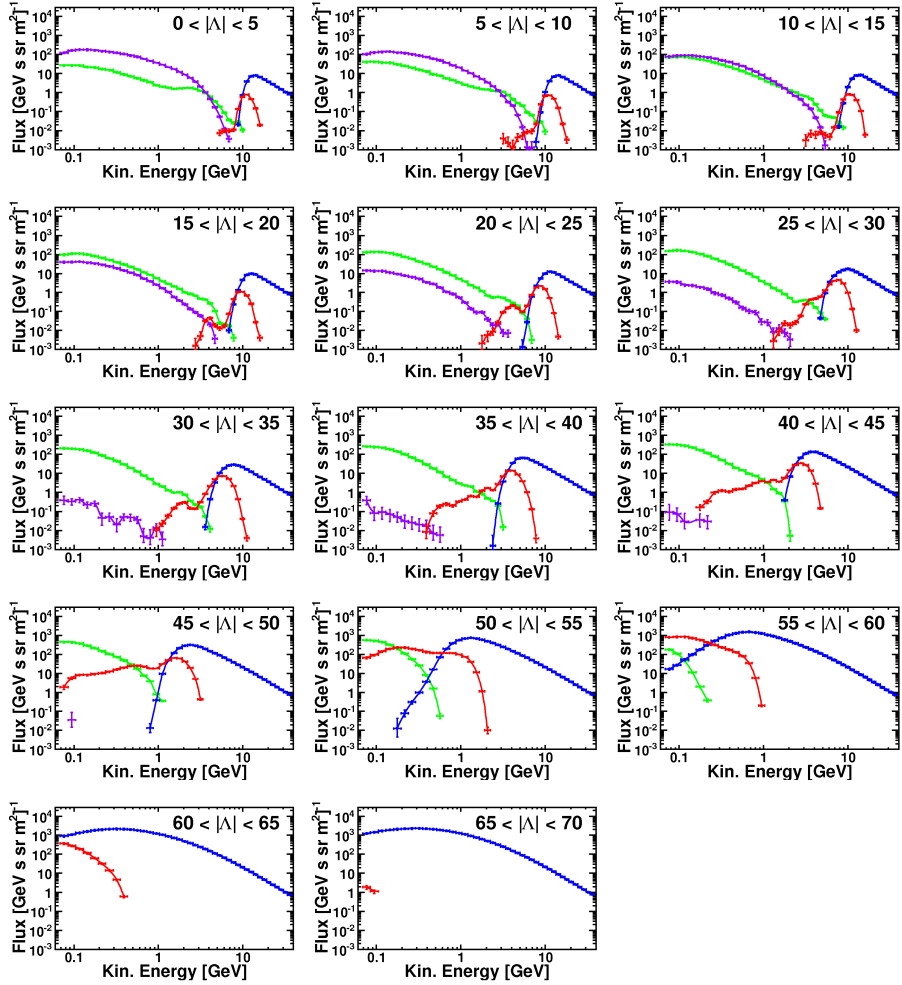}     
\caption{Differential energy spectra outside the SAA for different bins of AACGM la\-ti\-tude $|\Lambda|$ \cite{ref:Albedo2015}. 
Results for the several proton populations are shown: quasi-trapped (violet), precipitating (green), pseudo-trapped (red) and galactic (blue).}
\label{fig:albedo_protons}
\end{figure}

The Van Allen belts are regions of the Earth's magnetosphere where energetic charged particles experience long-term magnetic 
trapping. They constitute a well-known hazard to spacecraft systems, significantly constraining human activities in space. 
Speci\-fi\-cally, the inner belt is mainly populated by energetic protons, mostly originated by the decay of albedo neutrons
according to the so-called CRAND mechanism \cite{CRAND}. Despite the significant improvements made in the latest decades, the modeling of 
the trapped environment is still incomplete, with largest uncertainties affecting the high energy fluxes in the inner zone and 
the South Atlantic Anomaly (SAA), where the inner belt makes its closest approach to the Earth. This is exactly the observational
region explored by the PAMELA mission.
\begin{figure}[t]
\centering
\includegraphics[width=3.6in]{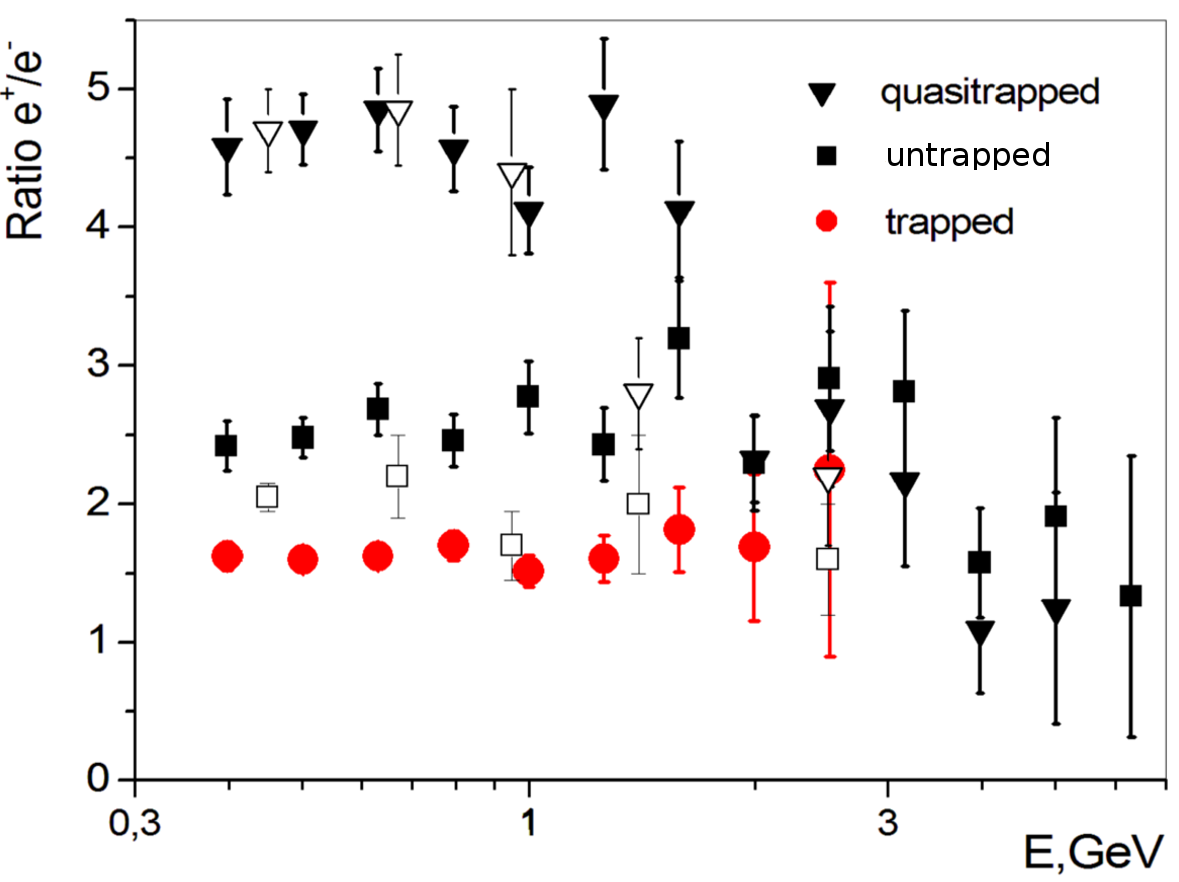}     
\caption{Positron/electron ratio as a function of energy for trapped, quasi-trapped and un-trapped 
components \cite{ref:TrappedPosit2016}. The AMS-01 data for long-lived and short-lived particles \cite{ref:Ams01} 
are also shown by open points for a comparison.}
\label{fig:trapped_pbarelepos1}
\end{figure}
PAMELA reported precise and detailed measurements of the geomagnetically trapped protons in the SAA, 
including energetic, angular and spatial distributions \cite{ref:TrappedProt2015}. Trapped fluxes were derived by
accounting for the strong pitch angle anisotropy due to the interaction with the atmosphere, and mapped as a function
of adiabatic invariant and geographic coordinates. PAMELA data extend the experimental observational range for trapped 
protons down to lower altitudes (McIlwain's $L$-shells $\sim$ 1.1 $R_{E}$), and up to the maximum energies cor\-res\-pon\-ding 
to the trapping limits (a few GeV). Figure \ref{fig:trapped_protons} compares the trapped proton spectrum measured by PAMELA 
(for illustrative equatorial pitch angle $\alpha_{\text{eq}}$ and $L$ values) with the predictions from two empirical models available in 
the same energy and altitude ranges: the AP8 \cite{ref:AP8} unidirectional (denoted with UP8) model for solar minimum conditions,
and the SAMPEX/PET PSB97 model \cite{ref:PSB97}. Model data were derived from SPENVIS (www.spenvis.oma.be). In general, the UP8 
model significantly overestimates PAMELA observations, while a better agreement can be observed with the PSB97 model. However, 
PAMELA fluxes do not show the same spectral structures present in the PSB97 predictions. A comparison study with the new AP9 model is in progress.

As shown in Figure \ref{fig:trapped_pbarelepos} PAMELA provided the first evidence of the existence of a stably-trapped antiproton population in the inner 
radiation belt \cite{ref:TrappedPbars2011}, mostly generated by the decay of albedo antineutrons (CRANbarD, in analogy to 
trapped protons). Such antiproton belt represents the major antiproton source near the Earth: 
the trapped flux (SAA) exceeds by three orders of magnitude the galactic flux at the present solar minimum, and by four 
orders of magnitude the re-entrant albedo flux.

\subsubsection{Re-entrant albedo particles}

\begin{figure}[t]
\centering
\includegraphics[width=5in]{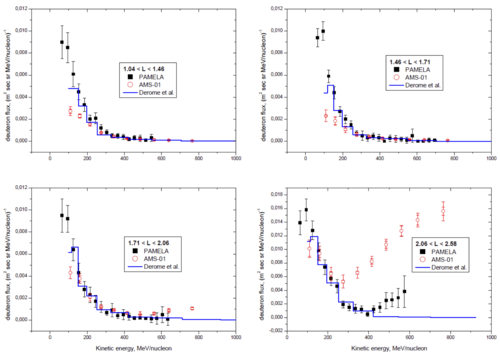}     
\caption{The albedo deuteron spectrum measured by PAMELA (black) for different geomagnetic latitudes. For comparison, 
the AMS-01 data (\cite{ref:lamanna2001}, red circles) are also reported.
Blue solid lines shows the calculation by \cite{ref:derome2001}.}
\label{fig:albedo_deuteron}
\end{figure}

\begin{figure}[t]
\centering
\includegraphics[width=4.5in]{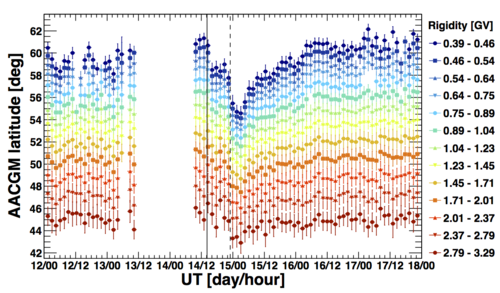}
\caption{Time profile of the geo\-ma\-gne\-tic cutoff latitudes measured by PAMELA, 
for different rigidity bins (color code), during the December $14^{\text{th}}$
2006 geomagnetic storm \cite{ref:Gstorm2016}. Vertical solid and dashed 
lines mark the shock and the magnetic cloud arrival, respectively.}
\label{fig:cutoff_var}
\end{figure}

The ma\-gne\-to\-sphe\-ric radiation also includes populations of albedo particles originated by the collisions of
CRs with the atmosphere. Using the back-tracing analysis, the detected sample was classified into quasi-trapped and un-trapped components:
the former consists of relatively long lived particles ($1-10$ seconds), detected in the near equatorial region, with trajectories similar to those of 
stably-trapped from the inner radiation belt; the latter was found to spread over all explored latitudes, including short-lived
(precipitating, $\sim 10^{-1}$ second)
particles together with a long-lived (pseudo-trapped, $\sim 1$ second) component constituted by particles with rigidities near the local geomagnetic cutoff and
characterized by a chaotic motion (non-adiabatic trajectories). Figure \ref{fig:albedo_protons} shows the spectra of the various albedo proton
components outside the SAA (B$>$0.23 G) measured at different AACGM latitudes\footnote{Altitude Adjusted Corrected GeoMagnetic (AACGM)
coordinates, developed to provide a more realistic description of high latitude regions by
accounting for the multipolar geomagnetic field \cite{AACGM}} $|\Lambda|$, along with the galactic component. Fluxes were 
averaged over longitudes.

PAMELA measured the fluxes of trapped and albedo electrons and positrons \cite{ref:TrappedPosit2016,ref:AlbedoEle2009}, originated  by the decay of 
charged pions (at region L-shell$\sim$1.2, magnetic field B$<$0.21 G inside SAA). Figure \ref{fig:trapped_pbarelepos1} 
displays the positron-to-electron ratio as a function of energy, for trapped, quasi-trapped and un-trapped populations.
The AMS-01 data \cite{ref:Ams01} for long-lived and short-lived albedo particles are also
shown by open points for a comparison (at L-shell$\sim$1.2, B$>$0.23 G near SAA). The East-West asymmetry for cosmic protons naturally 
explains the observed difference between the quasi-trapped and un-trupped populations.
The discrepancy with trapped population can be attributed to the different
production mechanisms for the two components (see \cite{ref:TrappedPosit2016} for details).

PAMELA albedo measurements were extended to deuterons \cite{ref:Deuteron2015}. 
Figure \ref{fig:albedo_deuteron} shows the measured spectra of albedo deuterons for different geomagnetic latitudes. The results 
of a calculation \cite{ref:derome2001} as well as the data of the AMS-01 data \cite{ref:lamanna2001} are reported 
for a comparison. The different galactic deuteron contribution in AMS-01 and PAMELA spectra results from the different 
geomagnetic cutoffs and acceptance of the two instruments (see \cite{ref:Deuteron2015} for details).

\subsubsection{Magnetospheric storms}
Finally, PAMELA data were used to measure the geomagnetic cutoff variation for high-energy ($\ge$80 MeV) protons during the 
December $14^{\text{th}}$ 2006  geomagnetic
storm \cite{ref:Gstorm2016}. The variations of the cutoff latitude as a function of rigidity were studied on relatively short timescales, 
corresponding to spacecraft orbital periods ($\sim$94 min). PAMELA results are the first direct measurement of these effects 
for protons with kinetic energies in the sub-GeV and GeV region.
Figure \ref{fig:cutoff_var} shows the geo\-ma\-gne\-tic cutoff AACGM latitudes measured by PAMELA as a function of time (December $12^{\text{th}}$ - 
December $18^{\text{th}}$ 2006) for different rigidity bins (color code). Each point denotes the cutoff latitude value averaged over a single spacecraft orbit; the 
error bars include the statistical uncertainties of the measurement. Data were missed from 1000 UT on December $13^{\text{th}}$ until 0914 UT on 
December $14^{\text{th}}$ 
because of an onboard system reset of the satellite. The evolution of the December $14^{\text{th}}$ magnetic
storm followed the typical scenario in which the cutoff latitudes move equatorward as a consequence of a CME impact on the ma\-gne\-to\-sphe\-re
with an associated transition to a southward IMF $B_{\text{z}}$ component. The re\-gi\-stered cutoff variation decreases with increasing rigidity, with 
a $\sim$7 deg maximum suppression at lowest rigidities (see Figure \ref{fig:cutoff_var}). The observed reduction in the geomagnetic shielding
and its temporal evolution were related to the changes in the magnetospheric configuration therefore allowing the investigation of
the role of IMF, solar wind, and 
geomagnetic parameters. 

\section{Conclusions and acknowledgments}

It was the 15th of June of 2006 when the PAMELA satellite-borne
experiment was launched from the Baikonur cosmodrome in
Kazakstan. Until 2016, PAMELA instrument 
had been making high-precision measurements of the charged component
of the cosmic radiation over four decades of energy with unprecedented statistics opening a new era of precision studies in cosmic rays.
 
The antiparticle components of the cosmic radiation were 
measured by the PAMELA instrument between a few tens of MeV and hundred of GeV, the most extended 
energy range ever achieved. The spectra of these particles show puzzling features that 
may be interpreted in terms of dark matter annihilation or 
pulsar contribution. 

Also, PAMELA measurements of the energy spectra of 
protons and helium nuclei pointed out spectral features that 
challenge our  
basic understanding of the mechanisms of production, acceleration and propagation 
of cosmic rays in the galaxy. A more sophisticated approach to the acceleration and propagation of 
cosmic rays is required to properly interpret these as well as PAMELA results on light nuclei and isotopes. 

The study 
of the time dependence of the various components of the cosmic
radiation measured by PAMELA from the unusual 23rd solar minimum through the following
period of solar maximum activity clearly shows solar modulation
effects as well as charge sign dependence introduced by particle drifts. 
PAMELA results provide the first clear observation of the relevance of drift effects and how these
unfolded with time from solar minimum to solar maximum, their dependence on the particle rigidity and the cyclic polarity of the solar magnetic field. 

PAMELA measurement of the energy spectra during more than 20 solar energetic particle 
events fills the existing energy gap between the highest energy
particles measured in space and the ground-based domain. Furthermore,
providing pitch angle measurements, it allows the study of the effects of
particle transport within interplanetary space over a broad range in
energy. Furthermore, PAMELA measurement of the 2012 May 17 SEP event provided the first comprehensive measurements of 
the effects of solar energetic particle transport in the Earth's magnetosheath. 

Finally, by sampling the particle radiation in different
regions of the magnetosphere, PAMELA data provided a detailed study
of the Earth s magnetosphere. As a results of this accurate study by the PAMELA experiment, an antiproton radiation belt was discovered 
around the Earth with flux exceeding  by three order of the magnitude the cosmic ray antiproton flux.

PAMELA operations were terminated in 2016 after almost 10 years of nearly uninterrupted measurements. 

\vspace{.7cm}

We would like to thank E. C. Christian, G. A. de Nolfo, M. S. Potgieter, J. M. Ryan and S. Stochaj for fruitful collaboration on the 
study of solar physics. 
We acknowledge partial financial support from The Italian Space Agency 
(ASI) under the program "Programma PAMELA - attivita' scientifica di analisi
dati in fase E", and support from our Universities and Institutes and from ASI, The
Deutsches Zentrum f\"{u}r Luft- und Raumfahrt (DLR), The
Swedish National Space Board, The Swedish Research Council, The Russian Space Agency
(Roscosmos), the Russian Ministry of Education and Science, project
 \textnumero$ \ 3.2131.2017$ and The Russian Foundation for Basic Research.


\begin{thebibliography}{99}


\bibitem{wiz1}		 \BY{Golden R.L.} et al.,   \IN{Astrophys. J.}{ 436}{1994}{ 769}.
\bibitem{wiz2}		 \BY{Hof M.} et al.,        \IN{Astrophys. J.}{ 467}{1996}{L33}.
\bibitem{wiz3}		 \BY{Golden R.L.} et al.,   \IN{Astrophys. J.}{ 457}{1996}{L103}.
\bibitem{Barbiellini}	 \BY{Barbiellini G.} et al.,\IN{Atron. \& Astrophys}{309}{1996}{L15}.
\bibitem{wiz4}		 \BY{Boezio M. } et al.,    \IN{Astrophys. J.}{ 487}{1997} {415}.
\bibitem{wiz5}		 \BY{Kremer J.} et al.,     \IN{Phys. Rev. Lett.}{83}{1999}{4241}.
\bibitem{Boezio}	 \BY{Boezio M.} et al.,     \IN{Phys. Rev. Lett.}{82}{1999}{4757}.
\bibitem{wiz6}		 \BY{Ambriola M.L.} et al., \IN{Nucl. Phys. B - Proc. Suppl.}{78}{1999}{32}.
\bibitem{Berg}		 \BY{Bergstr\"{o}m D.} et al., \IN{Astrophys. J.}{534}{2000}{L177}.
\bibitem{wiz7}		 \BY{Bidoli V.} et al.,     \IN{Astrophys. J. Suppl.}{132}{2001}{365}.
\bibitem{wiz8}		 \BY{Furano G.} et al.,     \IN{Adv. Space Res.}{31}{2003}{351}.
\bibitem{wiz9}		 \BY{Casolino M.} et al.,   \IN{Nature}{422}{2003}{680}.
\bibitem{wiz10}		 \BY{Narici L.} et al.,     \IN{Adv. Space Res.}{33}{2004}{1352}.
\bibitem{astroparticle}  \BY{Picozza P.} et al.,    \IN{Astropart. Phys.}{ 27}{2007}{ 296}.TV
\bibitem{ost04a}	 \BY{Osteria G.} et al.,    \IN{Nucl. Instrum. Meth. A}{ 535}{2004}{152}.
\bibitem{ors05a}	 \BY{Orsi S.} et al., 	    \IN{Proc. 29th Int. Cosmic Ray Conf. (Pune)} {vol. 3}{2005}{369}.
\bibitem{adr03}          \BY{Adriani O.} et al.,    \IN{Nucl. Instrum. Meth. A} {511}{2003}{72}.
\bibitem{calo}           \BY{Boezio M.} et al.,     \IN{Nucl. Instrum. Meth. A} {487}{2002}{407}.
\bibitem{nd}             \BY{Stozhkov Y.} et al.,   \IN{Internat. J. Modern Phys. A}{ 20}{2005}{6745}.
\bibitem{bos03}		 \BY{Boscherini M.} et al., \IN{Nucl. Instrum. Meth. A} {514}{2003}{112}.
\bibitem{sparvoli1}	 \BY{Sparvoli R.} et al.,   \IN{Adv. Space Res.}{ 37}{2006}{ 1841}.
\bibitem{gsi-campana}	 \BY{Campana D.} et al.,    \IN{Nucl. Instrum. Meth. A}{598}{2008}{696}.  


\bibitem{Adriani_PhysRep_2014}  \BY{Adriani O.} et al., \IN{Phys. Rep.}{544} {2014}{323}. 
 
\bibitem{Shong}         \BY{Shong J. A. D. and Hildebrand R. H. and  Meyer P.} et al.,                        \IN{Phys. Rev. Lett.} {12}{1964}{3}. 
\bibitem{Golden}         \BY{Golden R. L.} et al.,                        \IN{Phys. Rev. Lett.} {43}{1979}{1196}. 
\bibitem{Bogomolov}         \BY{Bogomolov E. A.} et al.,                        \IN{Proceeedings of the 16th International Cosmic Ray Conference (Kyoto)} {1}{1979}{330}. 
\bibitem{Adriani_positrons} \BY{Adriani O.} et al., 		                \IN{Nature} {458}{2009}{607}. 
\bibitem{Adriani_poslast}   \BY{Adriani O.} et al.,                             \IN{Phys. Rev. Lett.} {111} {2013} {081102}. 
\bibitem{Moskalenko09}      \BY{Moskalenko I. V. and Strong, A. W.}             \IN{Astrophys. J.}  {493} {1998}{694}.
\bibitem{Delahaye09}        \BY{Delahaye T.} et al.,		                \IN{ Astron. Astroph.} {501}{2009} {821}. 
\bibitem{Ackermann12}       \BY{Ackermann M.} et al.,		                \IN{ Phys. Rev. Lett.}  {108} {2012}{011103}.
\bibitem{AMS02-pos}	    \BY{Aguilar M.} et al.,      		        \IN{Phys. Rev. Lett.} {110} {2013} {141102}.  
\bibitem{Adriani_elelast}   \BY{Adriani O.} et al.,	  	 	        \IN{Astrophys. J.}  {810} {2015} {142}. 
\bibitem{Adriani_posfuture} \emph{Article in preparation}, 		
\bibitem{Adriani_ele}       \BY{Adriani O.} et al.,			        \IN{Phys. Rev. Lett.} {106} {2011} {201101}
\bibitem{AMS02-pos2}        \BY{Aguilar M.} et al. 			        \IN{Phys. Rev. Lett.} {113} {2014} {121101}. 
\bibitem{lipari}            \BY{Lipari P.} 					\IN{Phys. Rev. D.}    {95}{2017}{ 063009}
\bibitem{DMPos}             \BY{Tylka A. J.}					\IN{Phys. Rev. Lett.} {63} {1989} {840}. 
\bibitem{DMPos1}            \BY{Kaminkowski M. and  Turner M. S.}		\IN{Phys. Rev. D} {43} {1990} {1774}. 
\bibitem{DMPos2}            \BY{Cirelli M., Kadastik M., Raidal M. and Strumia A.}		\IN{Nucl. Phys. B} {813}{2008}{1}. 
\bibitem{DMPos3}            \BY{Cholis I. ,Dobler  G., Finkbeiner D. P., Goodenough L. and Weiner N.}		\IN{Phys. Rev. D} {80} {2009} {123518}. 

\bibitem{pulsars}           \BY{Atoyan A. M. , Aharonian F. A. and Volk H. J.}	\IN{Phys. Rev. D} {52} {1995} {3265}. 
\bibitem{remnants}          \BY{Blasi P.}					\IN{Phys. Rev. Lett.} {103} {2009} {051104}.
\bibitem{remnants1}         \BY{Ahlers M., Mertsch P. and  Sarkar S.}		\IN{Phys. Rev. D} {103} {2009} {123017}.
\bibitem{remnants2}         \BY{Fujita Y., Kohri K., Yamazaki R. and Ioka K.}	\IN{Phys. Rev. D} {80} {2009} {063003}.

\bibitem{Adriani_pbar} 	    \BY{Adriani O.} et al.,                     \IN{Phys. Rev. Lett.} {105} {2010} {121101}. 
\bibitem{AMS02-pbar}	    \BY{Aguilar M.}  et al., 			\IN{Phys. Rev. Lett.} {117} {2016} {091103}. 
\bibitem{BESS-Polar1-pbar}  \BY{Abe K.} et al., 			\IN{Phys. Lett. B} {670} {2008} {103}. 
\bibitem{BESSMOD}	    \BY{Asaoka Y.}  et al., 			\IN{Phys. Rev. Lett.} {88} {2002} {051101}. 
\bibitem{Cirelli}           \BY{Giesen G., Boudaud M., Génolini Y., Poulin V., Cirelli M., Salati P. and Serpico P.D.}  \IN{ JCAP }{1509}{2015} {023}
\bibitem{evoli}		    \BY{Evoli C.} et al.,			\IN{JCAP} {2015} {2015} {039}.
\bibitem{Adriani_Science}   \BY{Adriani O.} et al.,			\IN{Science} {332} {2011} {69}. 
\bibitem{AMS02-H}	    \BY{Aguilar M.} et al.,			\IN{Phys. Rev. Lett.} {114} {2015} {171103}. 
\bibitem{AMS02-He}	    \BY{Aguilar M.} et al.,			\IN{Phys. Rev. Lett.} {115} {2015} {211101}. 
\bibitem{BESS-Polar} 	    \BY{Abe K.} et al.,				\IN{Astrophys.J.} {822} {2016} {65}. 
\bibitem{ptuskin}   	    \BY{Ptuskin V. S.} et al.,			\IN{Astrophys.J.} {642} {2006} {902}.
\bibitem{blasi}	      	    \BY{Blasi P.} et al., 			\IN{Phys. Rev. Lett.} {109} {2012} {061101}.  
\bibitem{tomassetti1}	    \BY{Tomassetti N. and Donato F.} 		\IN{Astrophys. J. Lett.} {803} {2015}{L15}. 
\bibitem{tomassetti2}       \BY{Tomassetti N.}				\IN{Astrophys. J. Lett.} {815} {2015} {L1}. 
\bibitem{Adriani_BC}        \BY{Adriani O.} et al.,			\IN{Astrophys. J.} {791} {2014} {93}. 

\bibitem{AMS_BC} 						\IN{Proc. XXV European Cosmic Ray Symposium} {2016}.  

\bibitem{Adriani_HHeiso} \BY{Adriani O.} et al., \IN{ Astrophys. J.} {818}{2016} {68}
\bibitem{coste}		 \BY{Coste B.}  et al.,  \IN{Astron. Astroph.} {539} {2012} {88}. 

\bibitem{tomassetti} 	 \BY{Tomassetti N.} 	 \IN{N. Astrophys Space Sci} {342} {2012} {131}. 


 
\bibitem{Bruno14}     \BY{Bruno A.} et al.				     \IN{arXiv:1412.1765}{}{2014}.  
\bibitem{Adriani15}   \BY{Adriani O.}     			   	     \IN{Astrophys. J.}  {811}{2015} {21}. 
\bibitem{Panico15}    \BY{Panico B.} et al.				     \IN{Proc. 34th Int. Cosmic Ray Conf. (The Hague)}{410}{2015}.
\bibitem{Nagashima98} \BY{Nagashima K., Fujimoto K. and Jacklyn R. M.}       \IN{J. Geophys. Res.} {1031}{1998}{17429}. 
\bibitem{Amenomori06} \BY{Amenomori M.} et al.,			             \IN{Science} {314}{2006}{439}. 
\bibitem{Aglietta09}  \BY{Aglietta M.} et al., 		            	     \IN{Astrophys. J. Lett.} {692}{2009}{L130}.  
\bibitem{Linsley75}   \BY{Linsley J.}            	        	     \IN{Phys. Rev. Lett. } {34}{1975}{1530} 
\bibitem{BOOTSTRAP}   \BY{Efron B. and Tibshirani R. J.}   		     \IN{An introduction to the bootstrap method, Boca Raton, Chapman \& Hall} {}{1993}{}. 


\bibitem{DeJong11}   \BY{De Jong J. K.} for the MINOS collaboration,    \IN{Proc. 32nd Int. Cosmic Ray Conf. (Beijing)}{}{2011}.
\bibitem{Benzvi13}   \BY{Benzvi S., Fiorino D. and Sparks K.},          \IN{Proc. 33rd Int. Cosmic Ray Conf. (Rio De Janeiro)}{}{2013}.
\bibitem{J2000}      \BY{Kaplan G. H.},     				\IN{U.S. Naval Observatory Circular No. 163}{}{1982}.


\bibitem{gamma1} \BY{Adams F.C.} et al., 			  	\IN{Astrophys. J.}{491} {1997}{6}.
\bibitem{gamma2} \BY{Cohen A.G., Rujula A.D. and Glashow S.L.} 	\IN{Astrophys. J.} {495}{1998}{539}.
\bibitem{107}    \BY{Sakharov A.D.} 				  	\IN{JTEP Lett.} {5}{1967}{24}.
\bibitem{108}    \BY{Bambi C. and Dolgov A.D.}			  	\IN{Nuclear Phys. B} {784}{2007}{132}.
\bibitem{109}	 \BY{Antipov Y.M., Vishnevskii P.K. and Gorin Y.P.}   \IN{Sov. J. Nucl. Phys.} {13}{1971}{78}.
\bibitem{110}	 \BY{Agakishiev H.} et al. 				\IN{Nature} {473}{2011}{353}.
\bibitem{111}	 \BY{Allkofer O.C. and Brockhause D.}   		\IN{Astrophys. Space Science} {109}{1985}{145}.
\bibitem{112}	 \BY{Evenson P.} 					\IN{Astrophys. J.} {176}{1972}{797}.
\bibitem{113}	 \BY{Duperray R.} et al., 				\IN{Phys. Rev. D} {71}{2005}{083013}.
 
 
\bibitem{22}	 		\BY{Buffington A., Schindler S.M. and Pennypacker C.R.}   \IN{Astrophys. J.} {248}{1981}{1179}.
\bibitem{115}	 		\BY{Aizu H.} et al.					  \IN{Phys. Rev.} {121} {1961} {1206}.
\bibitem{116}	 		\BY{Smoot G.F., Buffington A. and Orth C.D.} 		  \IN{Phys. Rev. Lett.} {35}{1975}{258}.
\bibitem{117}	 		\BY{Badhwar G.D. and Golden R.L.}			  \IN{Nature} {274}{1978}{137}.
\bibitem{118}	 		\BY{Golden  R.L.} et al. 				  \IN{Astrophys. J.} {479}{1997}{992}.
\bibitem{119}	 		\BY{Sasaki  M.} et al.  				  \IN{Adv. Space Res.} {42}{2008}{450}.
\bibitem{120}	 		\BY{Abe K.}et al. 	 				  \IN{Phys. Rev. Lett.} {108}{2012}{131301}.
\bibitem{121}	 		\BY{Alcaraz J.} 	 				  \IN{Phys. Lett. B} {461}{1999}{387}. 
\bibitem{Mayorov_antihelium}    \BY{Mayorov A.G.} et al. 				  \IN{JETP Lett.} {93} {2011}{628}. 

\bibitem{1}	\BY{Witten E.}  		     		   \IN{Phys. Rev. D}{30}{1984} {272}.
\bibitem{2}	\BY{Madsen J.}  		     		   \IN{Phys. Rev. D} {71}{2005}{014026}.
\bibitem{3}	\BY{Madsen J.}  		     		   \IN{Lecture Notes in Physics} {516}{1999}{162}.
\bibitem{8}	\BY{Atreya A., Sarkar A. and Srivastava A. M.}     \IN{Phys. Rev. D} {90}{2014}{045010}.
\bibitem{sqm}   \BY{Adriani O.} et al. 		  	           \IN{Phys. Rev. Lett.} {115}{2015}{111101}.
  

\bibitem{Solar_Wind} 		 \BY{Parker E. N.} 		   \IN{Phys. Rev. Lett.} {} {1958}{}. 
\bibitem{GRL:GRL4201} 		 \BY{Jokipii J. R. and Kota J.}    \IN{Geophys. Res. Lett.} {16} {1989}{1}. 
\bibitem{munini:Parker1} 	 \BY{Parker E. N.}		   \IN{Planetary and Space Science} {30} {1965}{9}. 
\bibitem{Potgieter2014_LIS} 	 \BY{Potgieter M. S. }             \IN{Brazilian Journal of Physics} {44} {2014} {581}. 
\bibitem{lrsp-2013-3} 		 \BY{Potgieter M. S.}		   \IN{Living Reviews in Solar Physics}{10} {2013}. 
\bibitem{Bottino_2012}	         \BY{Bottino A., Fornengo N. and Scopel S.} 	        \IN{Phys. Rev. D} {85} {2012} {095013}. 
\bibitem{Cerde_o_2012} 		 \BY{Cerde{\~{n}}o D. G.,  Delahaye T. and Lavalle J. } \IN{Nucl. Phys. B} {854} {2012} {738}. 
\bibitem{Hooper_2015}		 \BY{Hooper D.,  Linden T. and Mertsch P. }		\IN{J. Cosmol. Astropart. Phys} {2015} {2015} {021}. 
\bibitem{proton_modulation}      \BY{Adriani O.} et al., 	   \IN{Astrophys. J.} {765} {2013}{91}. 
\bibitem{Potgieter_2014} 	 \BY{Potgieter M. S. } 	   \IN{Adv. Space Res.} {53} {2014} {1415}. 
\bibitem{0004-637X-810-2-142} 	 \BY{Adriani O.} et al., 	   \IN{Astrophys. J.} {810} {2015} {142}.
\bibitem{Munini}	 	 \BY{Munini R.} et al., 	   \IN{J.\ Phys.\ Conf.\ Ser.]}{632}{2015}{012073}. 
\bibitem{munini:deagostini} 	 \BY{D'Agostini G.} 		   \IN{arXiv:1010.0632v1 [physics.data-an]}{}{2010}{}. 
\bibitem{munini:phd} 		 \BY{Munini R.} 		   \IN{PhD thesis, Universit\'{a} degli Studi di Trieste, Italy} {}{2015} {}. 
\bibitem{0004-637X-810-2-141}    \BY{Potgieter M. S. } et al.,     \IN{Astrophys. J.} {810} {2015} {141}. 
\bibitem{Potgieter2014}          \BY{Potgieter M. S. } et al.,     \IN{Solar Physics} {289} {2014} {391}. 
\bibitem{munini:vos}             \BY{Vos  E. E.}                   \IN{Modelling charge-sign dependent modulation of cosmic rays in the heliosphere, PhD thesis, North-West University, Potchefstroom, South Africa}{} {2016} {}. 
\bibitem{Potgieter_2013}         \BY{Potgieter M. S. }             \IN{Liv. Rev. Sol. Phys.} {10} {2013}.
\bibitem{ep_ratio} 		 \BY{Di Felice V. , Munini R., Vos E. E. and Potgieter M. S.}      \IN{Astrophys. J.} {834} {2017} {89}. 
\bibitem{ep_ratio_modeling} 	 \BY{Potgieter M. S. and Vos E. E.}                                \IN{Astron. Astroph., in press} {}.   
\bibitem{Adriani_2016_elpos_mod} \BY{Adriani O.} et al.,  	                                   \IN{Phys. Rev. Lett. } {116} {2016} {241105}. 
\bibitem{reversal} 		 \BY{Sun X., Hoeksema J. T., Liu Y. and  Zhao J.}                  \IN{Phys. Rev. Lett. } {798} {2015} {114}. 

\bibitem{Mathews_1990} 		 \BY{Mathews T. and Venkatesan D.}\IN{Nature}{345} {1990} {600}.    
\bibitem{galina} 		 \BY{Bazilevskaya G. A.}\IN{Journal of Physics: Conference Series}{798} {2017} {012034}.    
\bibitem{Reames_2013}		 \BY{Reames D.~V.} 		  \IN{Space Science Reviews}    {175} {2013} {53}.        
\bibitem{Adriani_flare_2011}     \BY{Adriani O.} et al.,	  \IN{Astrophys. J.} {742} {2011}{102}.
\bibitem{NOAA} \IN{NOAA Space Environment Services Center}{} {http://umbra.nascom.nasa.gov/SEP/}
 
\bibitem{Vashenyuk_2006}  \BY{Vashenyuk E.~V. } et al.,  \IN{Ge\&Ae} {46} {2006} {L424}. 
\bibitem{McCracken_2008}  \BY{McCracken  K.~G. } et al., \IN{J. Geophys. Res.} {113} {2008} {12101}.
\bibitem{Adriani_ApjL_2015}      \BY{Adriani O.} et al., 	  \IN{Astrophys. J. Lett.} {801} {2015} {L3}. 
\bibitem{Earl_1976}       \BY{Earl J.~A.}, et al.,       \IN{Astrophys. J.}    {206} {1976} {301}. 
\bibitem{smart_shea_2000} \BY{Smart D.~F.} et al.,       \IN{Space Sci. Rev.} {93} {2000} {305}.      
\bibitem{IGRF}            \BY{Finlay C. C., Maus S. and Beggan, C.D.}  \IN{Geophysical Journal International}   {183} {2010} {12161230}.      
\bibitem{Tsyganenko_2007} \BY{Tsyganenko N.~A. } et al., \IN{J. Geophys. Res.}   {112} {2007} {6225}.  


\bibitem{ref:Bruno2016} 	\BY{Bruno A.} et al.,      	      \IN{Adv. Space Res.}{}{2016}{}, in press, doi: 10.1016/j.asr.2016.06.042. 
\bibitem{CRAND} 	        \BY{Singer S. F.}      	      \IN{Phys. Rev. Lett.}{1}{1958}{181}, 
\bibitem{ref:TrappedProt2015}   \BY{Adriani O.} et al.,		      \IN{Astrophys. J.}{L4}{2015}{799}.
\bibitem{ref:AP8} 		\BY{Sawyer D.~M. and Vette J.~I.}     \IN{NSSDC/WDC-A-R\&S}{76}{1976}{06}.
\bibitem{ref:PSB97}		\BY{Heynderickx D.} et al.,	      \IN{IEEE Trans. Nucl. Sci.}{46}{1999}{1475}.
\bibitem{ref:TrappedPbars2011}  \BY{Adriani O.} et al.,		      \IN{Astrophys. J.}{L29}{2011}{737}.
\bibitem{ref:Selesnick2007}	\BY{Selesnick R. S.} et al.,          \IN{Geophys. Res. Lett.}{34}{2007}{L20104}.
\bibitem{ref:Albedo2015}	\BY{Adriani O.} et al.,               \IN{J. Geophys. Res. Space Physics}{120}{2015}{5}.
\bibitem{AACGM}                 \BY{Baker K. B. and Wing, S.} 	      \IN{Geophys. Res.}{94}{1989}{9139}.

\bibitem{ref:TrappedPosit2016}  \BY{Mikhailov V.V.} et al.,	      \IN{J. Phys.: Conf. Ser.}{675}{2016}{032006}.
\bibitem{ref:AlbedoEle2009}	\BY{Adriani O.} et al.,               \IN{J. Geophys. Res. Space Physics}{114}{2009}{A12218}.

\bibitem{ref:Ams01}		\BY{Alcaraz J.} et al.,	              \IN{Phys. Lett. B}{484}{2000}{10}.

\bibitem{ref:Deuteron2015}	\BY{Koldobskiy S.~A.} et al.,	      \IN{Nuclear and Particle Physics Proceedings}{273-275}{2016}{2345}.
\bibitem{ref:derome2001}	\BY{Derome L. and Buenerd, M.}        \IN{Phys. Lett. B}{521}{2001}{139}
\bibitem{ref:lamanna2001}	\BY{Lamanna G.} et al.  	      \IN{Proc. of the 27th Intern. Cosmic Ray Conf.}{2001}{1614}.
\bibitem{ref:Gstorm2016}	\BY{Adriani O.} et al.,		      \IN{Space Weather}{14(3)}{2016}{210}.

\end{thebibliography}
\end{document}